\RequirePackage{rotating} 

\documentclass[fleqn,usenatbib]{mnras}
\pdfoutput=1
\usepackage{newtxtext,newtxmath}
\usepackage[T1]{fontenc}
\usepackage{ae,aecompl}
\usepackage{rotating} 
\usepackage{float}

\usepackage{graphicx}	
\usepackage{amsmath}	
\usepackage{pifont}
\usepackage{xspace}
\usepackage{subcaption} 
\usepackage{comment}
\usepackage{todonotes}
\usepackage{longtable} 
\usepackage{pdflscape} 
\usepackage{scalerel}
\usepackage{tikz}
\usetikzlibrary{svg.path}

\definecolor{orcidlogocol}{HTML}{A6CE39}
\tikzset{
  orcidlogo/.pic={
    \fill[orcidlogocol] svg{M256,128c0,70.7-57.3,128-128,128C57.3,256,0,198.7,0,128C0,57.3,57.3,0,128,0C198.7,0,256,57.3,256,128z};
    \fill[white] svg{M86.3,186.2H70.9V79.1h15.4v48.4V186.2z}
                 svg{M108.9,79.1h41.6c39.6,0,57,28.3,57,53.6c0,27.5-21.5,53.6-56.8,53.6h-41.8V79.1z M124.3,172.4h24.5c34.9,0,42.9-26.5,42.9-39.7c0-21.5-13.7-39.7-43.7-39.7h-23.7V172.4z}
                 svg{M88.7,56.8c0,5.5-4.5,10.1-10.1,10.1c-5.6,0-10.1-4.6-10.1-10.1c0-5.6,4.5-10.1,10.1-10.1C84.2,46.7,88.7,51.3,88.7,56.8z};
  }
}

\newcommand\orcid[1]{\href{https://orcid.org/#1}{\mbox{\scalerel*{
\begin{tikzpicture}[yscale=-1,transform shape]
\pic{orcidlogo};
\end{tikzpicture}
}{|}}}}

\usepackage{hyperref} 

\graphicspath{{figures/}}

\newcommand{\kms}{\mbox{km\,s$^{-1}$}\xspace}

\newcommand{\msun}{$M_{\odot}$\xspace}

\newcommand{\rsun}{$R_{\odot}$\xspace}

\newcommand{\mjup}{$M_{\rm JUP}$\xspace}

\newcommand{\mearth}{$M_\oplus$\xspace}
\newcommand{\rearth}{$R_\oplus$\xspace}

\newcommand{\feh}{\ensuremath{[\mbox{Fe}/\mbox{H}]}\xspace}

\newcommand{\Rmnum}[1]{\expandafter\@slowromancap\romannumeral #1@}

\newcommand{\isochrones}{\texttt{isochrones}\xspace}

\newcommand{\pytransit}{\texttt{PyTransit}\xspace}
\newcommand{\vespa}{\texttt{vespa}\xspace}
\newcommand{\batman}{\texttt{batman}\xspace}
\newcommand{\emcee}{\texttt{emcee}\xspace}
\newcommand{\lmfit}{\texttt{lmfit}\xspace}

\newcommand{\everest}{\texttt{EVEREST}\xspace}
\newcommand{\ktwosff}{\texttt{K2SFF}\xspace}
\newcommand{\ktwophot}{\texttt{K2phot}\xspace}
\newcommand{\spock}{\texttt{spock}\xspace}

\newcommand{\pyTTV}{\texttt{PyTTV}\xspace}
\newcommand{\celerite}{\texttt{celerite}\xspace}

\newcommand{\tls}{\texttt{transit-least-squares}\xspace}
\newcommand{\isoclassify}{\texttt{isoclassify}\xspace}

\newcommand{\mstar}{\ensuremath{M_{\star}}\xspace}
\newcommand{\rstar}{\ensuremath{R_{\star}}\xspace}

\newcommand{\teff}{\ensuremath{T_{\mathrm{eff}}}\xspace}  
\newcommand{\logg}{\ensuremath{\log g}\xspace}

\newcommand{\rhostar}{\ensuremath{\rho_{\star}}\xspace}

\newcommand{\Mp}{\ensuremath{M_{P}}\xspace} 
\newcommand{\Prot}{$P_{\rm{rot}}$\xspace}
\newcommand{\Rp}{$R_\mathrm{P}$\xspace}
\newcommand{\Porb}{$P_\mathrm{orb}$\xspace}
\newcommand{\RpRs}{$R_p/R_s$\xspace}
\newcommand{\aRs}{$a/R_s$\xspace}
\newcommand{\To}{$T_0$\xspace}
\newcommand{\imppar}{$b$\xspace}
\newcommand{\Teq}{$T_{\mathrm{eq}}$\xspace}

\newcommand{\kmax}{\RpRs$_{\mathrm{max}}$\xspace}

\newcommand{\kepler}{\textit{Kepler}\xspace} 
\newcommand{\ktwo}{\textit{K2}\xspace}

\newcommand{\tess}{\textit{TESS}\xspace}
\newcommand{\gaia}{\textit{Gaia}\xspace}

\newcommand{\poss}{\textit{POSS-1}\xspace}
\newcommand{\panstarrs}{\textit{PanSTARRS-1}\xspace}

\newcommand{\maxrad}{\texttt{maxrad}\xspace}
\newcommand{\secthresh}{\texttt{secthresh}\xspace}

\newcommand{\numstars}{68\xspace}
\newcommand{\numvp}{37\xspace}
\newcommand{\numvphost}{29\xspace}
\newcommand{\numpc}{28\xspace}

\newcommand{\numoldpc}{34\xspace}
\newcommand{\numnewpc}{3\xspace}
\newcommand{\numfp}{8\xspace}
\newcommand{\nummulti}{6\xspace}

\newcommand{\nummulticamp}{48\xspace}

\newcommand{\numtargetswithspeckle}{29\xspace} 
\newcommand{\numtargetswithspecklecompanions}{7\xspace} 
\newcommand{\numtargetswithspec}{58\xspace} %

\newcommand{\numprot}{42\xspace}
\newcommand{\minProt}{1.61\xspace}
\newcommand{\maxProt}{31.7\xspace}
\newcommand{\numprotwithplanets}{9\xspace}

\newcommand{\numstarsdave}{2\xspace}

\newcommand{\minP}{1.99\xspace}
\newcommand{\maxP}{52.71\xspace} 

\newcommand{\medianRp}{2.2~\rearth}

\newcommand{\numstarswithinaperture}{16\xspace}

\newcommand{\numstarspotentialNEBs}{9\xspace}

\title[\ktwo Cnc/Vir Planets]{\numvp New Validated Planets in Overlapping \ktwo Campaigns}

\author[J.~P.~de~Leon et al.]{J.~P.~de~Leon\orcid{0000-0002-6424-3410},$^{1}$\thanks{E-mail: jpdeleon@g.ecc.u-tokyo.ac.jp}
J.~Livingston\orcid{0000-0002-4881-3620},$^{1}$
M.~Endl,$^{2}$
W.~D.~Cochran\orcid{0000-0001-9662-3496},$^{2,3}$
T.~Hirano\orcid{0000-0003-3618-7535},$^{4}$
\newauthor
R.~A.~Garc\'\i a\orcid{0000-0002-8854-3776},$^{5}$
S.~Mathur\orcid{0000-0002-0129-0316},$^{6,7}$
K.~W.~F.~Lam\orcid{0000-0002-9910-6088},$^{8}$
J.~Korth\orcid{0000-0002-0076-6239},$^{9}$
A.~A.~Trani\orcid{0000-0001-5371-3432},$^{10}$
\newauthor
F.~Dai\orcid{0000-0002-8958-0683},$^{11}$
E.~D{\'\i}ez~Alonso\orcid{0000-0002-5826-9892},$^{12}$
A.~Castro-Gonz{\'a}lez\orcid{0000-0001-7439-3618},$^{12}$
M.~Fridlund\orcid{0000-0002-0855-8426},$^{14}$
\newauthor
A.~Fukui\orcid{0000-0002-4909-5763},$^{15,6}$
D.~Gandolfi\orcid{0000-0001-8627-9628},$^{16}$
P.~Kabath\orcid{0000-0002-1623-5352},$^{17}$
M.~Kuzuhara,$^{18}$
R.~Luque,$^{6,7}$
\newauthor
A.B.~Savel\orcid{0000-0002-2454-768X},$^{19}$
H.~Gill\orcid{0000-0001-6171-7951},$^{20}$
C.~Dressing\orcid{0000-0001-8189-0233},$^{20}$
S.~Giacalone\orcid{0000-0002-8965-3969},$^{20}$
N.~Narita\orcid{0000-0001-8511-2981},$^{15,6,21,22}$
\newauthor
E.~Palle\orcid{0000-0003-0987-1593},$^{6,7}$
V.~Van~Eylen\orcid{0000-0001-5542-8870},$^{23}$
and M.~Tamura\orcid{0000-0002-6510-0681}$^{1,21,24}$
\\
{\textit Affiliations are listed at the end of the paper}
}

\date{Accepted 2021 July 20. Received 2021 July 16; in original form 2020 October 15}

\pubyear{2021}

\begin{document}
\label{firstpage}
\pagerange{\pageref{firstpage}--\pageref{lastpage}}

\maketitle

\begin{abstract}
    We analysed \numstars candidate planetary systems first identified during Campaigns 5 and 6 (C5 and C6) of the NASA \ktwo mission. We set out to validate these systems by using a suite of follow-up observations, including adaptive optics, speckle imaging, and reconnaissance spectroscopy. The overlap between C5 with C16 and C18, and C6 with C17, yields lightcurves with long baselines that allow us to measure the transit ephemeris very precisely, revisit single transit candidates identified in earlier campaigns, and search for additional transiting planets with longer periods not detectable in previous works. Using \vespa, we compute false positive probabilities of less than 1\% for \numvp candidates orbiting \numvphost unique host stars and hence statistically validate them as planets. These planets have a typical size of \medianRp and orbital periods between \minP and \maxP days. We highlight interesting systems including a sub-Neptune with the longest period detected by \ktwo, sub-Saturns around F stars, several multi-planetary systems in a variety of architectures. These results show that a wealth of planetary systems still remains in the \ktwo data, some of which can be validated using minimal follow-up observations and taking advantage of analyses presented in previous catalogs.
    \end{abstract}
    
\begin{keywords}
exoplanets -- transits -- observations -- spectroscopy -- imaging
\end{keywords}


\section{Introduction}

    The \kepler \citep{2010BoruckiKepler} and \ktwo \citep{2014HowellK2} missions have brought many exciting exoplanet discoveries that yield new insights into the occurrence rate, formation and evolution of planets. 
    This success was driven primarily by the sustained efforts to homogeneously analyse ensembles of lightcurves to detect new candidate systems and consequently statistically validate or confirm their planetary nature
    aided by follow-up data. Here, "validation" is different from "confirmation", wherein the former means that there is overwhelming evidence that the transits must be explained by a planet, through elimination of all false positive scenarios, whereas the latter involves the determination that the planet's mass is in the substellar regime (\Mp$\lesssim$13\mjup). 
    Confirmation via radial velocity (RV) mass measurements have been conducted for planets around bright stars \citep[e.g., ][]{2017DaiK2-131b, 2017FridlundK2111, 2020LiloBoxRV} but it is usually impractical for faint or magnetically active stars, and it is prohibitively expensive for a large number of planet candidates detected by dedicated space missions, such as \kepler, \ktwo, and \tess \citep{2014RickerTESS}.

    A series of papers have so far presented catalogs of planet candidates and/or statistically validated planets in the following \ktwo campaigns: \citet{2015MontetC1} in Campaign 1 (C1), \citet{2016VanderburgC0to3} in C0-C3, \citet{2016AdamsC0to5USP} and \citet{2016CrossfieldC0to4} in C0-C4, \citet{2020ZinkC5} and \citet{2016NardielloC5} in C5, \citet{2017DressingC1to7planets} in C1-C7, \citet{2018PetiguraC5to8} and \citet{2018Livingston60planets} in C5-C8, \citet{2019KruseC0to8} in C0-C8, \citet{2018MayoC0to10} in C0-C10 except C9, \citet{2018Livingston44planets} in C10, \citet{2020WittenmyerC1to13} in C1-C13, \citet{2019ZinkZooniverse} in C0-C14 except C9 \& C11, \citet{2020CastroGonzalezC12to15} in C12-C15, \citet{2019DressingC1to17} in C1-C17, \citet{2018CrossfieldC17} in C17, and \citet{2018HiranoMdwarfs}, \citet{2019KostovDAVE}, \citet{2019DattiloDL}, and \citet{2019HellerTLS2} spanning almost the entire \ktwo mission (C0-C18). 
    Altogether these catalogs have increased the cumulative number of validated planets and planet candidates, especially those with sizes <4~\rearth orbiting stars relatively brighter than \kepler host stars. 
    Despite the overwhelming number of planets found 
    in the last few years, hundreds more remain to be discovered in the \ktwo mission alone. Based on the original \kepler catalog, \citet{2019DotsonK2candidates} predicted 1317$\pm$261 detectable exoplanets in the \ktwo data set but only 1/3 of this prediction are validated or confirmed \ktwo planets\footnote{426 at time of writing: \url{https://exoplanetarchive.ipac.caltech.edu/docs/counts_detail.html}}.
    Moreover, there are currently more than 800 planet candidates from the \ktwo mission alone that still await validation.
    
    Here we present follow-up observations of \numstars host stars 
    including reconnaissance spectroscopy and high-resolution imaging to measure host stars' properties and identify nearby stars; both of which are helpful in identification and ruling out false positive scenarios. The majority of these host stars were first observed by \ktwo either during Campaign 5 (hereafter C5) or C6, after which transiting candidates were reported by \citet{2016PopeC5to6}, \citet{2019DressingC1to17}, \citet{2018MayoC0to10}, and \citet{2019KruseC0to8}. Fifty stars were observed again in the succeeding overlapping \ktwo campaigns whereas eighteen stars were only observed in a single \ktwo Campaign.  Because C5 overlaps with C16 \& C18, and C6 overlaps with C17, this provides lightcurves with baselines as long as 3 years (i.e. for C5 \& C18; see Table~\ref{tab:k2_campaigns}). This allows us to measure the transit ephemeris very precisely, revisit single transit candidates identified in earlier campaigns, and search for additional transiting planets with longer periods leveraging multiple \ktwo campaigns for the first time. 
    We validated \numvp planet candidates, \numoldpc of which were also detected in previous catalogs and \numnewpc are new detections. We also measured rotation periods for \numprot stars in our sample, and searched for additional planets via transit timing variations. 
    This research was done as part of the KESPRINT collaboration\footnote{\url{https:www.iac.es/proyecto/kesprint}}, which has so far primarily focused on the detection of planets for the purpose of characterizing interesting individual systems in detail \citep[e.g.,][]{2016HiranoK2-34, 2018BarraganK2-141, 2018VanEylenHD89345, 2019Korth}; in this paper we present follow-up observations and statistical validation results for a large number of planet candidates found as part of this process, similar to \cite{2018Livingston44planets}.
    
    The paper is structured as follows: in \S\ref{sec:obs}, we present the observations and ancillary data we analyzed in this work, comprising \ktwo photometry, reconnaissance spectroscopy, adaptive optics, and speckle imaging. In \S\ref{sec:analysis} we describe our transit search for new candidates, characterization of the host stars, transit modeling, planet validation, stability analysis of multi-planet systems, and search for transit timing variations. In \S\ref{sec:results} we present the results of our analyses and discuss individually interesting systems, and we conclude with a summary in \S\ref{sec:summary}.

\section{Data and observations} \label{sec:obs}
    
    \subsection{\ktwo photometry} \label{sec:phot}
    
        The \ktwo mission observed a series of patches of the sky with an area of 100 square degrees along the ecliptic called "campaigns", lasting up to 83 days each. In a typical \ktwo campaign, the number of targets ranges in the tens of thousands with long-cadence observations (30-minute integration), and a few hundred with short-cadence observations (1-minute integration). There are occasional overlaps between campaign fields, especially between C5, C16, and C18, located in Cancer constellation as well as C6 and C17, located in Virgo. Table~\ref{tab:k2_campaigns} summarizes the start and end dates of observations and typical coordinate positions of each campaign. All of the \numstars stars in our sample were first observed in C5 and C6 in the long cadence mode. Of the \numstars stars in our sample, \nummulticamp stars were observed again in the succeeding \ktwo campaigns. 
        The \ktwo campaigns used for each target are listed in Table~\ref{tab:obs}.
    
    \begin{table}
        \centering
        \caption{\ktwo observations in Cancer and Virgo.}
\begin{tabular}{ccccc}
    \hline
    Campaign & Start & End & RA & Dec. \\
    \hline
    Cancer	& & & & \\
    5	& 2015 Apr 27 &	2015 Jul 10 & 08:40:38	& 16:49:47 \\
    16  & 2017 Dec 07 &	2018 Feb 25	& 08:54:50	& 18:31:31	\\
    18	& 2018 May 12 &	2018 Jul 02	& 08:40:39	& 16:49:40 \\
    Virgo	& & & & \\
    6	& 2015 Jul 14 & 2015 Sep 30	& 13:39:28	& -11:17:43	\\
    17  & 2018 Mar 01 &	2018 May 08	& 13:30:12	& -07:43:16	\\
    \hline
\end{tabular}
        \label{tab:k2_campaigns}
    \end{table}
    
    \begin{table}
        \scriptsize
        \caption{Summary of \ktwo and follow-up data used in this work. \ktwo/C=campaign; AO=Adaptive optics; SI=Speckle imaging; RS=Reconnaissance spectroscopy; D17=Dressing et al. 2017; P18=Petigura et al. 2018; M18=Mayo et al. 2018; Mat18=Matson et al. 2018}
\begin{tabular}{ccccc}
\hline
EPIC & K2/C & AO & SI & RS \\
\hline
211314705 & 5 & This work & This work & \\ 
211335816 & 5/18 & & & This work \\ 
211336288 & 5/18 & & & \\ 
211357309 & 5/18 & & & This work \\
211383821 & 5/18 & & & D17 \\
211399359 & 5/18 & D19 & & P18 \\ 
211401787 & 5/18 & & M18 & This work \\ 
211413752 & 5/16/18 & & This work & P18 \\ 
211439059 & 5/18 & & This work & P18 \\ 
211490999 & 5/16/18 & & This work & P18 \\ 

211502222 & 16 & & & This work \\ 
211578235 & 5/18 & This work & & P18 \\ 
211579112 & 5/18 & This work & This work & \\ 
211611158 & 5/16/18 & & & M18 \\ 
211645912 & 5/18 & & This work & P18 \\
211647930 & 16 & & & This work \\ 
211694226 & 5/16/18 & & & D17 \\ 
211730024 & 16 & & & This work \\ 
211731298 & 5/18 & & & \\ 
211743874 & 5/18 & & This work & P18 \\ 

211763214 & 5/16/18 & This work & This work & M18 \\ 
211762841 & 5/18 & & & D17 \\
211770696 & 5/18 & & M18 & P18 \\ 
211797637 & 5/18 & & & \\ 
211799258 & 5/18 & D17 & & D17 \\ 
211779390 & 5/18 & & & This work \\ 
211800191 & 5/18 & & M18 & This work \\ 
211817229 & 5/18 & & & D17 \\ 
211843564 & 5/18 & & & \\ 
211923431 & 5/18 & This work & & \\ 

211939692 & 5/18 & & This work & \\ 
211965883 & 5 & This work & This work & D17 \\
211978988 & 5/18 & & & M18 \\  
211987231 & 5 & & Mat18 & This work \\ 
211995398 & 5/18 & & & \\ 
211997641 & 5/16 & This work & & \\ 
212006318 & 5/18 & & This work & P18 \\ 
212009150 & 5/16/18 & & & \\ 
212040382 & 16 & This work & & This work \\ 
212041476 & 16 & & & This work \\ 

212058012 & 16 & & & This work \\ 
212072539 & 5/16/18 & & & \\ 
212081533 & 16 & This work & & This work \\ 
212088059 & 5/16/18 & This work & This work & D17 \\ 
212099230 & 5/16/18 & & Mat18 & This work \\ 
212132195 & 5/18 & This work & This work & M18 \\ 
212161956 & 5/18 & & This work & \\ 
212178066 & 16/18 & & & This work \\ 
212204403 & 16 & & & This work \\ 
212278644 & 6 & & & This work \\ 

212297394 & 6 & & This work & \\ 
212420823 & 6 & & This work & This work \\ 
212428509 & 6/17 & & This work & P18 \\ 
212435047 & 6/17 & & This work & P18 \\
212440430 & 6/17 & & This work & This work \\ 
212495601 & 6/17 & & This work & This work \\ 
212543933 & 6 & & This work & This work \\ 
212570977 & 6/17 & & This work & P18 \\
212563850 & 6/17 & & This work & This work \\ 
212587672 & 6/17 & & M18 & P18 \\ 

212628098 & 6/17 & & Mat18 & D17 \\ 
212628477 & 17 & & This work & This work \\ 
212634172 & 6/17 & & & D17 \\
212661144 & 6/17 & & This work & \\ 
212639319 & 6/17 & & M18 & P18 \\ 
212690867 & 6/17 & & & D17 \\
212797028 & 6/17 & & This work & P18 \\ 
251319382 & 16 & & This work & This work \\ 
251554286 & 17 & & This work & This work \\ 
\hline
\end{tabular}
        \label{tab:obs}
    \end{table}
    
    \subsection{\gaia DR2 photometry and astrometry}
        The presence of multiple unresolved stars in photometric observations of a transiting planetary system biases measurements of the planet's radius, mass, and atmospheric conditions \citep[e.g., ][]{2016SouthworthEvans}.
        For all our targets, we leverage \gaia Data Release 2 \citep[DR2, ][]{2019GaiaDR2} to search for direct and indirect evidence of potential contaminating sources in \ktwo observations. In our sample, we found \gaia DR2 sources separated from the target as close as $1\arcsec$. 
        \gaia DR2 can also be useful to look for hints of binarity. \citet{2018EvansGOF} proposed that systems with large Astrometric Goodness of Fit of the astrometric solution for the source in the Along-Scan direction ($\verb|GOF_AL|>20$) and Astrometric Excess Noise  significance ($\verb|D|>5$)\footnote{For details, see: \url{https://gea.esac.esa.int/archive/documentation/GDR2/Gaia_archive/chap_datamodel/sec_dm_main_tables/ssec_dm_gaia_source.html}} are plausibly poorly-resolved binaries. 
        We added these values in the last two columns in Table~\ref{tab:star}. 
        Stars that are exceptionally bright or have high proper motion are proposed to explain the large offset causing difficulties in modelling saturated or fast-moving stars, rather than unresolved binarity. We do not see this to be a concern however since EPIC~212178066, the brightest star (V=6.9 mag) in our sample with proper motions ($\mu_{\alpha}$,$\mu_{\delta}$)=(-47.30$\pm$0.06, -148.78$\pm$0.05) 
        has $\verb|GOF_AL|$=10.59 and $\verb|D|$=0.00.
        Such values are well below the aforementioned empirically-motivated cutoffs chosen for plausible unresolved binaries in \gaia DR2.

    \subsection{Speckle and AO imaging}  \label{sec:speckle}
    
        Adaptive optics (AO) and speckle imaging (SI) are useful to determine if any fainter point source exists closer to the target inside of \gaia's point-source detection limits. 
        We observed several of our targets with the NASA Exoplanet Star and Speckle Imager (NESSI) on the 3.5-m WIYN telescope at the Kitt Peak National Observatory. NESSI is an instrument that uses high-speed electron-multiplying CCDs (EMCCDs) to capture sequences of 40 ms exposures simultaneously in two bands \citep{2018ScottNESSI}. Data were collected following the procedures described by \citet{2011Howell}. We conducted all observations simultaneously in a 'blue' band centered at 562 nm with a width of 44 nm, and a 'red' band centered at 832 nm with a width of 40 nm. In total, 66 speckle images were collected for a distinct sample of \numtargetswithspeckle stars in our targets. These observations were made in 2016 October through November and 2017 March through May. All of our speckle imaging data are publicly available via the community portal ExoFOP\footnote{\url{https://exofop.ipac.caltech.edu/k2/}}. 
    
        We observed EPIC~211314705, EPIC~211579112, EPIC~211923431, EPIC~212040382, EPIC~211439059, and EPIC~211763214 using Infrared Camera and Spectrograph \citep[IRCS; ][]{2000KobayashiIRCS, 2008HayanoAO188} on the 8.2-m Subaru Telescope at the Mauna Kea Observatory to rule out false positives caused by an eclipsing binary as well as to search for potential (sub-)stellar companions within a few arcseconds from the target. For five stars in our sample (211314705, 211763214, 211965883, 212088059, 212132195), we obtained $H$-band images on UT 2016 November 6, which were reduced following the standard procedure described in \citet{2016HiranoK2-34}. EPIC~211579112 was very faint and its $R$-magnitude is close to the border magnitude for AO to work properly; we decided to take $K^\prime$-band images of this target, for which the natural seeing size is slightly better than in the $H$-band. 
        
        \begin{figure*}
            \includegraphics[clip,trim={0 0 0 0},width=0.8\textwidth]{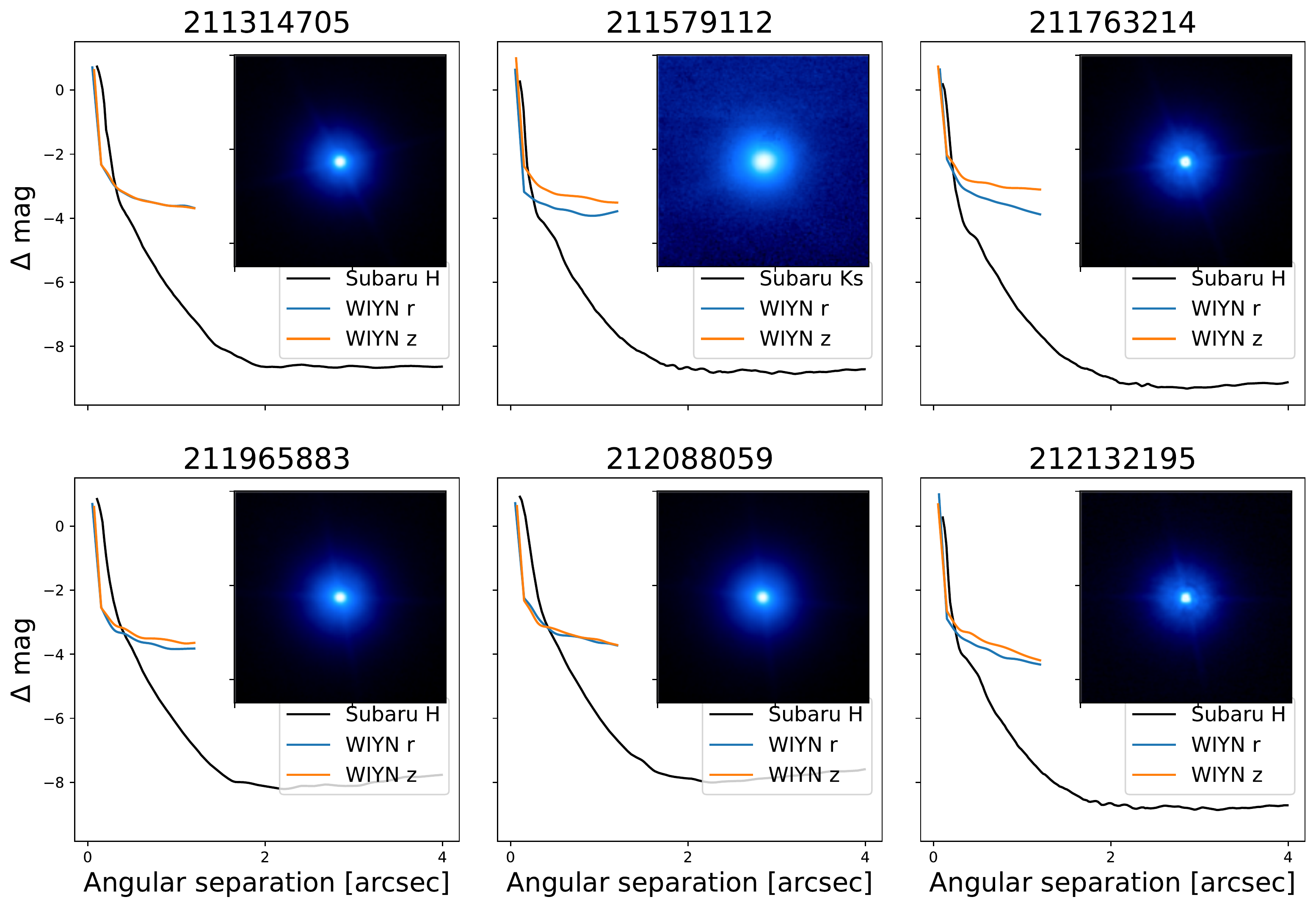}
            \caption{5-$\sigma$ contrast curves (black) extracted from the reduced Subaru/IRCS AO images (inset). For comparison, we also show the speckle imaging contrast curves from WIYN/NESSI; the complementary nature of the two techniques is especially pronounced in the case of EPIC~211579112. No companions were detected within 4$^{\prime\prime}$ down to a typical contrast of 8 magnitudes, and no bright close binary was seen with a resolution of 0.05$^{\prime\prime}$.}
            \label{fig:cc}
        \end{figure*}
        
        \begin{figure*}
            \begin{tabular}{ll}
                \includegraphics[clip,trim={10 30 0 10},width=0.7\columnwidth]{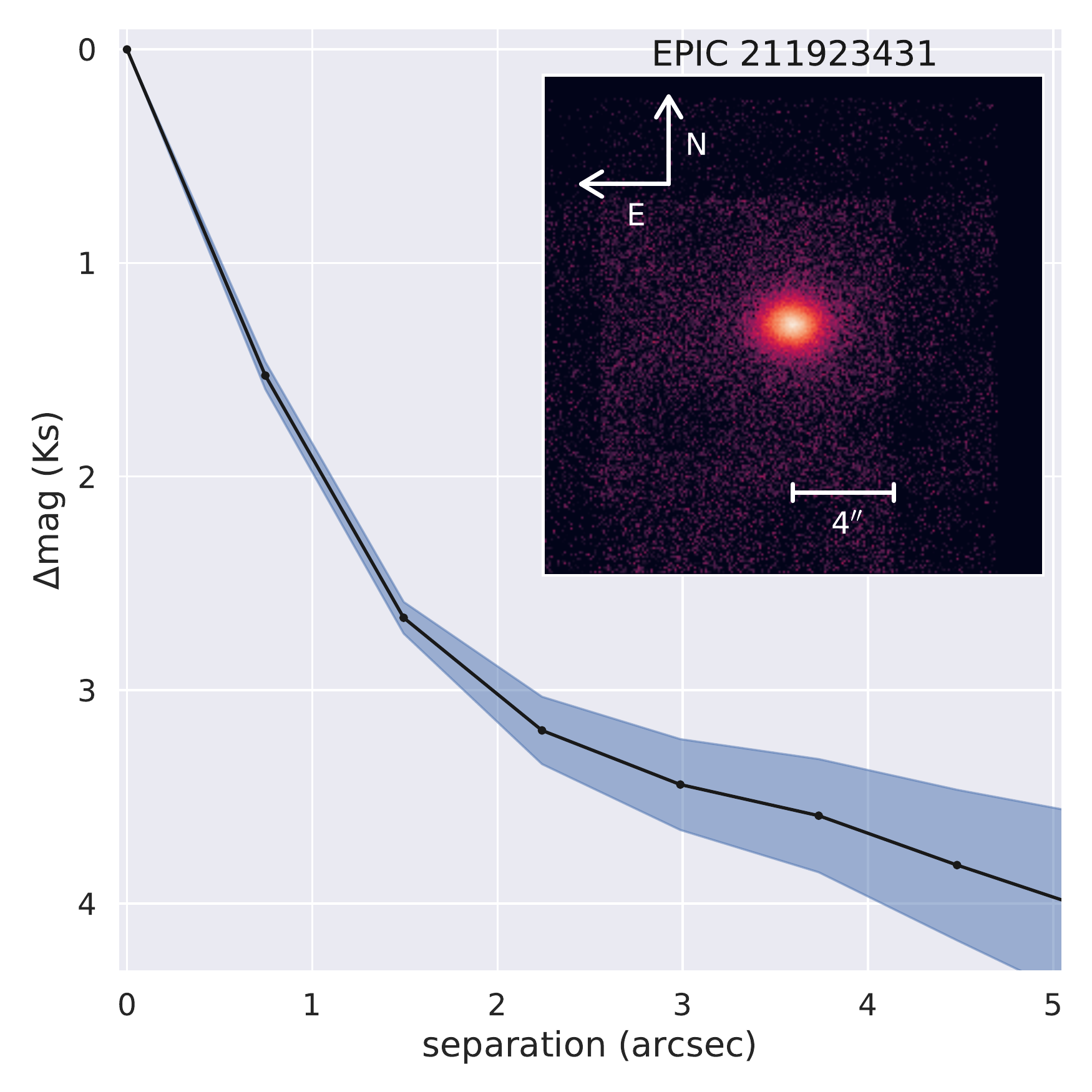}
                &
                \includegraphics[clip,trim={10 30 0 10},width=0.7\columnwidth]{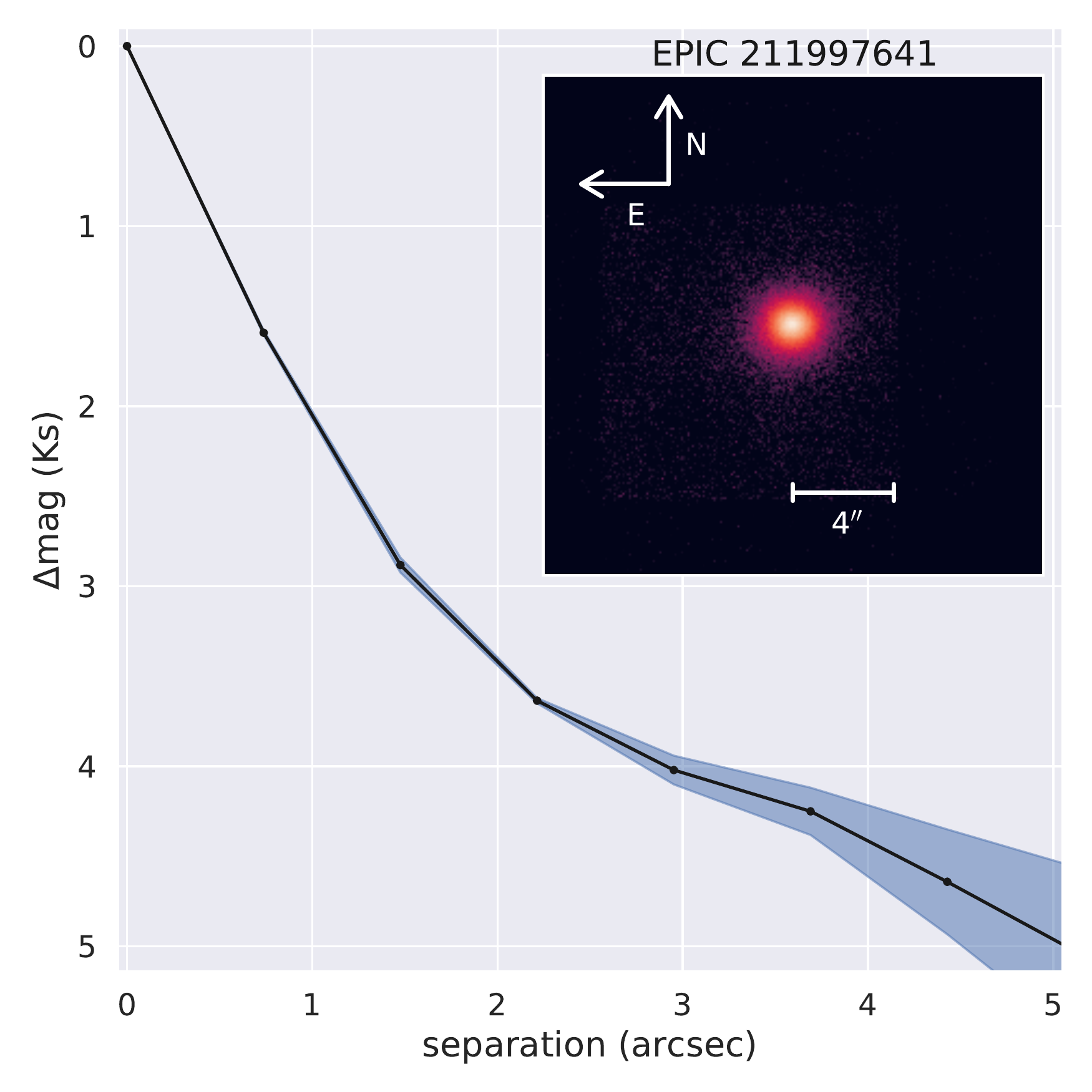} \\
                \includegraphics[clip,trim={10 0 0 10},width=0.7\columnwidth]{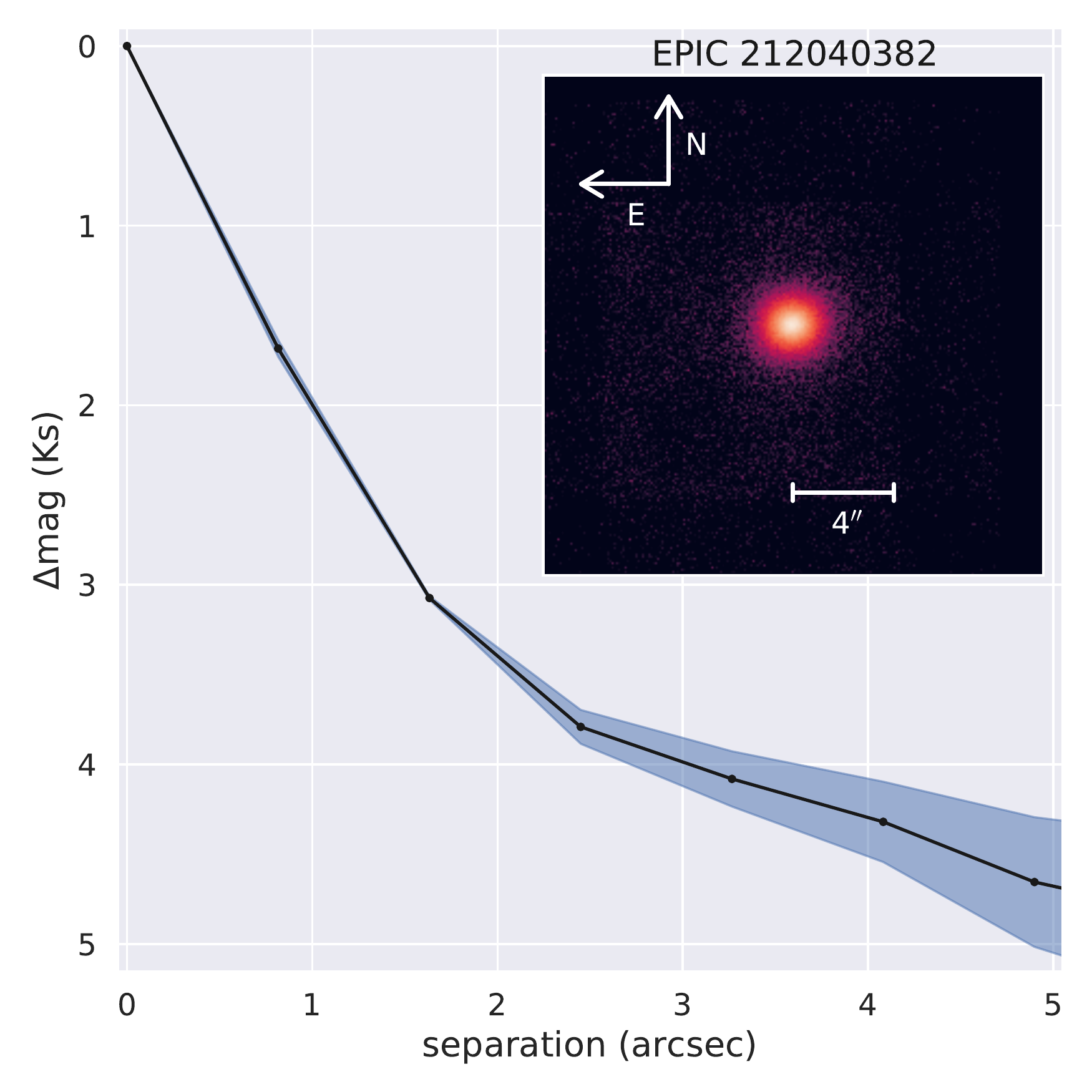}
                &
                \includegraphics[clip,trim={10 0 0 10},width=0.7\columnwidth]{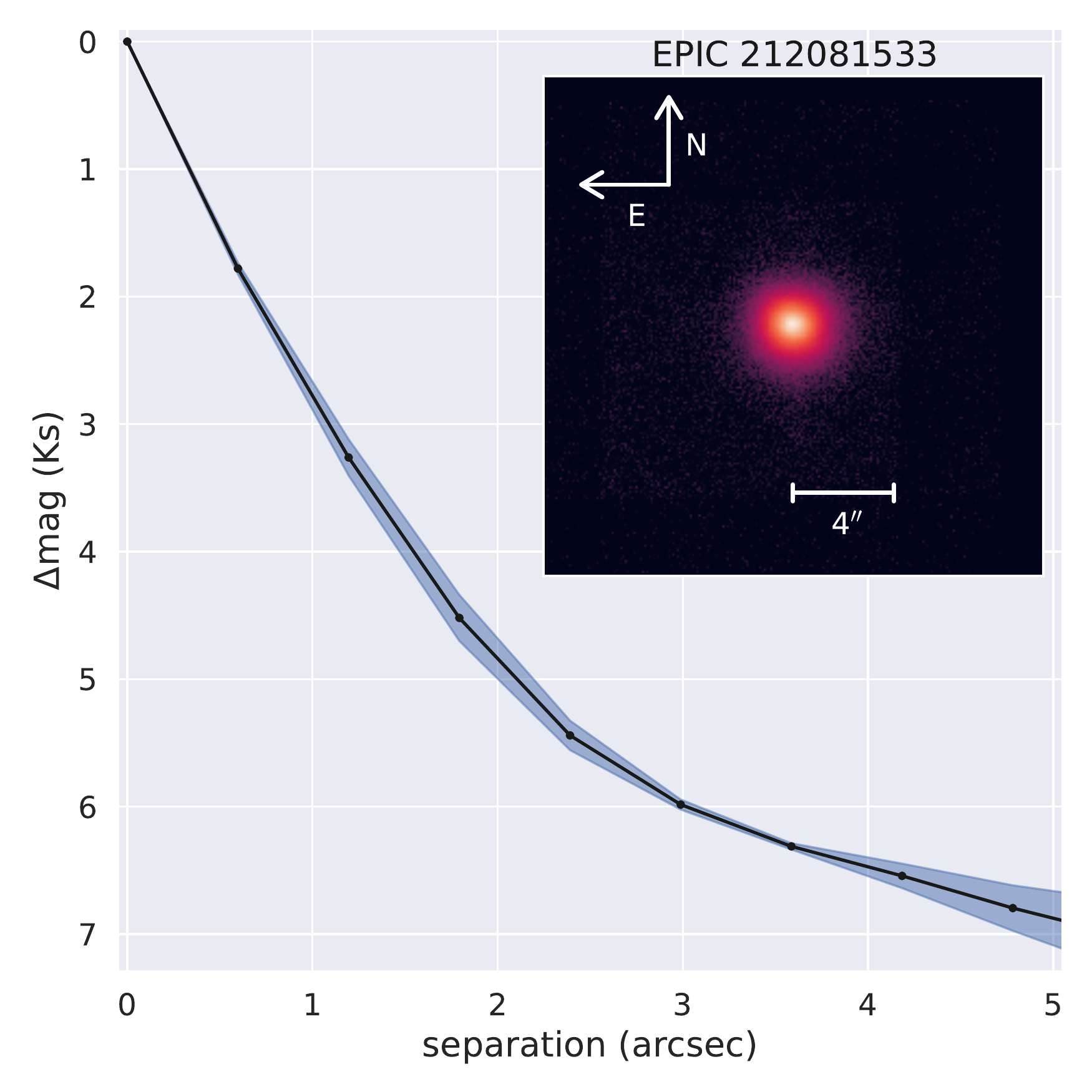} \\
            \end{tabular}
            \caption{Adaptive optics images of EPIC~211923431, EPIC~211997641, EPIC~212040382, and EPIC~212081533 taken with the ShARCS camera on the Shane 3-meter telescope at Lick Observatory. For each image, we also present a contrast curve generated by calculating the median values (solid lines) and root-mean-square errors (blue, shaded regions) in annuli centered on each target, where the bin width of each annulus is equal to the full width at half max of the point spread function.}
            \label{fig:ao2}
        \end{figure*}
        
        The reduced 
        Subaru/IRCS AO images are shown as inset in Figure~\ref{fig:cc} together with their corresponding contrast curves.
        Smooth contrast curves were produced from the reconstructed images by fitting a cubic spline to the 5-$\sigma$ sensitivity limits within a series of concentric annuli. 
        Also shown are the contrast curves from speckle-interferometric images taken with WIYN/NESSI. 
        The AO and speckle images and their corresponding contrast curves in Figure~\ref{fig:cc} illustrate that no companions were detected within a radius of $4\arcsec$ down to a contrast level of 8 magnitudes, and no bright close binary was seen with a resolution of $0.1\arcsec$. These observations sharply reduce the possibility that an unresolved background star is the source of the transits.
        There are \numtargetswithspecklecompanions stars in our sample however that have companions detected in speckle images. We report the separation $r$ and the magnitude difference between the brighter and the fainter star in \kepler band, $\Delta K_{\mathrm{p}}$ in Table~\ref{tab:gamma}.
        The contrast curves are also used as additional constraints for false positive calculation in \S\ref{sec:fpp}.
        
        We also observed EPIC~211923431, EPIC~211997641, EPIC~212040382, and EPIC~212081533 on UT 2019 January 24 using the ShARCS camera on the Shane 3-meter telescope at Lick Observatory \citep{2012Kupke, 2014Gavel, 2014McGurk}. Observations were taken with the Shane adaptive optics system in natural guide star and laser guide star modes (See \cite{2020Savel} for a detailed description of the observing strategy and reduction procedure). We collected all of our observations using a $Ks$ filter ($\lambda_0 = 2.15$ $\mu$m, $\Delta \lambda = 0.32$ $\mu$m). The AO images and their contrast curves are shown in Figure~\ref{fig:ao2}. We find no nearby stellar companions within our detection limits.

    \subsection{Reconnaissance spectroscopy} \label{sec:spec}
    
        Medium to high-resolution spectra enable precise physical characterization of the star and therefore planet. For \numtargetswithspec stars in our sample, we obtained over the course of four years (2015-2019) high-resolution spectra with the Tull Coud\'e cross-dispersed echelle spectrograph \citep{1995Tull} at the Harlan J. Smith 2.7-m telescope at the McDonald Observatory. Observations were conducted with the $1.2\arcsec \times 8.2\arcsec$ slit, yielding a resolving power of $R\sim$60, 000. The spectra cover 375--1020 nm, with increasingly larger inter-order gaps long ward of 570 nm. For each target star, we obtained three successive short exposures in order to allow removal of energetic particle hits on the CCD detector. We used an exposure meter to obtain an accurate flux-weighted barycentric correction and to give an exposure length that resulted in a signal-to-noise ratio (SNR) of about 30 per pixel. Bracketing exposures of a Th-Ar hollow cathode lamp were obtained in order to generate a wavelength calibration and to remove spectrograph drifts. This enabled calculation of absolute radial velocities from the spectra. 
        We traced the apertures for each spectral order and used an optimal extraction algorithm to obtain the detected stellar flux as a function of wavelength. We computed stellar parameters from our reconnaissance Tull spectra using Kea \citep{2016EndlCochranKea} dense grid.
        In brief, we used standard IRAF routines \citep{1986TodyIRAF} to perform flat fielding, bias subtraction, and order extraction, and we used a blaze function determined from high SNR flat field exposures to correct for curvature induced by the blaze. Kea is calibrated to stars in the \teff range 5000-6700~K and uses a large grid of synthetic model stellar spectra to compute stellar effective temperatures (\teff), surface gravities (\logg), and metallicities (\feh).
        The values and their formal 1-$\sigma$ uncertainties derived using Kea fine grid were used as spectroscopic constraints in stellar characterization using \isochrones in \S\ref{sec:isochrone}. 
        
    \subsection{Archival imaging}
    
        For the stars with AO/speckle non-detections, there is still the possibility that a background eclipsing binary star could be positioned behind the target star, evading detection. A few of the stars in our sample have proper motions $\geq$ 50 mas yr$^{-1}$, so they have moved across the sky by $>2\arcsec$ since they were imaged since the first Palomar Observatory Sky Survey (POSS1) in the 1950s. 
        For such stars, we downloaded the POSS1 images from Space Telescope Science Institute (STScI) Digitized Survey (DSS)\footnote{\url{http://archive.stsci.edu/cgi-bin/dss_form}}. 
        We expect a low probability of chance alignment with a foreground or background star given the high (28-50$^{\circ}$) galactic latitudes of the stars in our sample.
    
        \begin{figure*}
            \centering
            \includegraphics[clip,trim={20 40 20 40},width=0.9\textwidth]{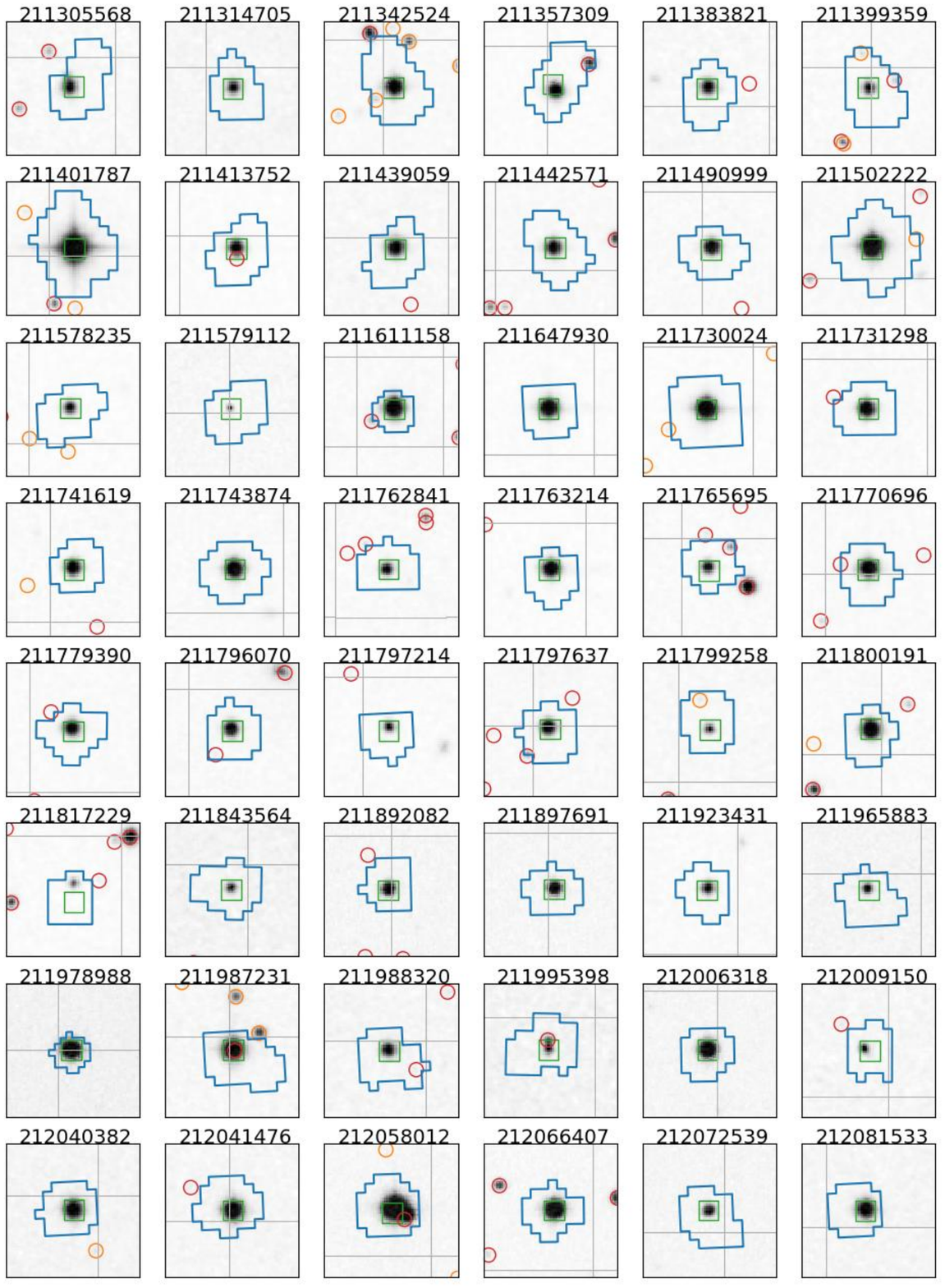}
            \caption{$1\arcmin\times1\arcmin$ cut-out images from the Digital Sky Survey 2 (DSS2) in red filter centered on the target superposed with photometric aperture outline used in the first campaign of the star was observed. The \gaia sources that are potential NEBs and those we ruled out are indicated as red and orange circles, respectively. All figures are aligned such that north is up and east to the left.}
            \label{fig:aper_grid}
        \end{figure*}
        
        \begin{figure*}
            \centering
            \includegraphics[clip,trim={20 40 20 40},width=0.95\textwidth]{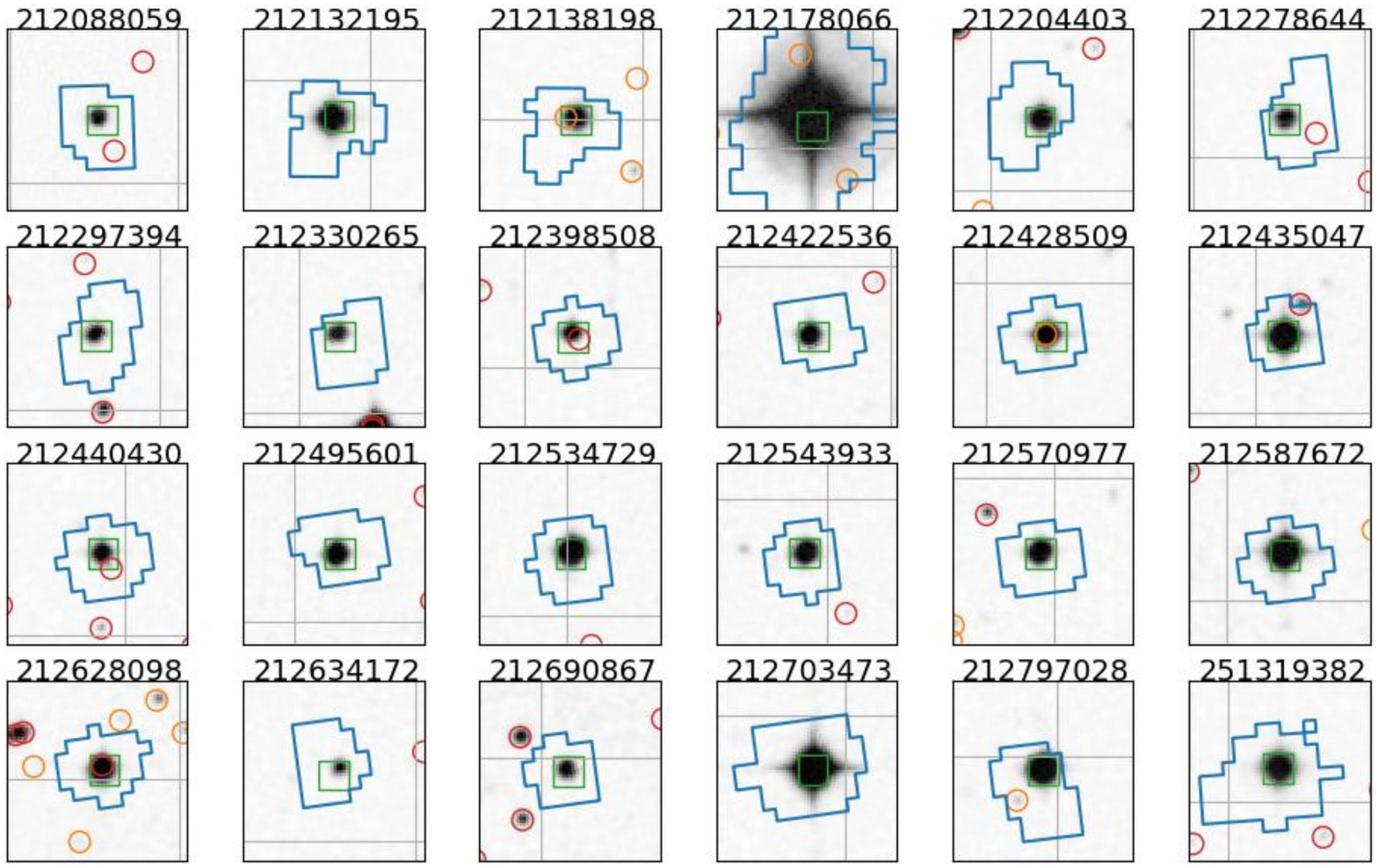}
            \contcaption{$1\arcmin\times1\arcmin$ cut-out images from the Digital Sky Survey 2 (DSS2) in red filter centered on the target superposed with photometric aperture outline (blue) used in the first \ktwo campaign of the star was observed. The \gaia sources that are potential NEBs and those we ruled out are indicated as red and orange circles, respectively. All figures are aligned such that north is up and east to the left.}
        \end{figure*}

\section{Analysis}
    \label{sec:analysis}
    
    \subsection{Vetting and transit search}
    We downloaded the lightcurves from Mikulski Archive for Space Telescopes (MAST),
    which were processed by the \everest\footnote{\url{https://archive.stsci.edu/hlsp/everest}} pipeline \citep{2016LugerEverest, 2018LugerEverest}. We also analyzed the lightcurves from the
    \ktwosff\footnote{\url{https://archive.stsci.edu/hlsp/k2sff}} pipeline \citep{2014VanderburgK2SFF} to facilitate cross-pipeline comparison and obtain reliable results. In most cases, the \everest lightcurves have relatively smaller out-of-transit scatter for the stars in our sample, except for EPIC~212178066 and EPIC~211502222. 
    We also found that the optimum photometric aperture for EPIC~211978988 selected by the \everest pipeline is too small to contain most of the flux from the star. 
    After removing all flagged cadences and other data points that are more than 5-$\sigma$ above the running mean, 
    we flattened and normalized the raw lightcurves by using a median filter with kernel size of 49 cadences, corresponding to $\sim$1 day\footnote{We checked that our choice on the kernel size neither affects the transit depth nor introduces edge effects in the lightcurves.}. For targets observed in multiple campaigns, we applied this process for each campaign before finally concatenating the data to form the final flattened lightcurve.
    Note that there is at least a 2.5 years gap between C5 \& C16 and C6 \& C17 data sets.
    We used the vetting results from DAVE pipeline\footnote{\url{http://keplertcert.seti.org/DAVE/}} \citep{2019KostovDAVE} when available and applied a similar method to the remaining targets to identify astrophysical false positives and instrumental false alarms.
    The tests include (a) photocenter analysis to
    rule out background eclipsing binaries and (b) flux time-series
    analysis to rule out odd-even differences, secondary eclipses,
    low-S/N events, variability other than a transit, and size of the transiting object. The candidates flagged after vetting were not included in the succeeding analyses. 
    We also ran EDI-Vetter\footnote{\url{https://github.com/jonzink/EDI-Vetter}} which is a similar tool to identify false positive transit signal in the \ktwo data set by detecting flux contamination, odd/even transits, non-unique signal, and secondary eclipse among others \citep{2020ZinkC5}.
    We found \numstarsdave stars (EPIC~211843564 \& EPIC~212428509) with secondary eclipses. 
    In particular, \citet{2018PetiguraC5to8} reported EPIC~212428509.01 to have a period of 3.02~d when in fact the primary and secondary eclipses with almost equal depths can be clearly seen if the lightcurve is phase-folded at twice this period. We excluded these targets in the succeeding analyses.
    
    We then did a blind search for transiting exoplanet candidates on the \everest and \ktwosff lightcurves 
    using the DST algorithm \citep{2012CabreraDST} which optimizes the fit to the transit shapes with a parabolic function. We also used the \tls algorithm \citep[TLS, ][]{2019HippkeTLS} which searches for transit-like features in an unbinned lightcurve using a transit model template 
    and restricting the trial transit durations to a smaller range that encompasses the periods of all known planets. 
    In general, all transit signals with reported ephemerides from \citet{2018MayoC0to10} and \citet{2019KruseC0to8} in \ktwo/C5 \& C6 and from \citet{2018YuC16} in C16 are recovered in the lightcurves using both pipelines, as well as from our own custom pipeline described in \citet{2017DaiK2-131b} which implements a similar approach as the transit search method in \citet{2014VanderburgK2SFF}. In the rare case when we did not detect the signal in one campaign, mainly due to the significantly larger scatter relative to the other campaigns, we did not include the data to avoid injecting unwanted noise and bias in the modeling results.
    The pre-processed lightcurves were then used for transit analyses presented in \S\ref{sec:transit_modeling}.
    
    \subsection{Stellar characterization}
    \label{sec:isochrone}
    
\begin{figure*}
        \begin{subfigure}[b]{0.48\linewidth}
            \includegraphics[clip,trim={20 20 20 20},
            width=\columnwidth]{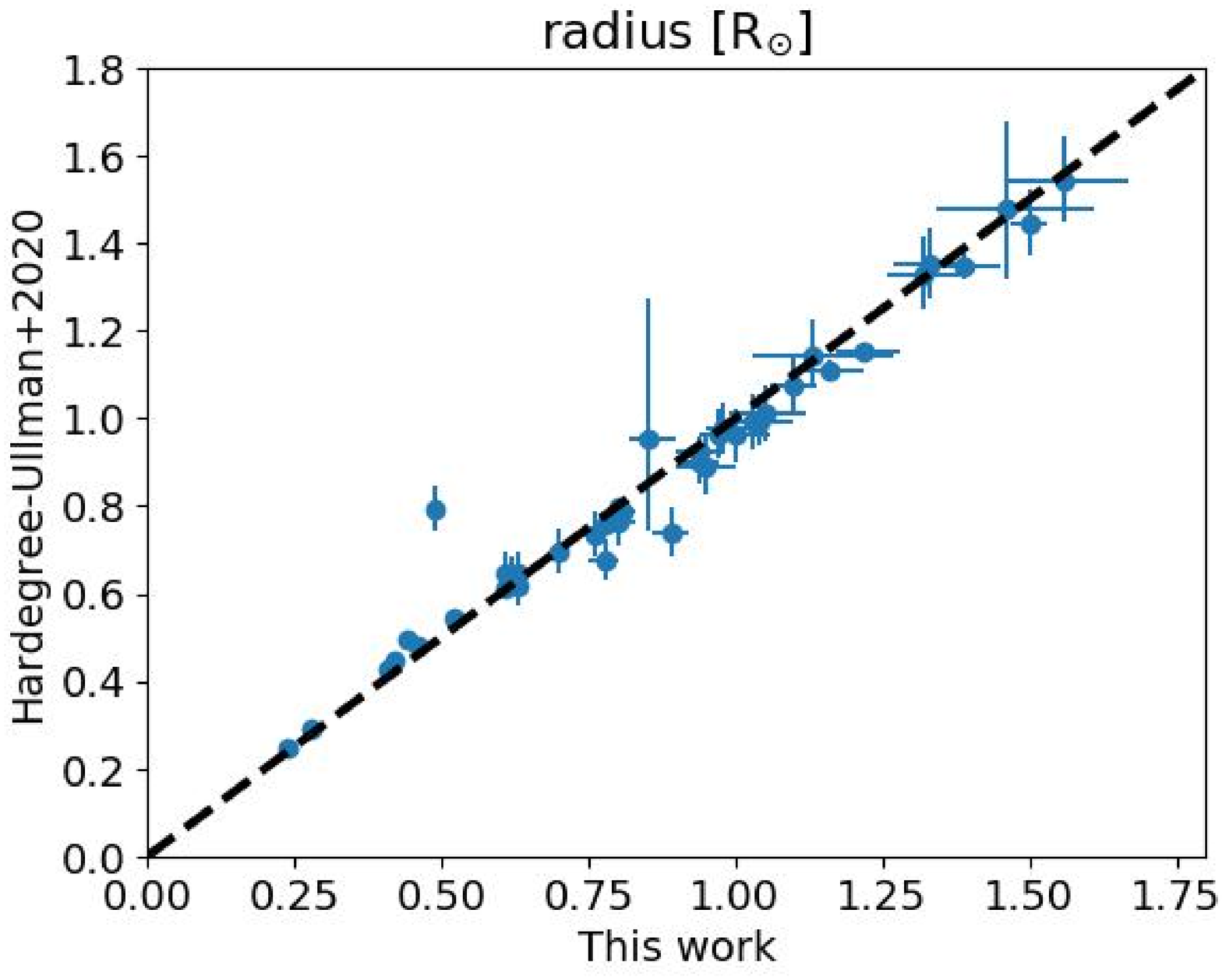}
            \caption{
            Stellar radius
            }
            \label{fig:radius}
        \end{subfigure}
        \begin{subfigure}[b]{0.48\linewidth}
            \includegraphics[clip,trim={20 20 20 20},
            width=\columnwidth]{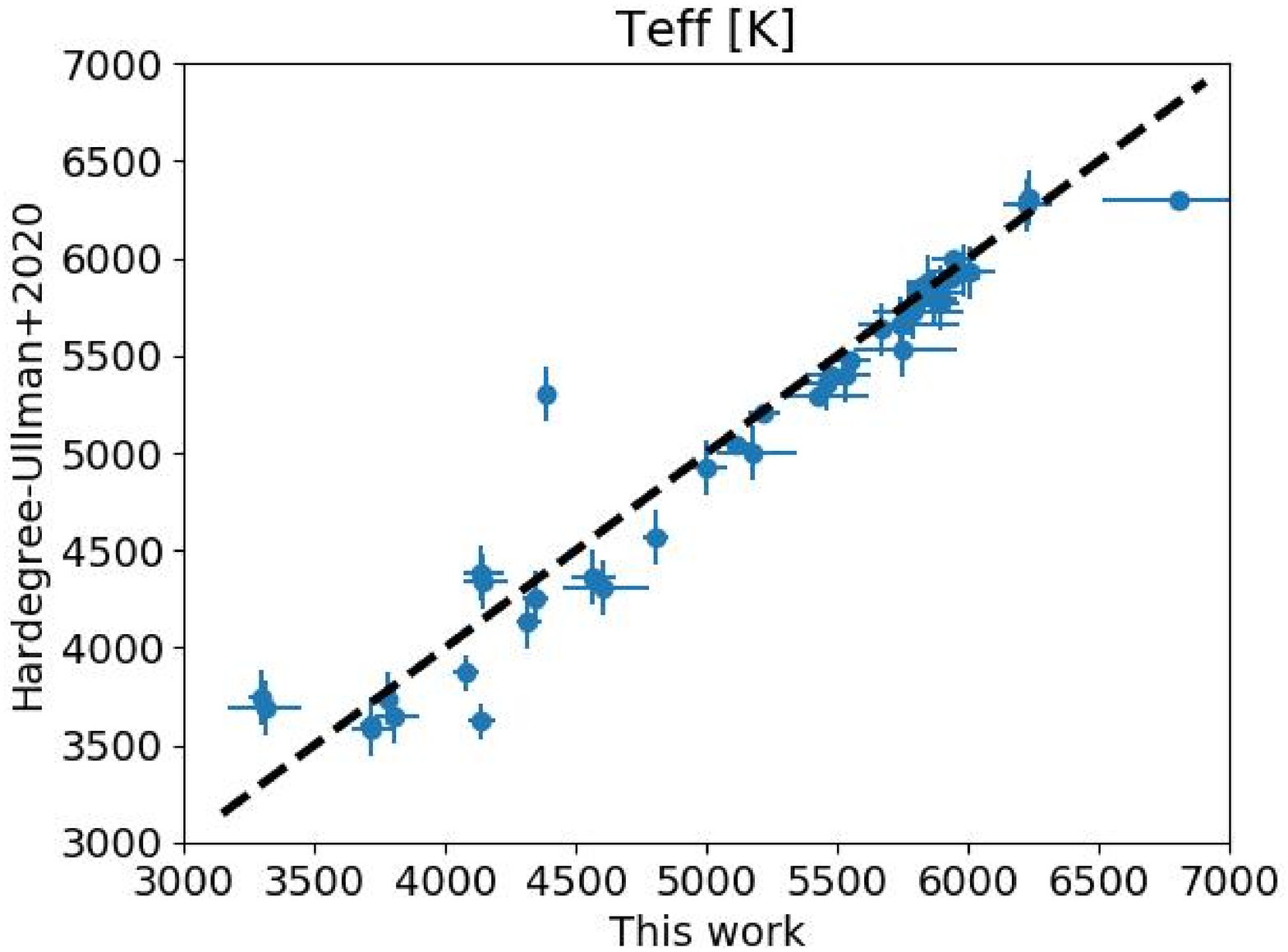}
            \caption{Effective temperature
            }
            \label{fig:teff}
        \end{subfigure}
        \caption{Comparison of stellar radius and \teff derived from this work 
        and reported in \citet{2020HardegreeUllmanK2host}. The black dashed lines represent one-to-one correspondence (x=y). Our results are consistent with those of  \citet{2020HardegreeUllmanK2host} at the 1-$\sigma$ level.}
        \label{fig:roc_curve}
    \end{figure*}

    We begin by characterizing the bulk properties of the host stars in our sample. The main parameters of interest include stellar radius, mass, effective temperature, surface gravity, and metallicity. Obtaining robust stellar parameters is important as the result of the subsequent analyses will be dependent on the derived stellar parameters. 
    To obtain these parameters, we utilized the Python package \isochrones \citep{2015MortonIsochrones}\footnote{\url{http://github.com/timothydmorton/isochrones}}, that relies on the
    MESA Isochrones \& Stellar Tracks \citep[MIST, ][]{2016DotterMIST0, 2016ChoiMIST1, 2015PaxtonMESA} grid to infer stellar parameters using a nested sampling scheme given photometric, spectroscopic data and other empirical constraints. We used \isochrones for stellar modeling in tandem with \vespa similar to previous catalogs \citep[e.g., ][]{2016MortonKeplerFPP, 2018MayoC0to10, 2018Livingston44planets, 2018Livingston60planets}. 
    In particular, we used 2MASS ($JHKs$) photometry \citep{2006Skrutskie2MASS} 
    along with \gaia DR2 parallax \citep{2019GaiaDR2} and extinction.
    We corrected the parallax for the offset found in \citet{2018StassunTorresGaiaOffset} while quadratically adding 0.1 mas to the uncertainty to account for systematics in the \gaia DR2 data \citep{2018LuriGaiaUnc}. 
    Additionally, we used \teff, \logg, and \feh derived from spectroscopy (see \S\ref{sec:spec}) or taken from the literature as additional priors. 
    We note that adding optical photometry to the aforementioned inputs did not change the results to within 1-$\sigma$ in all the stars in our sample. In fact, using 2MASS $JHKs$ photometry alone yields more reliable stellar parameters than the combined 2MASS+optical photometry, as previously observed by \citet{2018MayoC0to10}. Including photometry from additional surveys that are calibrated differently and have distinct systematic uncertainties could bias our results.
    The results of \isochrones fits are summarized in Table~\ref{tab:star}. 
    To check for consistency, we compared our derived values to the stellar parameters of \ktwo hosts derived by \citet{2020HardegreeUllmanK2host} if available. These parameters were inferred using photometric bands in combination with spectroscopic parameters (spectral type, \teff, \logg, \feh) derived from the Large Sky Area Multi-Object Fibre Spectroscopic Telescope \citep[LAMOST, ][]{2012CuiLAMOST} spectra. 
    We found that both derived parameters in our sample are typically consistent within 1-$\sigma$ as shown in Figure~\ref{fig:radius} and Figure~\ref{fig:teff}. Note however, that we found an apparent systematic bias of \teff=70 K in \isochrones when compared to the \teff reported in \citet{2020HardegreeUllmanK2host}, this bias is smaller than the typical uncertainties in our reported \teff. We also ran \isoclassify \footnote{\url{https://github.com/danxhuber/isoclassify}} to obtain stellar parameters of interest following the prescription of \citet{2017HuberIsoclassify} using identical inputs as in our \isochrones runs. We obtained results that are consistent to within 1-$\sigma$.
    
    \begin{table*}
    \caption{Summary of stellar parameters. (a) astrometric goodness of fit; (b) astrometric excess noise significance. }
\begin{tabular}{cccccccccc}
\hline
EPIC &          \rstar [\rsun] &          \mstar [\msun] &             \teff [K] &             \logg [cgs] &                \feh [dex] &  $K_p$ [mag] &  $\pi$ [mas] &  $\verb|GOF_AL|^a$ &  $\verb|D|^b$ \\
\hline
211314705 &  0.41$^{+0.01}_{-0.01}$ &  0.43$^{+0.01}_{-0.01}$ &    3669$^{+88}_{-82}$ &  4.86$^{+0.01}_{-0.01}$ &   -0.04$^{+0.2}_{-0.24}$ &    14.38 &        10.87 &    6.67 &    6.44 \\
211335816 &  1.65$^{+0.08}_{-0.07}$ &  1.29$^{+0.05}_{-0.05}$ &  6236$^{+140}_{-138}$ &  4.12$^{+0.04}_{-0.04}$ &   0.03$^{+0.09}_{-0.08}$ &    11.94 &         1.86 &    3.93 &    0.00 \\
211336288 &  0.56$^{+0.01}_{-0.01}$ &  0.58$^{+0.01}_{-0.01}$ &    4052$^{+42}_{-38}$ &   4.7$^{+0.01}_{-0.01}$ &  -0.07$^{+0.07}_{-0.07}$ &    14.56 &         4.95 &   -2.44 &    0.00 \\
211357309 &  0.44$^{+0.01}_{-0.01}$ &  0.47$^{+0.01}_{-0.01}$ &    4134$^{+60}_{-51}$ &  4.82$^{+0.01}_{-0.01}$ &   -0.9$^{+0.11}_{-0.15}$ &    13.08 &        14.49 &   15.83 &   13.48 \\
211383821 &  0.62$^{+0.01}_{-0.01}$ &  0.64$^{+0.01}_{-0.01}$ &    4343$^{+46}_{-45}$ &  4.65$^{+0.01}_{-0.01}$ &  -0.11$^{+0.07}_{-0.07}$ &    14.02 &         4.34 &   -3.23 &    0.00 \\
211399359 &  0.76$^{+0.02}_{-0.02}$ &  0.82$^{+0.02}_{-0.02}$ &    5000$^{+76}_{-71}$ &  4.58$^{+0.02}_{-0.03}$ &   0.03$^{+0.05}_{-0.06}$ &    14.39 &         2.23 &   -0.97 &    0.00 \\
211401787 &   1.5$^{+0.03}_{-0.03}$ &  1.23$^{+0.02}_{-0.02}$ &    6232$^{+33}_{-39}$ &  4.18$^{+0.02}_{-0.02}$ &  -0.03$^{+0.03}_{-0.03}$ &     9.51 &         6.22 &    9.52 &    0.00 \\
211413752 &  0.78$^{+0.02}_{-0.02}$ &  0.84$^{+0.02}_{-0.02}$ &    5106$^{+70}_{-61}$ &  4.58$^{+0.02}_{-0.03}$ &   0.03$^{+0.05}_{-0.05}$ &    13.56 &         3.02 &   -5.01 &    0.00 \\
211439059 &  0.85$^{+0.05}_{-0.03}$ &   0.9$^{+0.03}_{-0.03}$ &    5472$^{+85}_{-93}$ &  4.54$^{+0.03}_{-0.05}$ &  -0.01$^{+0.06}_{-0.06}$ &    13.03 &         3.04 &  264.08 & 5704.41 \\
211490999 &  0.94$^{+0.04}_{-0.04}$ &  0.91$^{+0.03}_{-0.03}$ &    5543$^{+80}_{-78}$ &  4.45$^{+0.05}_{-0.04}$ &  -0.01$^{+0.05}_{-0.05}$ &    13.44 &         2.13 &   -3.42 &    0.00 \\
211502222 &  1.06$^{+0.03}_{-0.02}$ &   1.1$^{+0.03}_{-0.04}$ &    5994$^{+93}_{-91}$ &  4.43$^{+0.02}_{-0.03}$ &   0.08$^{+0.08}_{-0.08}$ &    11.19 &         4.70 &   13.25 &    0.00 \\
211578235 &   1.15$^{+0.1}_{-0.08}$ &   0.9$^{+0.02}_{-0.02}$ &    5653$^{+55}_{-54}$ &  4.27$^{+0.06}_{-0.07}$ &  -0.12$^{+0.05}_{-0.05}$ &    14.33 &         1.28 &   -3.91 &    0.00 \\
211579112 &  0.28$^{+0.02}_{-0.01}$ &  0.27$^{+0.01}_{-0.02}$ &  3315$^{+137}_{-152}$ &  4.97$^{+0.03}_{-0.03}$ &   0.09$^{+0.15}_{-0.17}$ &    16.48 &         8.09 &    5.06 &    4.50 \\
211611158 &  0.94$^{+0.03}_{-0.03}$ &  0.96$^{+0.05}_{-0.07}$ &  5788$^{+190}_{-157}$ &  4.48$^{+0.03}_{-0.04}$ &    -0.1$^{+0.16}_{-0.2}$ &    12.06 &         3.70 &    8.70 &    0.00 \\
211645912 &  0.97$^{+0.03}_{-0.03}$ &  1.03$^{+0.03}_{-0.03}$ &    5892$^{+73}_{-71}$ &  4.48$^{+0.02}_{-0.03}$ &    0.0$^{+0.05}_{-0.05}$ &    12.47 &         2.84 &    3.36 &    0.00 \\
211647930 &  1.22$^{+0.05}_{-0.05}$ &  1.06$^{+0.04}_{-0.04}$ &    5880$^{+85}_{-82}$ &  4.29$^{+0.04}_{-0.04}$ &   0.07$^{+0.07}_{-0.07}$ &    11.99 &         2.88 &    4.93 &    0.00 \\
211730024 &  1.49$^{+0.07}_{-0.06}$ &  1.36$^{+0.05}_{-0.05}$ &  6502$^{+132}_{-130}$ &  4.22$^{+0.04}_{-0.04}$ &     0.17$^{+0.1}_{-0.1}$ &    11.35 &         2.66 &   13.35 &    1.92 \\
211743874 &  1.33$^{+0.07}_{-0.06}$ &  1.23$^{+0.04}_{-0.04}$ &    6222$^{+96}_{-91}$ &  4.28$^{+0.04}_{-0.04}$ &    0.1$^{+0.05}_{-0.05}$ &    12.47 &         1.76 &    3.08 &    0.00 \\
211762841 &  0.61$^{+0.01}_{-0.01}$ &  0.63$^{+0.02}_{-0.01}$ &    4079$^{+50}_{-48}$ &  4.68$^{+0.01}_{-0.01}$ &   0.13$^{+0.07}_{-0.07}$ &    14.79 &         4.14 &    0.58 &    0.00 \\
211763214 &   0.8$^{+0.01}_{-0.01}$ &  0.86$^{+0.03}_{-0.05}$ &  5424$^{+192}_{-144}$ &  4.56$^{+0.02}_{-0.03}$ &   -0.17$^{+0.16}_{-0.2}$ &    12.51 &         4.15 &    5.57 &    0.00 \\
211770696 &  1.32$^{+0.06}_{-0.06}$ &  0.94$^{+0.04}_{-0.03}$ &    5869$^{+88}_{-81}$ &  4.17$^{+0.04}_{-0.04}$ &  -0.27$^{+0.06}_{-0.06}$ &    12.23 &         2.34 &    4.06 &    0.00 \\
211779390 &  0.63$^{+0.01}_{-0.01}$ &  0.66$^{+0.02}_{-0.02}$ &    4558$^{+91}_{-81}$ &  4.65$^{+0.01}_{-0.01}$ &  -0.23$^{+0.12}_{-0.13}$ &    13.05 &         6.34 &    0.18 &    0.00 \\
211796070 &  0.89$^{+0.03}_{-0.03}$ &  0.15$^{+0.05}_{-0.03}$ &    4134$^{+88}_{-68}$ &   3.7$^{+0.15}_{-0.13}$ &  -2.93$^{+0.26}_{-0.25}$ &    13.88 &         3.68 &   -4.15 &    0.00 \\
211797637 &  0.78$^{+0.02}_{-0.03}$ &  0.17$^{+0.06}_{-0.04}$ &    4144$^{+93}_{-74}$ &  3.88$^{+0.16}_{-0.14}$ &  -2.78$^{+0.31}_{-0.26}$ &    13.69 &         4.49 &   -7.90 &    0.00 \\
211799258 &  0.44$^{+0.01}_{-0.01}$ &  0.47$^{+0.01}_{-0.01}$ &    3699$^{+66}_{-74}$ &  4.82$^{+0.01}_{-0.01}$ &   0.04$^{+0.17}_{-0.16}$ &    15.91 &         5.49 &    8.06 &    7.94 \\
211800191 &  1.22$^{+0.06}_{-0.05}$ &  0.94$^{+0.05}_{-0.04}$ &  5929$^{+120}_{-112}$ &  4.24$^{+0.04}_{-0.04}$ &  -0.28$^{+0.06}_{-0.06}$ &    12.44 &         2.47 &    5.03 &    0.00 \\
211817229 &    0.16$^{+0.0}_{-0.0}$ &    0.14$^{+0.0}_{-0.0}$ &    3246$^{+32}_{-37}$ &  5.17$^{+0.01}_{-0.01}$ &  -0.23$^{+0.06}_{-0.06}$ &    15.49 &        23.29 &   20.31 &   31.01 \\
211843564 &  0.59$^{+0.02}_{-0.02}$ &  0.61$^{+0.02}_{-0.02}$ &    3944$^{+42}_{-43}$ &  4.69$^{+0.02}_{-0.02}$ &   0.22$^{+0.13}_{-0.13}$ &    16.05 &         2.62 &   21.12 &   35.81 \\
211897691 &  0.72$^{+0.03}_{-0.03}$ &  0.74$^{+0.04}_{-0.03}$ &    4857$^{+86}_{-84}$ &  4.59$^{+0.02}_{-0.02}$ &  -0.11$^{+0.14}_{-0.14}$ &    14.34 &         2.75 &   -4.78 &    0.00 \\
211923431 &   1.13$^{+0.14}_{-0.1}$ &  0.93$^{+0.04}_{-0.04}$ &    5532$^{+90}_{-90}$ &   4.3$^{+0.07}_{-0.09}$ &   0.08$^{+0.13}_{-0.14}$ &    14.13 &         1.33 &   -3.26 &    0.00 \\
211939692 &  1.39$^{+0.06}_{-0.06}$ &  1.34$^{+0.08}_{-0.09}$ &  6806$^{+411}_{-293}$ &  4.28$^{+0.04}_{-0.05}$ &  -0.06$^{+0.17}_{-0.22}$ &    11.72 &         2.42 &    6.15 &    0.00 \\
211965883 &  0.61$^{+0.01}_{-0.01}$ &  0.63$^{+0.02}_{-0.01}$ &    4314$^{+50}_{-47}$ &  4.67$^{+0.01}_{-0.01}$ &  -0.14$^{+0.07}_{-0.07}$ &    14.09 &         4.74 &    0.02 &    0.00 \\
211978988 &  1.16$^{+0.06}_{-0.05}$ &  0.98$^{+0.07}_{-0.07}$ &    5817$^{+45}_{-48}$ &   4.3$^{+0.05}_{-0.05}$ &  -0.05$^{+0.18}_{-0.18}$ &    12.56 &         2.34 &    2.37 &    0.00 \\
211987231 &  1.46$^{+0.15}_{-0.12}$ &  1.11$^{+0.09}_{-0.07}$ &  5980$^{+118}_{-133}$ &  4.16$^{+0.07}_{-0.07}$ &  -0.01$^{+0.08}_{-0.09}$ &    11.89 &         2.60 &   50.13 &   99.62 \\
211995398 &  1.16$^{+0.15}_{-0.13}$ &  0.13$^{+0.03}_{-0.02}$ &    3921$^{+91}_{-83}$ &  3.43$^{+0.08}_{-0.09}$ &  -2.17$^{+0.29}_{-0.29}$ &    16.70 &         1.33 &    1.78 &    1.06 \\
211997641 &  2.54$^{+0.35}_{-0.28}$ &  1.65$^{+0.15}_{-0.16}$ &  6591$^{+338}_{-304}$ &  3.84$^{+0.08}_{-0.09}$ &   0.07$^{+0.16}_{-0.17}$ &    12.87 &         0.96 &   13.09 &    3.46 \\
212006318 &   1.56$^{+0.11}_{-0.1}$ &  1.11$^{+0.07}_{-0.05}$ &    5891$^{+89}_{-88}$ &   4.1$^{+0.05}_{-0.05}$ &   0.03$^{+0.06}_{-0.06}$ &    12.96 &         1.41 &    6.10 &    0.00 \\
212009150 &  0.24$^{+0.01}_{-0.01}$ &  0.22$^{+0.01}_{-0.01}$ &    3293$^{+46}_{-45}$ &  5.03$^{+0.01}_{-0.01}$ &   0.02$^{+0.05}_{-0.06}$ &    16.28 &         9.92 &    8.93 &    8.91 \\
212036875 &  1.47$^{+0.05}_{-0.05}$ &  1.22$^{+0.02}_{-0.02}$ &    6394$^{+57}_{-51}$ &  4.19$^{+0.02}_{-0.02}$ &  -0.21$^{+0.03}_{-0.03}$ &    10.91 &         3.23 &   12.44 &    1.37 \\
212040382 &   2.33$^{+0.25}_{-0.2}$ &  1.32$^{+0.11}_{-0.07}$ &  6310$^{+146}_{-141}$ &  3.83$^{+0.07}_{-0.07}$ &  -0.21$^{+0.13}_{-0.12}$ &    12.51 &         1.05 &    4.56 &    0.00 \\
212041476 &  0.97$^{+0.03}_{-0.03}$ &  1.01$^{+0.03}_{-0.04}$ &    5791$^{+74}_{-78}$ &  4.47$^{+0.03}_{-0.03}$ &   0.03$^{+0.06}_{-0.06}$ &    12.09 &         3.60 &    8.84 &    0.00 \\
212058012 &  1.09$^{+0.03}_{-0.03}$ &  1.01$^{+0.05}_{-0.05}$ &  5920$^{+104}_{-104}$ &  4.36$^{+0.03}_{-0.03}$ &  -0.06$^{+0.08}_{-0.09}$ &    11.07 &         4.77 &   12.98 &    0.00 \\
212066407 &   2.0$^{+0.18}_{-0.16}$ &   1.2$^{+0.07}_{-0.06}$ &    5943$^{+88}_{-87}$ &  3.92$^{+0.05}_{-0.05}$ &  -0.08$^{+0.06}_{-0.05}$ &    12.27 &         1.07 &   40.79 &   48.41 \\
212072539 &  0.46$^{+0.01}_{-0.01}$ &  0.49$^{+0.01}_{-0.01}$ &    3804$^{+93}_{-74}$ &   4.8$^{+0.01}_{-0.01}$ &   -0.1$^{+0.18}_{-0.25}$ &    15.13 &         5.99 &    2.77 &    0.69 \\
212081533 &  0.49$^{+0.01}_{-0.01}$ &  0.51$^{+0.01}_{-0.01}$ &    4374$^{+38}_{-34}$ &  4.76$^{+0.01}_{-0.01}$ &  -0.95$^{+0.02}_{-0.02}$ &    12.74 &        13.27 &    7.29 &    0.00 \\
212088059 &  0.52$^{+0.01}_{-0.01}$ &  0.56$^{+0.01}_{-0.01}$ &    3779$^{+30}_{-26}$ &  4.74$^{+0.01}_{-0.01}$ &   0.26$^{+0.07}_{-0.07}$ &    14.70 &         6.13 &    4.22 &    2.45 \\
212099230 &  0.98$^{+0.02}_{-0.02}$ &   0.9$^{+0.03}_{-0.02}$ &    5469$^{+63}_{-57}$ &   4.4$^{+0.02}_{-0.02}$ &   0.05$^{+0.08}_{-0.08}$ &    10.52 &         8.06 &   10.17 &    0.00 \\
212132195 &   0.7$^{+0.01}_{-0.01}$ &  0.71$^{+0.02}_{-0.02}$ &    4801$^{+49}_{-49}$ &   4.6$^{+0.01}_{-0.01}$ &  -0.18$^{+0.08}_{-0.08}$ &    11.68 &         9.44 &    4.98 &    0.00 \\
212161956 &  0.63$^{+0.02}_{-0.02}$ &  0.66$^{+0.03}_{-0.03}$ &  4599$^{+178}_{-152}$ &  4.65$^{+0.02}_{-0.02}$ &  -0.26$^{+0.17}_{-0.21}$ &    14.81 &         3.09 &   -1.92 &    0.00 \\
212178066 &   1.2$^{+0.02}_{-0.01}$ &   1.2$^{+0.04}_{-0.04}$ &    6243$^{+89}_{-94}$ &  4.36$^{+0.02}_{-0.02}$ &   0.08$^{+0.08}_{-0.08}$ &     6.75 &        28.33 &   10.59 &    0.00 \\
212204403 &  0.85$^{+0.02}_{-0.02}$ &  0.83$^{+0.02}_{-0.01}$ &    5077$^{+40}_{-39}$ &   4.5$^{+0.02}_{-0.01}$ &   0.13$^{+0.05}_{-0.05}$ &    12.33 &         4.93 &    6.94 &    0.00 \\
212278644 &  1.47$^{+0.14}_{-0.13}$ &  1.14$^{+0.06}_{-0.05}$ &    5978$^{+64}_{-65}$ &  4.16$^{+0.07}_{-0.07}$ &   0.03$^{+0.02}_{-0.02}$ &    14.00 &         1.04 &   -1.75 &    0.00 \\
212297394 &   0.8$^{+0.03}_{-0.03}$ &  0.83$^{+0.05}_{-0.04}$ &  5171$^{+172}_{-130}$ &  4.55$^{+0.03}_{-0.03}$ &  -0.03$^{+0.15}_{-0.16}$ &    14.19 &         2.36 &   -2.45 &    0.00 \\
212420823 &    0.49$^{+0.0}_{-0.0}$ &  0.54$^{+0.01}_{-0.01}$ &    4385$^{+29}_{-31}$ &  4.78$^{+0.01}_{-0.01}$ &  -0.66$^{+0.04}_{-0.04}$ &    14.18 &         2.18 &   -3.28 &    0.00 \\
212428509 &  1.29$^{+0.06}_{-0.06}$ &  0.88$^{+0.03}_{-0.02}$ &    5834$^{+68}_{-54}$ &  4.17$^{+0.04}_{-0.04}$ &  -0.37$^{+0.06}_{-0.06}$ &    12.57 &         2.24 &    2.34 &    0.00 \\
212435047 &   1.1$^{+0.04}_{-0.04}$ &  1.01$^{+0.04}_{-0.04}$ &    5842$^{+85}_{-81}$ &  4.36$^{+0.04}_{-0.04}$ &   0.01$^{+0.06}_{-0.06}$ &    12.35 &         2.74 &    4.57 &    0.00 \\
212440430 &  1.04$^{+0.06}_{-0.05}$ &  0.98$^{+0.02}_{-0.02}$ &    5789$^{+46}_{-50}$ &  4.39$^{+0.05}_{-0.04}$ &  -0.02$^{+0.03}_{-0.03}$ &    13.31 &         2.01 &   -2.14 &    0.00 \\
212495601 &  1.03$^{+0.05}_{-0.05}$ &  0.87$^{+0.02}_{-0.01}$ &    5666$^{+46}_{-47}$ &  4.35$^{+0.04}_{-0.04}$ &  -0.19$^{+0.02}_{-0.02}$ &    13.82 &         1.59 &   -2.90 &    0.00 \\
\hline
\end{tabular}

    \label{tab:star}
    \end{table*}
    
    \begin{table*}
    \contcaption{Summary of stellar parameters.}
\begin{tabular}{cccccccccc}
\hline
EPIC &          \rstar [\rsun] &          \mstar [\msun] &             \teff [K] &             \logg [cgs] &                \feh [dex] &  $K_p$ [mag] &  $\pi$ [mas] &  $\verb|GOF_AL|^a$ &  $\verb|D|^b$ \\
\hline
212543933 &  1.05$^{+0.07}_{-0.06}$ &  1.02$^{+0.02}_{-0.02}$ &    5769$^{+39}_{-37}$ &   4.4$^{+0.06}_{-0.06}$ &   0.08$^{+0.02}_{-0.02}$ &    13.99 &         1.41 &   -2.83 &    0.00 \\
212570977 &  1.12$^{+0.07}_{-0.06}$ &  1.05$^{+0.04}_{-0.04}$ &    5698$^{+87}_{-91}$ &  4.36$^{+0.06}_{-0.05}$ &   0.24$^{+0.05}_{-0.05}$ &    13.94 &         1.40 &   -2.37 &    0.00 \\
212587672 &  0.98$^{+0.04}_{-0.03}$ &  0.99$^{+0.03}_{-0.04}$ &    6004$^{+77}_{-78}$ &  4.45$^{+0.03}_{-0.04}$ &  -0.18$^{+0.06}_{-0.06}$ &    12.20 &         3.10 &    3.61 &    0.00 \\
212628098 &  0.88$^{+0.03}_{-0.03}$ &  0.77$^{+0.02}_{-0.03}$ &   4109$^{+85}_{-115}$ &  4.44$^{+0.03}_{-0.05}$ &  -0.01$^{+0.09}_{-0.09}$ &    13.47 &         4.67 &   -5.25 &    0.00 \\
212628477 &  1.35$^{+0.07}_{-0.06}$ &    1.0$^{+0.2}_{-0.06}$ &    5715$^{+95}_{-97}$ &   4.2$^{+0.03}_{-0.04}$ &  -0.09$^{+0.07}_{-0.07}$ &    12.62 &         2.35 &    1.58 &    0.00 \\
212634172 &  0.39$^{+0.01}_{-0.01}$ &  0.41$^{+0.01}_{-0.01}$ &    3431$^{+41}_{-48}$ &  4.86$^{+0.01}_{-0.01}$ &   0.32$^{+0.08}_{-0.07}$ &    15.26 &         9.89 &   12.53 &   15.65 \\
212639319 &  2.56$^{+0.22}_{-0.18}$ &  1.39$^{+0.08}_{-0.07}$ &    5456$^{+95}_{-93}$ &  3.77$^{+0.04}_{-0.05}$ &   0.26$^{+0.05}_{-0.05}$ &    12.42 &         1.29 &   15.23 &    4.68 \\
212661144 &   1.0$^{+0.06}_{-0.05}$ &  0.96$^{+0.08}_{-0.07}$ &  5749$^{+206}_{-191}$ &  4.43$^{+0.05}_{-0.06}$ &  -0.03$^{+0.17}_{-0.19}$ &    13.74 &         1.86 &   -3.08 &    0.00 \\
212690867 &  0.41$^{+0.01}_{-0.01}$ &  0.43$^{+0.01}_{-0.01}$ &    3713$^{+37}_{-37}$ &  4.85$^{+0.01}_{-0.01}$ &  -0.13$^{+0.09}_{-0.09}$ &    15.30 &         6.34 &    3.76 &    1.88 \\
212797028 &  1.77$^{+0.13}_{-0.11}$ &  1.14$^{+0.05}_{-0.05}$ &    5767$^{+87}_{-78}$ &   4.0$^{+0.05}_{-0.05}$ &    0.1$^{+0.06}_{-0.06}$ &    13.11 &         1.15 &   -2.08 &    0.00 \\
251319382 &  0.95$^{+0.02}_{-0.02}$ &  0.98$^{+0.04}_{-0.04}$ &    5791$^{+81}_{-81}$ &  4.47$^{+0.03}_{-0.03}$ &  -0.05$^{+0.07}_{-0.07}$ &    11.11 &         5.65 &   16.02 &    7.46 \\
251554286 &  0.99$^{+0.03}_{-0.03}$ &  0.87$^{+0.02}_{-0.02}$ &    5698$^{+55}_{-55}$ &  4.38$^{+0.03}_{-0.03}$ &  -0.21$^{+0.04}_{-0.04}$ &    12.10 &         3.70 &    2.84 &    0.00 \\
\hline
\end{tabular}
    \end{table*}
    
    \subsection{Stellar multiplicity} \label{sec:dilution}
    
        Figure~\ref{fig:aper_grid} shows the $1\arcmin\times1\arcmin$ images from the Digital Sky Survey 2 (DSS2) taken in red filter centered on the target (green square). Also superposed are the \gaia sources (circles) within the field of view and the photometric aperture (blue polygon) chosen for the campaign when the target was first observed by \ktwo. The optimum aperture determined by the chosen pipeline changes depending on the campaign. For the \everest pipeline, the aperture radius typically spans 3 - 5 \kepler pixels in radius except for the bright star EPIC~212178066 which has a radius of 7 pixels. 
        In total, \numstarswithinaperture stars in our sample have at least one \gaia source within the aperture. 
        For each case, we compute the amount by which the target is diluted by the flux from the neighboring \gaia source using $\gamma = 1+10^{0.4 \Delta m}$, where $\Delta m$ is the difference in magnitude in the \kepler bandpass \citep[Eq.~1, ][]{2018Livingston60planets}.
        Assuming the signal originates from (i.e. planet orbits) the fainter secondary star, then the observed transit depth, $\delta '$ is scaled by $\gamma$ to obtain the true transit depth, $\delta = \delta'\gamma = \delta'(1+10^{0.4 \Delta m}$).
        
        \begin{figure*}
            \includegraphics[clip,trim={0 0 0 0},width=\textwidth]{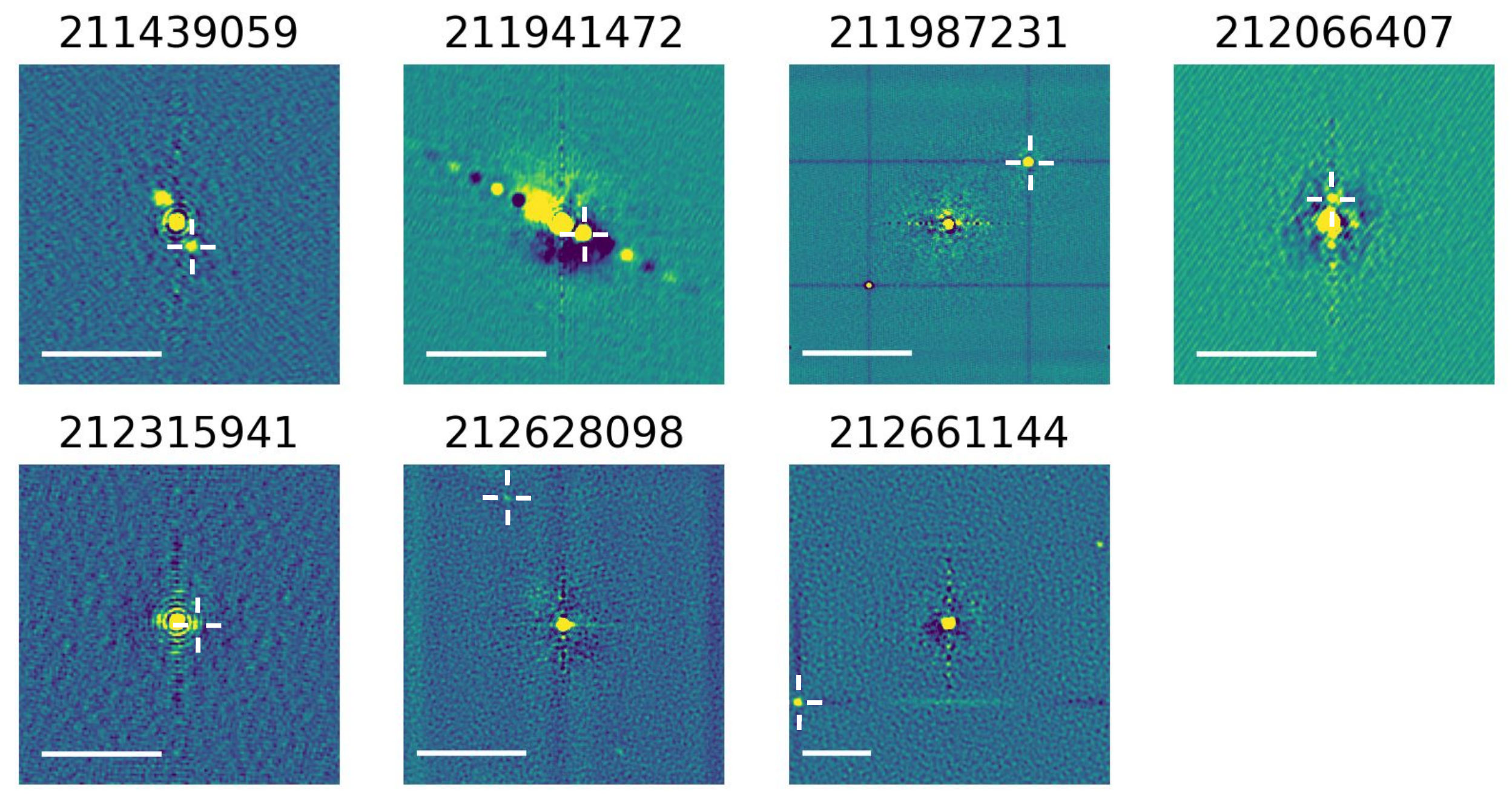}
            \caption{Reconstructed 832nm WIYN speckle images centered on the target stars with detected companions shown as a pair of bright points 180$^{\circ}$ apart. The white reticles mark the position angle of the companions but the actual position angle could be off by exactly 180$^{\circ}$. North is at the top and east to the left. White bar in the lower left corner corresponds to $1\arcsec$.}
        \label{fig:speckle}
        \end{figure*}
    
        Table~\ref{tab:gamma} lists the dilution factors $\gamma_{\mathrm{pri}}$ and $\gamma_{\mathrm{sec}}$, which assumed the signal originates from the primary or secondary star, respectively. 
        Then, we identified which \gaia sources surrounding the target are bright enough to reproduce the measured transit depths. A common false positive scenario involves a faint (blended) eclipsing binary whose eclipses are diluted to levels that match the transit depth.
        To be conservative, we first assume a maximum eclipse depth of 100\% (i.e., $\delta=1$), so if $\delta'> \gamma^{-1}$, then the observed depth is too deep to have originated from the secondary star. If this conservative criterion is not satisfied, we then use Equation~21 in \citet{2003Seager} to compute the maximum radius ratio \kmax based on the transit shape and compare it with undiluted \RpRs. If \RpRs$>1\sigma$ \kmax, then the observed radius is non-physical and hence the secondary star being the origin of the signal is ruled out. 
        In Figure~\ref{fig:aper_grid}, the sources that are potential nearby eclipsing binaries (NEBs) and those we ruled out are indicated as red and orange circles, respectively. Of the \numstarswithinaperture stars with nearby stars within the aperture, NEB scenarios in \numstarspotentialNEBs stars are not completely ruled out.
        In such cases we cannot rule out NEBs, we performed pixel level multi-aperture analysis to determine the actual source of the signal. For each target, we compared the transit depths of the phase-folded lightcurves created using apertures with different sizes available in the \ktwosff pipeline. 
        The power of pixel level multi-aperture analysis was also demonstrated by \citet{2017Cabrera3FP}, where they found two previously validated planets are actually false positives since they exhibited increased transit depths for larger aperture masks, suggesting that a nearby star was responsible for the eclipses. 
        In some cases, a nearby fainter star a few pixels away from the target can be essentially excluded in the photometry using a small enough aperture. In general, we find no apparent difference between the transit depths using a large and a small aperture, implying that the target star is the source of the signal. In these cases, we used the lightcurves with large aperture including the nearby star in transit modeling (taking into account dilution) due to its higher photometric precision compared to the lightcurve produced using a small aperture excluding the nearby star. 
        Similar to host stars with hints of binarity (see \S\ref{sec:isochrone}), we do not validate any planet candidates for which we cannot rule out all detected companions (either from \gaia or AO/speckle imaging) as the source of the signal. 
        Finally, the companion radius reported in Table~\ref{tab:planet} is corrected for dilution using $\gamma_{\mathrm{pri}}$ in Table~\ref{tab:gamma}.
        
        \begin{table}
            \centering
            \caption{Systems with detected companions in AO/speckle (top), and \gaia DR2 (bottom) within or near the photometric aperture boundary ($r\lesssim 30\arcsec$).}
%
\begin{tabular}{lllll}
\hline
      EPIC & $\Delta_{K_p}$ & $r$ [$\arcsec$] & $\gamma_\mathrm{pri}$ & $\gamma_\mathrm{sec}$ \\
\hline
 211439059 &  1.07 &  0.23 &       1.37 &       3.68 \\
 211941472 &  0.32 &  0.20 &       1.74 &       2.34 \\
 211987231 &  1.70 &  0.94 &       1.21 &       5.77 \\
 212066407 &  4.06 &  0.22 &       1.02 &      43.07 \\
 212315941 &  1.29 &  0.07 &       1.30 &       4.28 \\
 212628098 &  2.39 &  1.65 &       1.11 &      10.05 \\
 212661144 &  2.85 &  2.94 &       1.07 &      14.87 \\
 \hline
 211335816 &  3.24 &  7.65 &       1.05 &      20.75 \\
 211357309 &  2.25 & 18.94 &       1.13 &       8.92 \\
 211383821 &  6.86 & 19.15 &       1.00 &     554.16 \\
 211399359 &  3.83 & 12.35 &       1.03 &      35.01 \\
 211401787 &  6.81 & 26.31 &       1.00 &     529.45 \\
 211413752 &  5.83 &  4.70 &       1.00 &     215.07 \\
 211502222 &  9.62 & 20.22 &       1.00 &    7015.58 \\
 211578235 &  5.58 & 19.61 &       1.01 &     171.32 \\
 211611158 &  5.97 & 11.68 &       1.00 &     246.44 \\
 211765695 &  2.90 & 13.67 &       1.07 &      15.48 \\
 211770696 &  7.17 & 12.29 &       1.00 &     735.68 \\
 211797637 &  4.10 & 15.55 &       1.02 &      44.83 \\
 211826814 &  4.84 & 14.76 &       1.01 &      87.03 \\
 211995398 &  0.60 &  4.17 &       1.58 &       2.74 \\
 212058012 &  2.58 &  6.49 &       1.09 &      11.75 \\
 212088059 &  6.28 & 11.77 &       1.00 &     326.98 \\
 212099230 &  5.29 & 12.30 &       1.01 &     131.12 \\
 212161956 &  3.53 &  8.85 &       1.04 &      26.91 \\
 212178066 & 13.38 & 21.56 &       1.00 &  225299.47 \\
 212278644 &  6.88 & 11.67 &       1.00 &     564.19 \\
 212297394 &  6.21 & 24.20 &       1.00 &     305.79 \\
 212428509 &  3.33 &  1.09 &       1.05 &      22.52 \\
 212435047 &  7.08 & 12.00 &       1.00 &     678.40 \\
 212440430 &  6.08 &  6.00 &       1.00 &     271.20 \\
 212563850 &  1.77 &  9.67 &       1.20 &       6.11 \\
 212690867 &  0.88 & 18.93 &       1.44 &       3.26 \\
 212797028 &  5.99 & 13.27 &       1.00 &     249.67 \\
 251554286 &  4.41 &  9.85 &       1.02 &      59.12 \\
 \hline
\end{tabular}
            \label{tab:gamma}
        \end{table}
        
    \subsection{Transit modeling} 
    \label{sec:transit_modeling}
    
        After the pre-processing step described in \S\ref{sec:phot}, we model the lightcurves similar to the procedure detailed in \citet{2018Livingston44planets} which we briefly summarize here. We adopted the analytic transit model \citep{2002MandelAgol} as implemented in the Python package \batman \citep{2015KriedbergBatman} with a Gaussian likelihood function, and assuming a linear ephemeris and quadratic limb darkening. We set the following as free parameters: the orbital period \Porb, mid-transit time \To, scaled semi-major axis \aRs, impact parameter \imppar, and quadratic limb darkening coefficients in q-space ($q_1$ and $q_2$) as prescribed by \citet{2013KippingLimbdark}. We also fit the logarithm of the Gaussian errors ($\log \sigma$) and a constant out-of-transit baseline offset. 
        We imposed Gaussian priors on $q_1$ and $q_2$, with mean and standard deviation determined by Monte Carlo sampling an interpolated grid of the theoretical limb darkening coefficients (in the \kepler bandpass) tabulated by \citet{2012ClaretLimbDarkening} given \teff, \feh, and \logg of the host stars. 
        This allows the uncertainties in host star \teff, \logg, and \feh (see Table~\ref{tab:star}) to propagate in the final estimate. 
        
        We used the Python package \emcee~\citep{2013DFMemcee} for Markov Chain Monte Carlo (MCMC) exploration of the posterior probability distribution using 100 "walkers" in a Gaussian ball around the least squares solution determined using the Python package \lmfit \citep{2016NewvilleLMFit}. We ran MCMC for at least 2$\times$10$^4$ steps and discarded the first 10$^3$ steps as "burn-in". To assess convergence, we checked that the acceptance fraction is between 0.01 and 0.4. We also estimated the integrated autocorrelation time ($\tau_\mathrm{acf}$) of the ensemble and verified that it is appropriate for the chain length. Finally, we visually inspect the MCMC chains in the trace plot and the posterior distributions of each model parameter to make sure they are well-mixed and uni-modal, respectively.
        We computed the autocorrelation time of each parameter to ensure that we collected thousands of effectively independent samples after discarding the burn-in steps. We report the median and 68\% credible interval of the resulting marginalized posterior distributions in Table~\ref{tab:planet}. We also computed the planet's equilibrium temperature (\Teq) using 
        the MCMC samples of the host star and planet directly, and assuming bond albedo=0.3 applicable for Neptune-like planets.
        
        We also computed the stellar density by using Equation~4 of \citet{2014Kipping}, assuming circular orbits and $M_p \ll M_s$, where $M_p$ and $M_s$ are the masses of the planet and star, respectively. This is useful to check consistency with the bulk density computed using the stellar mass and radius derived in \S\ref{sec:isochrone}. 
        More importantly, agreement between these two results is a sign that the transit signal comes from a planet, and it is not an astrophysical false positive. 
        We also checked the effect of using the stellar density as a prior on the transit modeling and confirmed that adding it did not generally bias the resulting best-fit transit parameters.
        The \ktwo lightcurves with the best-fit transit model are shown in Figure~\ref{fig:lc-grid}.
        
        \begin{figure*}
            \includegraphics[clip,trim={10 60 10 60},width=\textwidth]{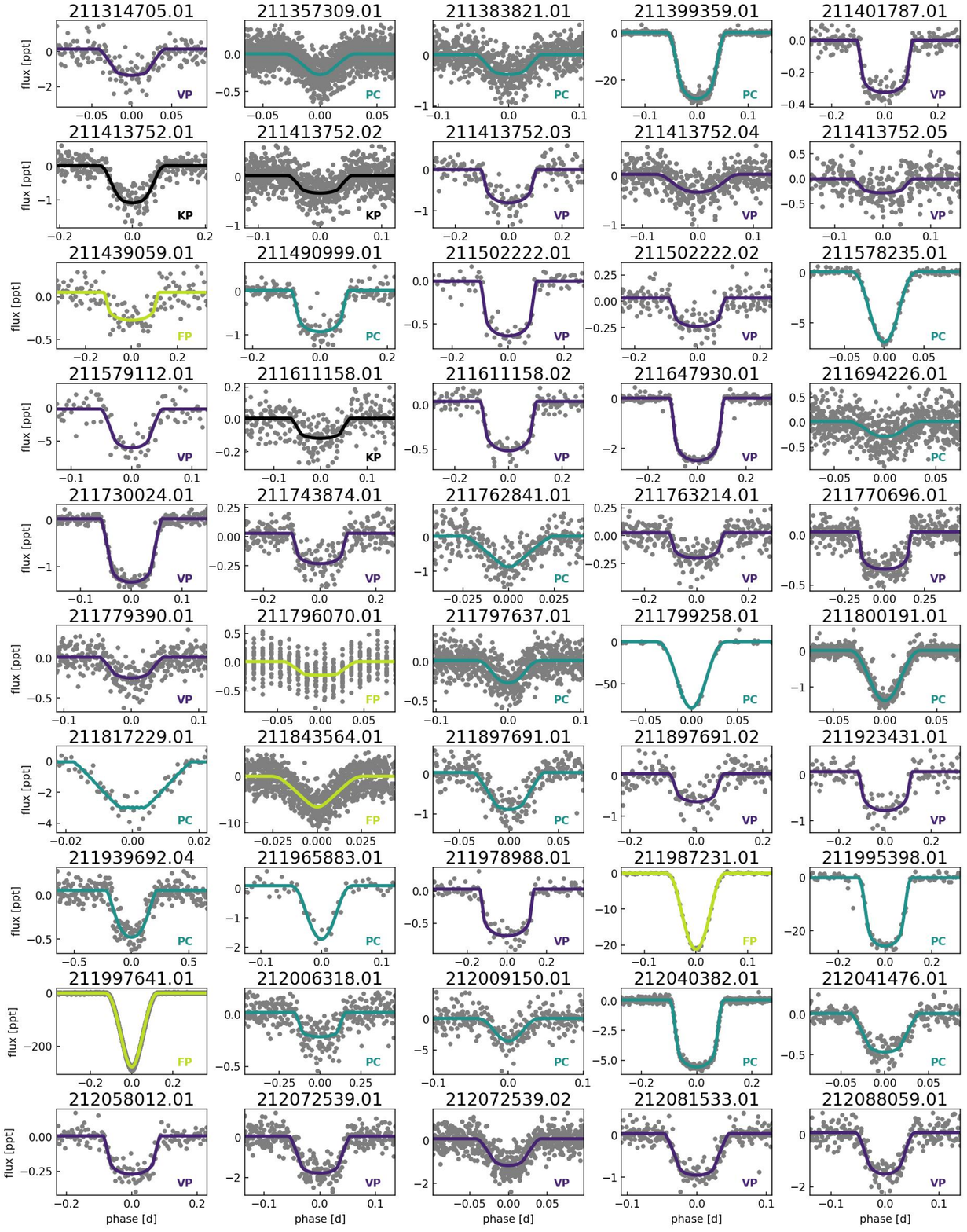}
            \caption{Best-fit transit model (colored line) 
            superposed on the phase-folded lightcurve (gray points). Final dispositions in the lower right corner (VP=validated planet; PC=planet candidate; FP=false positive; KP=known planet). 
            }
            \label{fig:lc-grid}
        \end{figure*}
        
        \begin{figure*}
            \includegraphics[clip,trim={80 10 70 20},width=\textwidth]{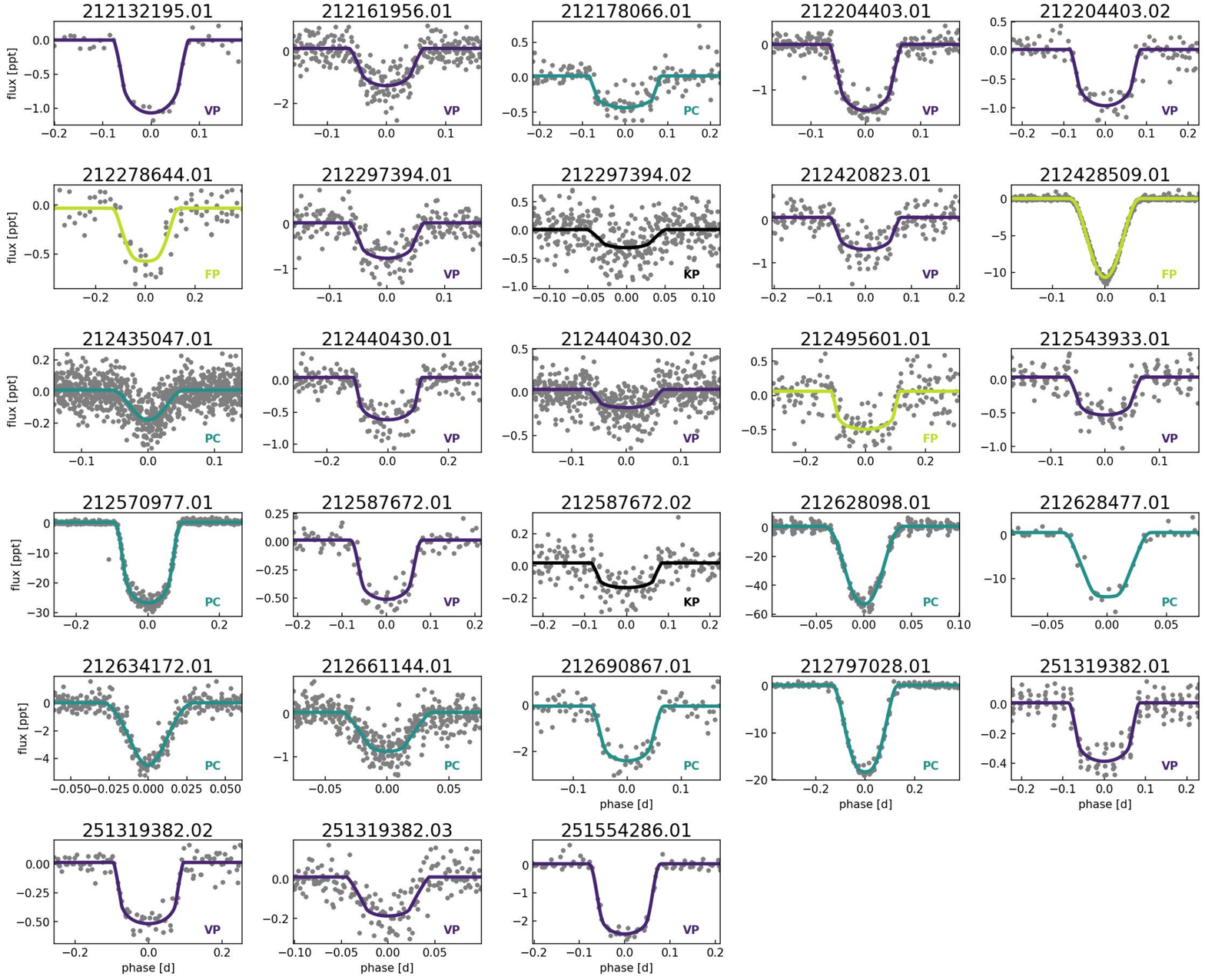}
            \contcaption{Best-fit transit model (colored line) 
            superposed on the phase-folded lightcurve (gray points). Final dispositions in the lower right corner (VP=validated planet; PC=planet candidate; FP=false positive; KP=known planet). 
            }
        \end{figure*}

    \subsection{False positive probabilities} 
    \label{sec:fpp}
    
    The concept of validation has been developed and calibrated over the years \citep[e.g., ][]{2011Torres, 2012Morton, 2014DiazPASTIS, 2015SanternePASTIS, 2016MortonKeplerFPP, 2020GiacaloneDressing}. 
    At its core, validating a transiting planet means statistically arguing that the data are much more likely to be explained by a planet than by an astrophysical false positive. 
    Here we quantify the false positive probability (FPP) of each candidate by using the Python package \vespa\footnote{\url{https://github.com/timothydmorton/VESPA}} \citep{2015MortonVespa}, which was developed as a tool for robust statistical validation of planet candidates identified by the \kepler mission \citep[e.g.,][]{2012Morton} and its successor \ktwo \citep[e.g., ][]{2016CrossfieldC0to4, 2018Livingston60planets, 2018MayoC0to10}. \vespa compares the likelihood of planetary scenario to the likelihoods of several astrophysical false positive scenarios involving eclipsing binaries (EBs), hierarchical triple systems (HEBs), background eclipsing binaries (BEBs), and the double-period cases of all these scenarios. The likelihoods and priors for each scenario are based on the shape of the transit signal, the star's location in the Galaxy, and single-, binary-, and triple-star model fits to the observed photometric and spectroscopic properties of the star generated using \isochrones. 
    
\begin{figure}
        \includegraphics[clip,trim={0 0 105 0},width=0.47\textwidth]{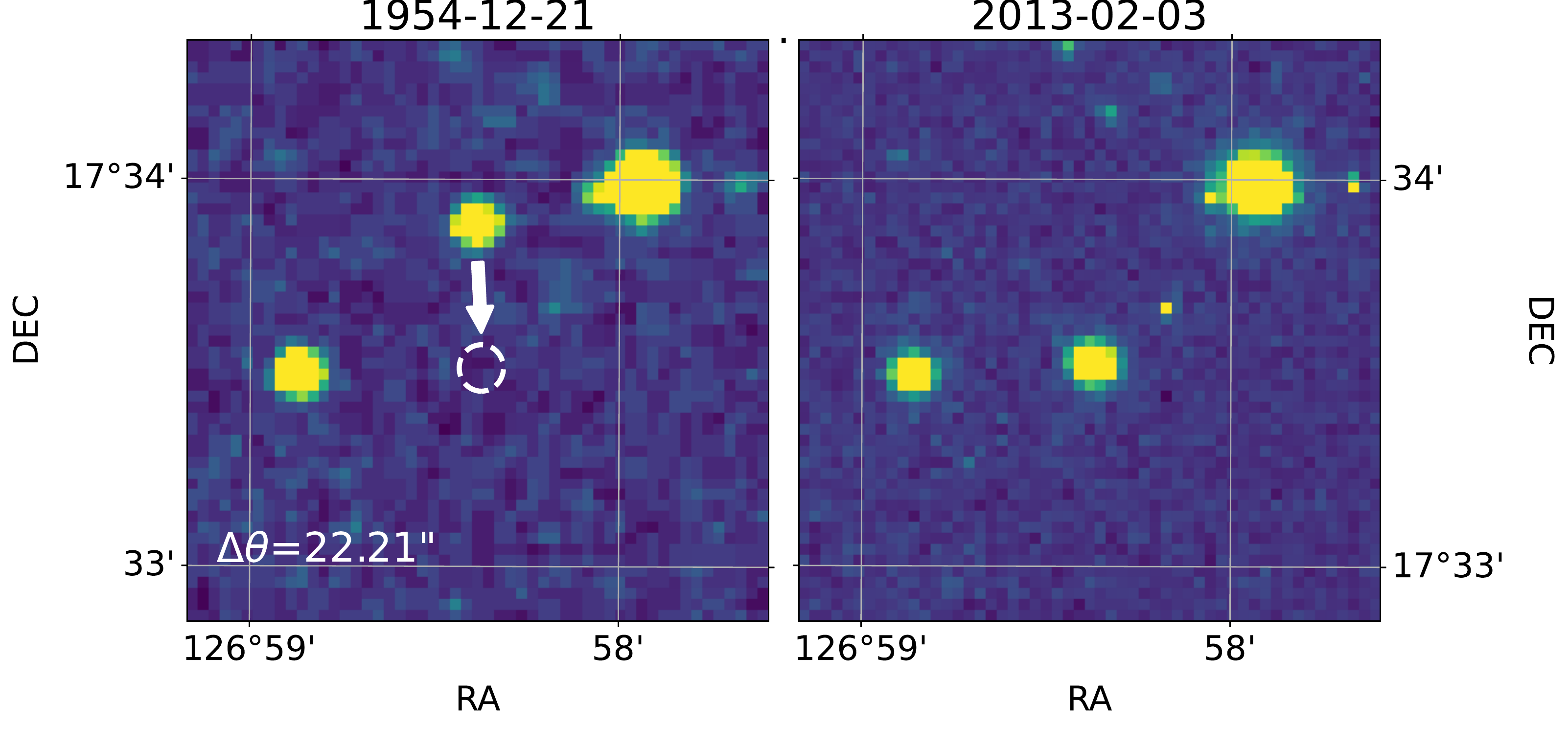}
        \caption{\poss (left) and \panstarrs (right) sky survey images taken 59 years apart, long enough for EPIC~211817229 to have moved about d$\theta \approx 22\arcsec$ along the direction of the white arrow. The white circle shows a clear view along the line of sight to the position of the target in the right image, helpful to rule out a background eclipsing binary scenario.}
        \label{fig:poss}
    \end{figure}

    As additional constraints, we used the available AO/speckle contrast curves described in \S\ref{sec:speckle}, the maximum aperture radius (\maxrad)--interior to which the transit signal must be produced--and the maximum allowed depth of potential secondary eclipse (\secthresh) estimated from the given lightcurves. 
    Similar to \citet{2018MayoC0to10}, we computed \secthresh by binning the phase-folded lightcurves by measuring the transit duration and taking thrice the value of the standard deviation of the mean in each bin. Effectively, we are asserting that we did not detect a secondary eclipse at any phase (not only at phase=0.5) at 3-$\sigma$ level.
    We also experimented with the choice of \maxrad between the largest and smallest aperture radius used for stars observed in multiple campaigns. We found that bigger \maxrad results in higher probabilities for BEB likelihoods but ultimately did not significantly affect our final FPP. We list the likelihoods of false positive scenarios considered by \vespa in Table~\ref{tab:fpp}. 
    For a few targets with large proper motions, such as EPIC~211827229 shown in  Figure~\ref{fig:poss}, archival images are helpful to rule out background eclipsing binary as the origin of the signal.
    
    Because the FPPs from \vespa do not reflect multiplicity, we applied a "multiplicity boost", which effectively reduces the FPP for each candidate in a multi-planet system. 
    Equations 8 \& 9 in \citet{2011Lissauer} introduce a factor of 25 to the planet scenario prior for systems with two planets and a factor of 50 for systems of three or more candidates. These factors are based on the observed false positive rate for the \kepler field that is also applied in boosting FPP of multi-planet candidates found in \ktwo fields \citep[e.g, ][]{2016VanderburgC0to3, 2018MayoC0to10}. Although \citet{2016Sinukoff} argues that such factor cannot be assumed to be the same as that for \ktwo, given the different Galactic backgrounds and pointing characteristics of the observations, \cite{2020CastroGonzalezC12to15} computed very similar values between 28-40 based on early \ktwo campaigns. Thus, we adapted a factor of 25 and 50 for two-planet and three or more planet systems, respectively. Such factors are already reflected into the final FPP in Table~\ref{tab:planet}. We note, however, that none of the multi-planet candidates we validate in this work require this boost in order to meet our validation criterion of FPP<1\%. 
        
    \subsection{Candidate dispositions} 
    \label{sec:disposition}
    
    We followed a decision tree to assign the final disposition for each candidate. 
    We began by checking if the signal is on-target using the dilution and the multi-aperture analyses (\S\ref{sec:dilution}), and if there is no hint of binarity (\S\ref{sec:isochrone}). If there exists any nearby star that cannot be ruled out as a potential NEB, and if there is a hint of aperture-dependent depth variation, then the candidate is categorized as a planet candidate (PC). If the source of the signal is identified to be the nearby star as demonstrated in Figure~\ref{fig:PLA_9230}, then the undiluted depth and hence true radius of the companion is derived using the actual host star radius (if known) and then its disposition is evaluated in a similar fashion.
    
    As mentioned in \S\ref{sec:fpp}, host stars with large Astrometric Goodness of Fit in the Along-Scan direction ($\verb|GOF_AL|>20$) and Astrometric Excess Noise ($\verb|D|>5$) are designated as false positive (FP) as in the cases of EPIC~211439059, EPIC~212534729, and EPIC~212703473 despite their final FPP<1\%. This is motivated by the fact that the presence of multiple stars biases the measurements of the planet's derived properties. 
    We then took the final FPP (accounted for multiplicity) and adopted the standard criteria of <1\% and >99\% as potentially validated planet (VP) and FP, respectively. A candidate with 1\% < FPP < 99\% is designated as neither validated nor false positive, and thus remains a planet candidate similar to previous works \citep[e.g., ][]{2015MontetC1, 2016CrossfieldC0to4, 2016MortonKeplerFPP, 2017DressingC1to7planets, 2018Livingston44planets}.
    Because giant planets, brown dwarfs, and low-mass stars are typically indistinguishable based on radius alone, we used a radius upper limit of \Rp$<8$~\rearth to avoid validating any of the common false positives, similar to previous studies \citep[e.g., ][]{2018MayoC0to10,2020GiacaloneTOI}. 
    This radius roughly corresponds to the minimum radius of a brown dwarf \citep[e.g., ][]{2013SorahanaBD,2020Carmichael} or an eclipsing dwarf star \citep[e.g., ][]{2017Shporer3FP}.
    As a final check, we designate VP only to those candidates that have stellar density derived from transit modeling to be consistent within 3-$\sigma$ with stellar density derived from \isochrones. Those that are inconsistent by more than 3-$\sigma$ are noted with "rho" in Table~\ref{tab:planet}. 
    The final disposition for each planet candidate is indicated in the last column of Table~\ref{tab:planet}. 
    
    \subsection{Multi-planet system stability} 
    \label{sec:stability}
    
        We validate the orbital solutions of the \nummulti multi-planet systems discussed in \S\ref{sec:multi} by analyzing their orbital stability. For simplicity, we assumed the best-case scenario, namely zero (mutual) inclinations and zero eccentricities of the planetary orbits. 
        We estimate the planetary masses from the observed radius using the \textsc{MRExo} package\footnote{\url{https://github.com/shbhuk/mrexo}}, which performs non-parametric fitting of the mass-radius relation \citep{2018Ning,2019Kanodia}. Our dynamical stability pipeline is described as follows. For coplanar, nearly-circular two-planet systems, analytic tools provide sufficient understanding of dynamical stability to render $N$-body simulations unnecessary. Particularly, here we adopt the Hill stability criterion of \citet{1993Icar..106..247G}, which has been extensively validated by direct integration. 
        
    \begin{landscape}
        \begin{table}
        \centering
        \caption{Candidate parameters and dispositions. (VP=validated planet; KP=known planet; PC=planet candidate; FP=false positive; LR=large radius; AO=bright nearby star detected in AO/speckle; GB=\gaia binary; rho=discrepancy in derived stellar density; A16=\citet{2016AdamsC0to5USP}; B16=\citet{2016BarrosC1to6}; P16=\citet{2016PopeC5to6}; D17=\citet{2017DressingC1to7planets}; L18=\citet{2018Livingston60planets}; M18=\citet{2018MayoC0to10}; P18=\citet{2018PetiguraC5to8}; Y18=\citet{2018YuC16}; H19=\citet{2019HellerTLS2}; K19=\citet{2019KruseC0to8})
        \label{tab:planet}}
        \renewcommand{\arraystretch}{1.5}
\begin{tabular}{lllllllllrrllll}
\hline
EPIC &             Name &            $T_{0}$ [BKJD] &                              $P$ [d] &      $R_p/R_{\star}$ [\%] &              $a/R_{\star}$ &                     $b$ &            $T_{14}$ [hr] &            \Teq [K] &             $R_p$ [\rearth] &     SNR &  FPP &  notes & disposition &           ref \\
\hline
211314705.01 & K2-329 b &  2307.225970$^{+0.004008}_{-0.003737}$ &   3.793306$^{+0.000324}_{-0.000343}$ &   3.53$^{+0.17}_{-0.17}$ &    19.05$^{+0.21}_{-0.21}$ &  0.24$^{+0.17}_{-0.16}$ &   1.53$^{+0.04}_{-0.08}$ &       543$^{+8}_{-9}$ &      1.56$^{+0.08}_{-0.08}$ &   20.49 & 0.00 &            &          VP &        P16 \\
211357309.01 & &  2306.751359$^{+0.000665}_{-0.000677}$ &   0.463975$^{+0.000000}_{-0.000000}$ &   1.59$^{+0.12}_{-0.05}$ &     3.78$^{+0.47}_{-1.04}$ &  0.47$^{+0.32}_{-0.32}$ &   0.86$^{+0.05}_{-0.04}$ &    1269$^{+18}_{-15}$ &      0.86$^{+0.07}_{-0.03}$ &   72.92 & 0.05 &            &          PC &        A16 \\
211383821.01 & &  2307.154894$^{+0.001782}_{-0.001753}$ &   1.567125$^{+0.000003}_{-0.000003}$ &   1.81$^{+0.21}_{-0.08}$ &     6.41$^{+0.67}_{-1.77}$ &  0.42$^{+0.35}_{-0.29}$ &   1.74$^{+0.09}_{-0.08}$ &    1005$^{+15}_{-13}$ &      1.24$^{+0.14}_{-0.06}$ &   25.67 & 0.03 &            &          PC &        P16 \\
211399359.01 & &  2308.417493$^{+0.000125}_{-0.000118}$ &   3.114897$^{+0.000001}_{-0.000001}$ &  15.06$^{+0.09}_{-0.08}$ &    11.64$^{+0.09}_{-0.16}$ &  0.10$^{+0.10}_{-0.07}$ &   2.34$^{+0.01}_{-0.01}$ &     979$^{+20}_{-22}$ &     12.94$^{+0.40}_{-0.34}$ &  479.01 & 0.00 &         LR &          PC &        P16 \\
211401787.01 & K2-330 b &  2318.064162$^{+0.001517}_{-0.001464}$ &  13.774798$^{+0.000028}_{-0.000028}$ &   1.69$^{+0.08}_{-0.03}$ &    21.76$^{+1.56}_{-4.85}$ &  0.38$^{+0.32}_{-0.26}$ &   4.57$^{+0.08}_{-0.07}$ &       969$^{+9}_{-9}$ &      2.77$^{+0.14}_{-0.09}$ &   57.88 & 0.00 &            &          VP &        P16 \\
211413752.01 & K2-268 b &  2307.846650$^{+0.002133}_{-0.002101}$ &   9.327527$^{+0.000021}_{-0.000020}$ &   3.12$^{+0.91}_{-0.22}$ &    17.17$^{+4.82}_{-9.46}$ &  0.65$^{+0.30}_{-0.45}$ &   3.35$^{+0.62}_{-0.11}$ &     696$^{+12}_{-13}$ &      2.69$^{+0.77}_{-0.21}$ &  141.74 &   -- &            &          KP &  L18 \\
211413752.02 & K2-268 c &  2310.654276$^{+0.002437}_{-0.002523}$ &   2.151676$^{+0.000005}_{-0.000006}$ &   1.75$^{+0.14}_{-0.06}$ &     7.76$^{+0.72}_{-1.84}$ &  0.42$^{+0.33}_{-0.30}$ &   1.97$^{+0.06}_{-0.05}$ &    1136$^{+18}_{-19}$ &      1.50$^{+0.12}_{-0.07}$ &   34.08 &   -- &            &          KP &  L18 \\
211413752.03 & K2-268 f &  2309.191739$^{+0.003964}_{-0.003853}$ &  26.270570$^{+0.000105}_{-0.000109}$ &   2.59$^{+0.17}_{-0.08}$ &    41.41$^{+3.15}_{-8.33}$ &  0.39$^{+0.29}_{-0.27}$ &   4.62$^{+0.13}_{-0.11}$ &      492$^{+8}_{-10}$ &      2.23$^{+0.15}_{-0.09}$ &   27.64 & 0.00 &            &          VP &        K19 \\
211413752.04 & K2-268 d &  2310.974654$^{+0.003182}_{-0.003300}$ &   4.528598$^{+0.000016}_{-0.000015}$ &   1.73$^{+0.16}_{-0.06}$ &    12.57$^{+1.18}_{-3.32}$ &  0.40$^{+0.35}_{-0.28}$ &   2.57$^{+0.12}_{-0.10}$ &     888$^{+16}_{-15}$ &      1.49$^{+0.14}_{-0.07}$ &   74.65 & 0.05 &            &          VP &        K19 \\
211413752.05 & K2-268 e &  2309.343959$^{+0.004850}_{-0.004706}$ &   6.131243$^{+0.000032}_{-0.000033}$ &   1.55$^{+0.13}_{-0.08}$ &    16.93$^{+1.89}_{-4.06}$ &  0.42$^{+0.32}_{-0.28}$ &   2.54$^{+0.14}_{-0.13}$ &     801$^{+13}_{-13}$ &      1.33$^{+0.11}_{-0.08}$ &   24.63 & 0.00 &            &          VP &        K19 \\
211439059.01 & &  2313.521148$^{+0.007066}_{-0.008427}$ &  18.637056$^{+0.000183}_{-0.000174}$ &   1.68$^{+0.14}_{-0.09}$ &    24.57$^{+2.88}_{-6.31}$ &  0.42$^{+0.33}_{-0.29}$ &   5.32$^{+0.33}_{-0.28}$ &     610$^{+18}_{-21}$ &      2.15$^{+0.21}_{-0.16}$ &   13.76 & 0.00 &         GB &          FP &        P16 \\
211490999.01 & &  2313.329630$^{+0.002489}_{-0.002464}$ &   9.844401$^{+0.000706}_{-0.000686}$ &   2.86$^{+0.18}_{-0.09}$ &    19.68$^{+1.63}_{-4.63}$ &  0.40$^{+0.32}_{-0.27}$ &   3.64$^{+0.12}_{-0.10}$ &     799$^{+19}_{-24}$ &      2.96$^{+0.21}_{-0.16}$ &   36.86 & 0.01 &            &          PC &        P16 \\
211502222.01 & K2-331 c &  3280.304754$^{+0.002330}_{-0.002335}$ &  22.996591$^{+0.001848}_{-0.001859}$ &   2.35$^{+0.11}_{-0.07}$ &    37.87$^{+2.65}_{-7.38}$ &  0.37$^{+0.30}_{-0.25}$ &   4.43$^{+0.10}_{-0.09}$ &     673$^{+12}_{-11}$ &      2.72$^{+0.14}_{-0.10}$ &   33.14 & 0.00 &            &          VP &        Y18 \\
211502222.02 & K2-331 b &  3267.920192$^{+0.005312}_{-0.005582}$ &   9.398977$^{+0.001506}_{-0.001348}$ &   1.55$^{+0.12}_{-0.07}$ &    16.76$^{+1.93}_{-4.78}$ &  0.43$^{+0.34}_{-0.29}$ &   3.92$^{+0.22}_{-0.19}$ &     909$^{+14}_{-14}$ &      1.79$^{+0.14}_{-0.10}$ &   20.63 & 0.00 &            &          VP &  This work \\
211578235.01 & &  2314.979746$^{+0.000259}_{-0.000261}$ &  11.007605$^{+0.000004}_{-0.000004}$ &  12.95$^{+5.24}_{-3.07}$ &    30.60$^{+2.90}_{-1.51}$ &  0.98$^{+0.07}_{-0.06}$ &   1.56$^{+0.03}_{-0.05}$ &     874$^{+34}_{-35}$ &     16.49$^{+6.44}_{-4.00}$ &  266.82 & 0.71 &         LR &          PC &        B16 \\
211579112.01 & &  2323.420746$^{+0.002160}_{-0.002168}$ &  17.706320$^{+0.000063}_{-0.000063}$ &   7.07$^{+0.33}_{-0.31}$ &    66.30$^{+2.39}_{-2.60}$ &  0.39$^{+0.13}_{-0.20}$ &   2.03$^{+0.09}_{-0.10}$ &       266$^{+5}_{-8}$ &      2.20$^{+0.19}_{-0.15}$ &   17.04 & 0.00 &            &          VP &        P16 \\
211611158.01 & K2-185 b &  2311.727092$^{+0.005898}_{-0.005466}$ &  10.616384$^{+0.000069}_{-0.000074}$ &   1.11$^{+0.06}_{-0.06}$ &    21.50$^{+0.85}_{-0.92}$ &  0.74$^{+0.05}_{-0.05}$ &   2.60$^{+0.17}_{-0.18}$ &     809$^{+22}_{-25}$ &      1.15$^{+0.07}_{-0.07}$ &   16.71 &   -- &            &          KP &  M18 \\
211611158.02 & K2-185 c &  2326.157391$^{+0.002498}_{-0.002358}$ &  52.713494$^{+0.000155}_{-0.000164}$ &   2.32$^{+0.05}_{-0.05}$ &    62.62$^{+2.50}_{-2.69}$ &  0.73$^{+0.03}_{-0.03}$ &   4.64$^{+0.08}_{-0.08}$ &     477$^{+14}_{-17}$ &      2.39$^{+0.09}_{-0.09}$ &   48.91 & 0.02 &            &          VP &        M18 \\
211647930.01 & K2-333 b &  3264.395983$^{+0.000766}_{-0.000749}$ &  14.759287$^{+0.000243}_{-0.000240}$ &   4.61$^{+0.10}_{-0.05}$ &    23.06$^{+0.72}_{-2.14}$ &  0.26$^{+0.23}_{-0.18}$ &   4.96$^{+0.05}_{-0.04}$ &     826$^{+14}_{-17}$ &      6.18$^{+0.28}_{-0.25}$ &  113.06 & 0.00 &            &          VP &        Y18 \\
211694226.01 & &  2342.946754$^{+0.002181}_{-0.002209}$ &   1.918518$^{+0.000006}_{-0.000006}$ &   1.61$^{+0.18}_{-0.13}$ &    12.01$^{+2.49}_{-3.32}$ &  0.42$^{+0.35}_{-0.30}$ &   1.09$^{+0.14}_{-0.14}$ &     816$^{+16}_{-23}$ &      1.08$^{+0.13}_{-0.10}$ &   13.54 & 0.09 &            &          PC &        D17 \\
211730024.01 & K2-334 b &  3263.810228$^{+0.000550}_{-0.000536}$ &   5.113981$^{+0.000061}_{-0.000062}$ &   3.43$^{+0.18}_{-0.05}$ &    15.35$^{+1.09}_{-3.58}$ &  0.38$^{+0.33}_{-0.26}$ &   2.46$^{+0.06}_{-0.03}$ &    1380$^{+34}_{-30}$ &      5.65$^{+0.36}_{-0.28}$ &  133.98 & 0.00 &            &          VP &        Y18 \\
211743874.01 & K2-335 b &  2315.209626$^{+0.002700}_{-0.002774}$ &  12.283211$^{+0.000051}_{-0.000051}$ &   1.51$^{+0.08}_{-0.05}$ &    20.30$^{+1.76}_{-4.76}$ &  0.40$^{+0.31}_{-0.28}$ &   4.31$^{+0.13}_{-0.12}$ &     949$^{+21}_{-26}$ &      2.21$^{+0.18}_{-0.14}$ &   24.89 & 0.00 &            &          VP &        P16 \\
211762841.01 & &  2307.265782$^{+0.000812}_{-0.000828}$ &   1.565010$^{+0.000002}_{-0.000002}$ &   2.89$^{+0.29}_{-0.16}$ &    16.94$^{+2.87}_{-4.05}$ &  0.43$^{+0.31}_{-0.28}$ &   0.65$^{+0.06}_{-0.08}$ &     931$^{+13}_{-13}$ &      1.91$^{+0.19}_{-0.12}$ &   21.45 & 0.13 &            &          PC &        D17 \\
211763214.01 & K2-336 b &  2313.585567$^{+0.004276}_{-0.004476}$ &  21.194733$^{+0.000108}_{-0.000107}$ &   1.40$^{+0.10}_{-0.05}$ &    32.37$^{+2.92}_{-8.35}$ &  0.41$^{+0.33}_{-0.28}$ &   4.65$^{+0.16}_{-0.14}$ &     569$^{+15}_{-22}$ &      1.22$^{+0.09}_{-0.05}$ &   22.34 & 0.00 &            &          VP &        P16 \\
211770696.01 & K2-337 b &  2312.963531$^{+0.002514}_{-0.002528}$ &  16.273563$^{+0.000054}_{-0.000053}$ &   1.80$^{+0.10}_{-0.04}$ &    15.39$^{+1.13}_{-3.59}$ &  0.38$^{+0.32}_{-0.26}$ &   7.64$^{+0.14}_{-0.11}$ &     850$^{+19}_{-18}$ &      2.62$^{+0.19}_{-0.14}$ &   41.74 & 0.00 &            &          VP &        P16 \\
211779390.01 & K2-338 b &  2308.526349$^{+0.001912}_{-0.001839}$ &   3.850614$^{+0.000012}_{-0.000012}$ &   1.48$^{+0.15}_{-0.07}$ &    15.13$^{+1.75}_{-4.06}$ &  0.42$^{+0.34}_{-0.29}$ &   1.78$^{+0.11}_{-0.10}$ &     783$^{+12}_{-14}$ &      1.03$^{+0.11}_{-0.05}$ &   21.05 & 0.00 &            &          VP &        P16 \\
211796070.01 & &  2307.731014$^{+0.002660}_{-0.002404}$ &   1.889933$^{+0.000007}_{-0.000007}$ &   1.44$^{+0.09}_{-0.07}$ &     8.83$^{+1.23}_{-2.28}$ &  0.42$^{+0.33}_{-0.29}$ &   1.50$^{+0.13}_{-0.12}$ &    1371$^{+57}_{-63}$ &      1.40$^{+0.10}_{-0.08}$ &   15.16 & 1.00 &            &          FP &        B16 \\
211797637.01 & &  2306.788585$^{+0.002729}_{-0.002963}$ &   1.640772$^{+0.000112}_{-0.000102}$ &   1.62$^{+0.13}_{-0.09}$ &    10.09$^{+2.04}_{-3.24}$ &  0.45$^{+0.35}_{-0.31}$ &   1.11$^{+0.15}_{-0.14}$ &    1315$^{+59}_{-52}$ &      1.34$^{+0.15}_{-0.08}$ &   53.49 & 0.34 &            &          PC &        B16 \\
211799258.01 & &  2320.146470$^{+0.000338}_{-0.000330}$ &  19.533884$^{+0.000009}_{-0.000009}$ &  27.02$^{+1.82}_{-1.52}$ &  125.14$^{+16.96}_{-9.74}$ &  0.59$^{+0.11}_{-0.25}$ &   1.34$^{+0.06}_{-0.07}$ &       326$^{+5}_{-7}$ &     13.04$^{+0.95}_{-0.82}$ &  112.72 & 0.71 &         LR &          PC &        D17 \\
\hline
\end{tabular}
        \end{table}
    \end{landscape}
    
    \begin{landscape}
        \begin{table}
        \centering
        \contcaption{Candidate parameters and dispositions. }
\renewcommand{\arraystretch}{1.5}
\begin{tabular}{lllllllllrrllll}
\hline
EPIC &          Name &               $T_{0}$ [BKJD] &                              $P$ [d] &      $R_p/R_{\star}$ [\%] &              $a/R_{\star}$ &                     $b$ &            $T_{14}$ [hr] &            \Teq [K] &             $R_p$ [\rearth] &     SNR &  FPP &  notes & disposition &           ref \\
\hline
211800191.01 & &  2307.749483$^{+0.000190}_{-0.000208}$ &   1.106170$^{+0.000000}_{-0.000000}$ &   4.79$^{+2.51}_{-0.46}$ &     3.01$^{+0.47}_{-0.32}$ &  0.96$^{+0.04}_{-0.02}$ &   1.22$^{+0.04}_{-0.08}$ &    2011$^{+42}_{-46}$ &      6.42$^{+3.40}_{-0.75}$ &  329.65 & 0.07 &            &          PC &        P16 \\
211817229.01 & &  2307.694394$^{+0.000563}_{-0.000557}$ &   4.353783$^{+0.000003}_{-0.000003}$ &   6.55$^{+0.58}_{-0.41}$ &  86.19$^{+16.61}_{-22.23}$ &  0.41$^{+0.33}_{-0.29}$ &   0.37$^{+0.06}_{-0.05}$ &       348$^{+2}_{-2}$ &      1.14$^{+0.10}_{-0.07}$ &   37.80 & 0.09 &            &          PC &        D17 \\
211843564.01 & &  2307.077984$^{+0.000425}_{-0.000489}$ &   0.452018$^{+0.000000}_{-0.000000}$ &   7.93$^{+2.11}_{-0.42}$ &     4.96$^{+0.74}_{-1.95}$ &  0.49$^{+0.40}_{-0.35}$ &   0.68$^{+0.10}_{-0.05}$ &    1345$^{+23}_{-31}$ &      5.12$^{+1.30}_{-0.37}$ &   70.96 & 0.20 &         GB &          FP &        K19 \\
211897691.01 & &  2309.493095$^{+0.001192}_{-0.001177}$ &   5.750534$^{+0.000007}_{-0.000007}$ &   2.81$^{+0.25}_{-0.09}$ &    33.89$^{+3.12}_{-8.35}$ &  0.41$^{+0.32}_{-0.29}$ &   1.22$^{+0.05}_{-0.04}$ &     761$^{+14}_{-20}$ &      2.23$^{+0.20}_{-0.12}$ &   44.28 & 0.07 &        rho &          PC &        P16 \\
211897691.02 & K2-339 b &  2320.004667$^{+0.004642}_{-0.005128}$ &  19.507428$^{+0.000116}_{-0.000113}$ &   2.44$^{+0.23}_{-0.12}$ &   40.71$^{+4.50}_{-11.08}$ &  0.44$^{+0.33}_{-0.30}$ &   3.40$^{+0.18}_{-0.16}$ &     508$^{+12}_{-12}$ &      1.92$^{+0.19}_{-0.12}$ &   34.81 & 0.10 &            &          VP &        K19 \\
211923431.01 & K2-340 b &  2310.815863$^{+0.002975}_{-0.003102}$ &  29.740451$^{+0.000169}_{-0.000161}$ &   2.68$^{+0.17}_{-0.10}$ &    40.80$^{+3.40}_{-9.36}$ &  0.39$^{+0.31}_{-0.27}$ &   5.29$^{+0.19}_{-0.15}$ &     606$^{+31}_{-36}$ &      3.36$^{+0.44}_{-0.32}$ &   23.52 & 0.00 &            &          VP &        P16 \\
211939692.04 & &  2333.057118$^{+0.006513}_{-0.006699}$ &  26.855455$^{+0.000193}_{-0.000186}$ &   3.24$^{+1.31}_{-0.50}$ &     5.96$^{+0.73}_{-0.55}$ &  0.98$^{+0.02}_{-0.01}$ &  10.77$^{+0.39}_{-0.48}$ &     804$^{+32}_{-42}$ &      4.91$^{+1.97}_{-0.76}$ &   95.22 & 0.86 &            &          PC &        K19 \\
211965883.01 & &  2313.496997$^{+0.002648}_{-0.002570}$ &  21.110323$^{+0.001461}_{-0.001397}$ &   3.86$^{+2.36}_{-0.27}$ &  90.71$^{+19.45}_{-55.17}$ &  0.57$^{+0.40}_{-0.39}$ &   1.58$^{+0.37}_{-0.14}$ &       414$^{+5}_{-6}$ &      2.56$^{+1.62}_{-0.19}$ &  121.03 & 0.20 &            &          PC &        P16 \\
211978988.01 & K2-341 b &  2319.708694$^{+0.002195}_{-0.002083}$ &  36.552551$^{+0.000127}_{-0.000122}$ &   2.51$^{+0.17}_{-0.07}$ &   42.13$^{+3.55}_{-10.63}$ &  0.41$^{+0.32}_{-0.28}$ &   6.25$^{+0.17}_{-0.11}$ &     598$^{+16}_{-15}$ &      3.21$^{+0.25}_{-0.19}$ &   45.54 & 0.00 &            &          VP &        M18 \\
211987231.01 & &  2308.813645$^{+0.000170}_{-0.000172}$ &  17.035141$^{+0.000070}_{-0.000069}$ &  29.39$^{+3.60}_{-4.71}$ &    40.32$^{+0.36}_{-0.27}$ &  1.07$^{+0.04}_{-0.06}$ &   2.35$^{+0.02}_{-0.02}$ &     866$^{+30}_{-45}$ &     56.30$^{+8.74}_{-9.62}$ & 3162.20 & 0.99 &      LR,GB &          FP &        B16 \\
211995398.01 & &  2336.854100$^{+0.001240}_{-0.001254}$ &  32.579267$^{+0.000066}_{-0.000067}$ &  14.96$^{+0.50}_{-0.45}$ &    51.19$^{+4.75}_{-5.01}$ &  0.45$^{+0.15}_{-0.26}$ &   5.14$^{+0.17}_{-0.14}$ &     580$^{+28}_{-38}$ &     29.88$^{+3.95}_{-3.35}$ &   67.94 & 0.00 &         LR &          PC &        P16 \\
211997641.01 & &  3263.517490$^{+0.000156}_{-0.000099}$ &   1.744545$^{+0.000000}_{-0.000000}$ &  63.11$^{+8.98}_{-3.22}$ &     3.46$^{+0.15}_{-0.05}$ &  0.75$^{+0.13}_{-0.05}$ &   5.92$^{+0.08}_{-0.09}$ &  2534$^{+154}_{-178}$ &  179.37$^{+31.30}_{-24.23}$ & 8429.32 & 0.99 &         LR &          FP &        Y18 \\
212006318.01 & &  2314.327580$^{+0.006214}_{-0.007226}$ &  14.457821$^{+0.000149}_{-0.000146}$ &   1.41$^{+0.12}_{-0.07}$ &    15.20$^{+1.72}_{-4.36}$ &  0.44$^{+0.33}_{-0.30}$ &   6.65$^{+0.34}_{-0.33}$ &     936$^{+30}_{-34}$ &      2.41$^{+0.28}_{-0.20}$ &   33.81 & 0.02 &            &          PC &        P16 \\
212009150.01 & &  2312.162629$^{+0.002507}_{-0.002465}$ &   6.833191$^{+0.000022}_{-0.000021}$ &   5.60$^{+1.07}_{-0.37}$ &   40.03$^{+8.40}_{-18.34}$ &  0.49$^{+0.40}_{-0.34}$ &   1.22$^{+0.22}_{-0.15}$ &       344$^{+3}_{-4}$ &      1.47$^{+0.30}_{-0.11}$ &   49.97 & 0.10 &            &          PC &        K19 \\
212040382.01 & &  3266.349509$^{+0.000271}_{-0.000270}$ &   4.445602$^{+0.000028}_{-0.000029}$ &   7.32$^{+0.06}_{-0.07}$ &     6.54$^{+0.28}_{-0.23}$ &  0.68$^{+0.03}_{-0.04}$ &   4.37$^{+0.03}_{-0.04}$ &    1764$^{+74}_{-84}$ &     18.58$^{+1.97}_{-1.64}$ &  451.26 & 0.02 &         LR &          PC &        Y18 \\
212041476.01 & &  3262.559797$^{+0.001203}_{-0.001199}$ &   2.783676$^{+0.000073}_{-0.000072}$ &   2.03$^{+0.12}_{-0.06}$ &    14.50$^{+1.22}_{-3.43}$ &  0.40$^{+0.32}_{-0.27}$ &   1.38$^{+0.05}_{-0.04}$ &    1280$^{+22}_{-23}$ &      2.16$^{+0.14}_{-0.09}$ &   42.77 & 0.00 &        rho &          PC &        Y18 \\
212058012.01 & K2-342 b &  3266.107260$^{+0.002390}_{-0.002546}$ &  11.561052$^{+0.000690}_{-0.000668}$ &   1.55$^{+0.10}_{-0.04}$ &    21.96$^{+2.02}_{-5.36}$ &  0.42$^{+0.31}_{-0.29}$ &   3.73$^{+0.10}_{-0.08}$ &     861$^{+14}_{-15}$ &      2.03$^{+0.13}_{-0.09}$ &   37.15 & 0.01 &            &          VP &        Y18 \\
212072539.01 & K2-343 c &  2311.624554$^{+0.001285}_{-0.001309}$ &   7.676972$^{+0.000012}_{-0.000012}$ &   4.02$^{+0.19}_{-0.10}$ &    26.11$^{+1.88}_{-5.92}$ &  0.36$^{+0.33}_{-0.25}$ &   2.20$^{+0.08}_{-0.07}$ &       465$^{+7}_{-7}$ &      2.02$^{+0.10}_{-0.07}$ &   50.40 & 0.00 &            &          VP &        Y18 \\
212072539.02 & K2-343 b &  2308.324970$^{+0.001208}_{-0.001231}$ &   2.787174$^{+0.000004}_{-0.000004}$ &   3.29$^{+0.30}_{-0.11}$ &    13.47$^{+1.69}_{-4.14}$ &  0.48$^{+0.33}_{-0.32}$ &   1.46$^{+0.08}_{-0.05}$ &     653$^{+10}_{-12}$ &      1.65$^{+0.16}_{-0.08}$ &   62.62 & 0.04 &            &          VP &        K19 \\
212081533.01 & K2-344 b &  3262.747731$^{+0.001269}_{-0.001250}$ &   3.355850$^{+0.000091}_{-0.000093}$ &   2.95$^{+0.19}_{-0.09}$ &    13.79$^{+1.25}_{-3.53}$ &  0.41$^{+0.33}_{-0.28}$ &   1.76$^{+0.07}_{-0.05}$ &      722$^{+7}_{-10}$ &      1.59$^{+0.10}_{-0.05}$ &   43.44 & 0.00 &            &          VP &        Y18 \\
212088059.01 & K2-345 b &  2308.710348$^{+0.001423}_{-0.001468}$ &  10.367437$^{+0.000020}_{-0.000019}$ &   3.67$^{+0.35}_{-0.13}$ &   36.12$^{+3.76}_{-10.25}$ &  0.43$^{+0.34}_{-0.30}$ &   2.08$^{+0.11}_{-0.08}$ &       437$^{+4}_{-5}$ &      2.11$^{+0.20}_{-0.09}$ &   50.61 & 0.00 &            &          VP &        P16 \\
212132195.01 & K2-346 b &  2331.390197$^{+0.002081}_{-0.002164}$ &  26.201446$^{+0.003331}_{-0.003124}$ &   2.97$^{+0.25}_{-0.11}$ &   56.68$^{+5.27}_{-13.31}$ &  0.42$^{+0.31}_{-0.28}$ &   3.33$^{+0.12}_{-0.10}$ &       450$^{+5}_{-5}$ &      2.26$^{+0.19}_{-0.10}$ &   25.59 & 0.00 &            &          VP &        P16 \\
212161956.01 & K2-347 b &  2307.699277$^{+0.001927}_{-0.001812}$ &   7.187257$^{+0.000020}_{-0.000021}$ &   3.35$^{+0.09}_{-0.09}$ &    21.52$^{+0.37}_{-0.37}$ &  0.21$^{+0.13}_{-0.14}$ &   2.57$^{+0.06}_{-0.07}$ &     640$^{+20}_{-21}$ &      2.41$^{+0.11}_{-0.10}$ &   27.59 & 0.00 &            &          VP &        P16 \\
212178066.01 & &  3262.901443$^{+0.003040}_{-0.003115}$ &  15.611913$^{+0.000393}_{-0.000412}$ &   1.98$^{+0.15}_{-0.09}$ &    29.57$^{+3.14}_{-7.87}$ &  0.43$^{+0.32}_{-0.30}$ &   3.72$^{+0.16}_{-0.14}$ &       835$^{+8}_{-9}$ &      2.97$^{+0.29}_{-0.17}$ &   20.02 & 0.00 &  saturate &          PC &        Y18 \\
212204403.01 & K2-348 b &  3263.716772$^{+0.000995}_{-0.001053}$ &   4.688418$^{+0.000119}_{-0.000117}$ &   3.50$^{+0.25}_{-0.08}$ &    12.55$^{+0.93}_{-2.62}$ &  0.39$^{+0.30}_{-0.27}$ &   2.75$^{+0.07}_{-0.04}$ &     908$^{+12}_{-11}$ &      3.26$^{+0.22}_{-0.11}$ &   75.18 & 0.00 &            &          VP &        Y18 \\
212204403.02 & K2-348 c &  3271.435651$^{+0.003152}_{-0.003022}$ &  12.550171$^{+0.001018}_{-0.001057}$ &   2.87$^{+0.24}_{-0.11}$ &    24.47$^{+2.35}_{-6.03}$ &  0.43$^{+0.31}_{-0.30}$ &   3.68$^{+0.14}_{-0.11}$ &       655$^{+7}_{-9}$ &      2.67$^{+0.23}_{-0.11}$ &   28.30 & 0.00 &            &          VP &        Y18 \\
212278644.01 & &  2394.558248$^{+0.014831}_{-0.014570}$ &  12.421322$^{+0.005559}_{-0.004657}$ &   2.11$^{+0.27}_{-0.14}$ &    17.50$^{+3.29}_{-7.33}$ &  0.49$^{+0.38}_{-0.34}$ &   4.84$^{+0.64}_{-0.46}$ &     965$^{+34}_{-39}$ &      3.42$^{+0.60}_{-0.39}$ &   38.97 & 1.00 &            &          FP &        P16 \\
212297394.01 & K2-304 c & 2389.478879$^{+0.003648}_{-0.003472}$ &   5.213965$^{+0.000442}_{-0.000442}$ &   2.58$^{+0.18}_{-0.11}$ &    14.45$^{+1.25}_{-3.23}$ &  0.39$^{+0.32}_{-0.27}$ &   2.62$^{+0.11}_{-0.11}$ &     866$^{+27}_{-37}$ &      2.27$^{+0.19}_{-0.14}$ &   21.14 & 0.00 &            &          VP &        P16 \\
212297394.02 & K2-304 b &  2384.963190$^{+0.004892}_{-0.004738}$ &   2.289363$^{+0.000232}_{-0.000251}$ &   1.69$^{+0.12}_{-0.12}$ &     8.53$^{+1.18}_{-1.79}$ &  0.41$^{+0.29}_{-0.26}$ &   1.88$^{+0.17}_{-0.16}$ &    1149$^{+43}_{-44}$ &      1.48$^{+0.13}_{-0.11}$ &   11.89 &   -- &            &          KP &  H19 \\

\hline
\end{tabular}
        \end{table}
    \end{landscape}
    
    \begin{landscape}
        \begin{table}
        \centering
        \contcaption{Candidate parameters and dispositions. }
        \renewcommand{\arraystretch}{1.5}
\begin{tabular}{lllllllllrrllll}
\hline
EPIC &          Name &               $T_{0}$ [BKJD] &                              $P$ [d] &      $R_p/R_{\star}$ [\%] &              $a/R_{\star}$ &                     $b$ &            $T_{14}$ [hr] &            \Teq [K] &             $R_p$ [\rearth] &     SNR &  FPP &  notes & disposition &           ref \\
\hline
212420823.01 & K2-349 b &  2386.127431$^{+0.003865}_{-0.003894}$ &   9.032178$^{+0.000874}_{-0.000874}$ &   2.55$^{+0.18}_{-0.12}$ &    19.99$^{+1.94}_{-4.52}$ &  0.40$^{+0.31}_{-0.27}$ &   3.25$^{+0.15}_{-0.14}$ &       518$^{+5}_{-4}$ &      1.38$^{+0.10}_{-0.07}$ &   23.75 & 0.00 &            &          VP &        P16 \\
212428509.01 & &  2386.832645$^{+0.000130}_{-0.000131}$ &   5.335929$^{+0.000001}_{-0.000001}$ &  23.57$^{+0.72}_{-1.28}$ &     8.66$^{+0.04}_{-0.04}$ &  1.08$^{+0.01}_{-0.02}$ &   2.84$^{+0.01}_{-0.01}$ &    1218$^{+28}_{-28}$ &     34.37$^{+2.11}_{-2.28}$ & 3989.14 & 0.99 &         LR &          FP &        P18 \\
212435047.01 & &  2385.443969$^{+0.002187}_{-0.002349}$ &   1.115494$^{+0.000063}_{-0.000061}$ &   1.26$^{+1.21}_{-0.08}$ &     4.33$^{+1.08}_{-2.98}$ &  0.62$^{+0.38}_{-0.42}$ &   1.65$^{+0.57}_{-0.14}$ &    1860$^{+33}_{-35}$ &      1.54$^{+1.49}_{-0.13}$ &  172.91 & 0.13 &            &          PC &        P16 \\
212440430.01 & K2-350 c &  2395.164968$^{+0.002851}_{-0.002858}$ &  19.991944$^{+0.000123}_{-0.000194}$ &   2.39$^{+0.20}_{-0.09}$ &    28.71$^{+2.95}_{-8.21}$ &  0.45$^{+0.33}_{-0.29}$ &   4.94$^{+0.19}_{-0.20}$ &     687$^{+16}_{-18}$ &      2.74$^{+0.25}_{-0.18}$ &   33.77 & 0.00 &            &          VP &        P16 \\
212440430.02 & K2-350 b &  2386.277628$^{+0.003444}_{-0.003308}$ &   4.163873$^{+0.000022}_{-0.000023}$ &   1.35$^{+0.10}_{-0.08}$ &    10.68$^{+1.20}_{-2.62}$ &  0.41$^{+0.32}_{-0.28}$ &   2.75$^{+0.16}_{-0.16}$ &    1158$^{+29}_{-34}$ &      1.54$^{+0.15}_{-0.11}$ &   16.67 & 0.00 &            &          VP &  This work \\
212495601.01 & &  2396.654065$^{+0.004748}_{-0.004464}$ &  21.674345$^{+0.000160}_{-0.000165}$ &   2.19$^{+0.15}_{-0.10}$ &    30.45$^{+2.90}_{-7.50}$ &  0.41$^{+0.32}_{-0.28}$ &   5.10$^{+0.22}_{-0.20}$ &     665$^{+16}_{-19}$ &      2.47$^{+0.21}_{-0.17}$ &   15.25 & 1.00 &            &          FP &        P16 \\
212543933.01 & K2-351 b &  2390.495191$^{+0.002614}_{-0.002604}$ &   7.806164$^{+0.000673}_{-0.000623}$ &   2.21$^{+0.13}_{-0.09}$ &    20.20$^{+1.80}_{-4.36}$ &  0.39$^{+0.30}_{-0.27}$ &   2.78$^{+0.12}_{-0.11}$ &     934$^{+28}_{-32}$ &      2.54$^{+0.25}_{-0.19}$ &   23.08 & 0.00 &            &          VP &        P16 \\
212570977.01 & &  2390.894185$^{+0.000333}_{-0.000329}$ &   8.853066$^{+0.000004}_{-0.000004}$ &  14.98$^{+0.20}_{-0.16}$ &    17.95$^{+0.60}_{-0.69}$ &  0.29$^{+0.11}_{-0.16}$ &   4.20$^{+0.05}_{-0.04}$ &     912$^{+32}_{-29}$ &     18.34$^{+1.22}_{-1.07}$ &  295.58 & 0.20 &         LR &          PC &        P16 \\
212587672.01 & K2-307 c &  2404.042492$^{+0.001902}_{-0.001818}$ &  23.228555$^{+0.000068}_{-0.000071}$ &   2.15$^{+0.17}_{-0.07}$ &   50.74$^{+5.07}_{-15.28}$ &  0.43$^{+0.35}_{-0.30}$ &   3.25$^{+0.13}_{-0.09}$ &     658$^{+13}_{-14}$ &      2.32$^{+0.19}_{-0.12}$ &   46.10 & 0.01 &            &          VP &        P16 \\
212587672.02 & K2-307 b &  2394.644991$^{+0.004543}_{-0.004448}$ &  15.280780$^{+0.000121}_{-0.000116}$ &   1.17$^{+0.09}_{-0.07}$ &    30.75$^{+3.74}_{-8.15}$ &  0.43$^{+0.33}_{-0.29}$ &   3.46$^{+0.20}_{-0.21}$ &     756$^{+14}_{-15}$ &      1.26$^{+0.11}_{-0.09}$ &   14.56 &   -- &            &          KP &  H19 \\
212628098.01 & &  2390.347813$^{+0.000248}_{-0.000257}$ &   4.352495$^{+0.000002}_{-0.000002}$ &  23.27$^{+1.27}_{-0.87}$ &    20.86$^{+1.34}_{-1.22}$ &  0.69$^{+0.07}_{-0.07}$ &   1.63$^{+0.05}_{-0.06}$ &     777$^{+11}_{-10}$ &     24.80$^{+1.81}_{-1.30}$ &  229.66 & 0.61 &         LR &          PC &        P18 \\
212628477.01 & &  3347.727411$^{+0.001377}_{-0.001341}$ &  15.423327$^{+0.000580}_{-0.000550}$ &  11.62$^{+1.08}_{-0.71}$ &  90.89$^{+11.67}_{-23.65}$ &  0.43$^{+0.33}_{-0.29}$ &   1.33$^{+0.16}_{-0.10}$ &     835$^{+18}_{-20}$ &     17.17$^{+1.98}_{-1.39}$ &   47.84 & 0.53 &         LR &          PC &  This work \\
212634172.01 & &  2384.597110$^{+0.000405}_{-0.000407}$ &   2.851687$^{+0.000002}_{-0.000002}$ &   6.39$^{+0.78}_{-0.19}$ &    27.20$^{+2.96}_{-8.99}$ &  0.44$^{+0.37}_{-0.31}$ &   0.78$^{+0.07}_{-0.04}$ &       555$^{+5}_{-5}$ &      2.75$^{+0.30}_{-0.11}$ &  198.53 & 0.19 &            &          PC &        K19 \\
212661144.01 & &  2385.909593$^{+0.001217}_{-0.001215}$ &   2.458749$^{+0.000004}_{-0.000004}$ &   2.81$^{+0.18}_{-0.10}$ &    14.78$^{+1.45}_{-3.74}$ &  0.40$^{+0.34}_{-0.28}$ &   1.20$^{+0.06}_{-0.06}$ &    1357$^{+60}_{-58}$ &      3.31$^{+0.30}_{-0.22}$ &   43.10 & 0.01 &            &          PC &        D17 \\
212690867.01 & &  2396.603761$^{+0.002469}_{-0.002550}$ &  25.856312$^{+0.000090}_{-0.000088}$ &   4.53$^{+0.26}_{-0.16}$ &   68.44$^{+6.84}_{-15.84}$ &  0.42$^{+0.30}_{-0.29}$ &   2.77$^{+0.14}_{-0.12}$ &       290$^{+3}_{-3}$ &      2.91$^{+0.18}_{-0.12}$ &   28.58 & 0.59 &            &          PC &        D17 \\
212797028.01 & &  2397.464819$^{+0.000578}_{-0.000580}$ &  29.982306$^{+0.000027}_{-0.000028}$ &  14.37$^{+0.15}_{-0.13}$ &    28.98$^{+0.41}_{-0.41}$ &  0.84$^{+0.01}_{-0.01}$ &   6.12$^{+0.06}_{-0.06}$ &     763$^{+26}_{-26}$ &     27.82$^{+2.08}_{-1.81}$ &  528.48 & 0.55 &         LR &          PC &        P18 \\
251319382.01 & K2-352 c &  3265.716929$^{+0.002711}_{-0.002885}$ &   8.234885$^{+0.000508}_{-0.000475}$ &   1.85$^{+0.12}_{-0.05}$ &    16.35$^{+1.46}_{-4.07}$ &  0.41$^{+0.32}_{-0.28}$ &   3.59$^{+0.11}_{-0.09}$ &     885$^{+12}_{-14}$ &      1.92$^{+0.13}_{-0.07}$ &   54.86 & 0.00 &            &          VP &        Y18 \\
251319382.02 & K2-352 d &  3270.622842$^{+0.002276}_{-0.002256}$ &  14.871387$^{+0.000916}_{-0.000936}$ &   2.14$^{+0.12}_{-0.06}$ &    25.94$^{+2.11}_{-5.93}$ &  0.39$^{+0.31}_{-0.27}$ &   4.14$^{+0.11}_{-0.09}$ &     727$^{+10}_{-10}$ &      2.23$^{+0.13}_{-0.09}$ &   36.20 & 0.00 &            &          VP &        Y18 \\
251319382.03 & K2-352 b &  3265.635282$^{+0.003388}_{-0.003276}$ &   3.665912$^{+0.000273}_{-0.000295}$ &   1.32$^{+0.10}_{-0.06}$ &    15.66$^{+2.02}_{-4.09}$ &  0.42$^{+0.33}_{-0.29}$ &   1.63$^{+0.12}_{-0.11}$ &    1160$^{+16}_{-15}$ &      1.37$^{+0.10}_{-0.07}$ &   20.01 & 0.04 &            &          VP &  This work \\
251554286.01 & K2-353 b &  3356.851818$^{+0.000977}_{-0.000988}$ &  15.466805$^{+0.000572}_{-0.000565}$ &   5.03$^{+0.11}_{-0.09}$ &    25.06$^{+0.61}_{-0.61}$ &  0.77$^{+0.02}_{-0.02}$ &   3.39$^{+0.07}_{-0.06}$ &     735$^{+13}_{-14}$ &      5.55$^{+0.22}_{-0.20}$ &   74.92 & 0.00 &            &          VP &  This work \\
\hline
\end{tabular}
        \end{table}
    \end{landscape}
        
        On the other hand, $N$-body simulations are rather necessary to assess the stability of a multi-planet system. Our stability pipeline can employ either direct $N$-body simulations or the recently published machine learning model \spock \citep{2020TamayoSpock}. 
        \spock uses a combination of $N$-body simulations and machine learning to classify the stability of multi-planet systems, assigning a stability probability to a specific configuration of a planetary system. Given a set of configurations and their stability probability, it is thus possible to generate a posterior distribution of orbital parameters and planetary masses. Since here we are not interested in obtaining a posterior distribution from stability constraints, we opt to run simpler $N$-body simulations. We run the simulations with \texttt{REBOUND}'s integrator \texttt{WHFAST} \citep{2015MNRAS.452..376R}. For each system, we run 1000 realizations by varying the initial true longitude of the planets. We consider a simulation dynamically unstable if two particles come closer than the sum of their Hill radii. Each system is simulated for 10$^6$ orbits of the inner planets. All the \nummulti multi-planet systems were found to be dynamically stable according to the criteria described above.

    \subsection{Stellar rotation periods}
    
        Stellar variability can masquerade as transiting planets. \citep[e.g., ][]{2018Hatzes}. 
        Thus, it is important to vet candidates with orbital period that are synchronized or in resonance with the stellar rotation period to eliminate potential false positives. 
        To measure rotation period robustly, several methods are applied \citep[e.g., ][]{2014GarciaCS, 2019Santos}. The first one consists of doing a time-period analysis based on wavelets \citep{1998TorrenceCompoWavelet} and projecting the result into the period axis to get the Global Wavelet Power Spectrum \citep[see for more details, ][]{2010MathurWavelets}. The main peaks are then fitted in an iterative way using Gaussian functions. The second one calculates the auto-correlation function \citep[ACF, ][]{2014MacquillanACF}. The third one is a combination of the first two, called composite spectrum \citep[CS, ][]{2017Ceillier}. 
        
        We used the \ktwo \everest lightcurves where the transits were masked. We then corrected for instrumental problems and drifts following \citet{2011Garcia}. We removed outliers, jumps, and filled gaps using the inpainting technique \citep{2014GarciaInpainting, 2015PiresInpainting}. We finally concatenated the different campaigns (when several were available). Moreover, we divided each available campaign by its median and checked for the continuity. Alternatively, we transformed the original flux, $F$, into ppm by dividing by a triangular filter of 55 d width ($F_{55}$) for each campaign and subtracting one, i.e., $F/F_{55}$-1. The results per campaign have zero mean. Both methods provide similar results for all the stars studied in this work. The second method removes all instrumental drifts of periods longer than the filter width. To avoid any border effects at the extremes introduced by the filter, the lightcurve is extended by mirroring the beginning and the end by half of the size of the filter (27.5 d). 
    
     \subsection{Transit timing variations}
     
         We searched for evidence of additional non-transiting planets by measuring transit timing variations (TTVs) in all lightcurves in our sample.
         We took the pre-processed lightcurves (\S\ref{sec:phot}) containing one or more campaigns for each target and searched for TTVs using the Python Tool for Transit Variations \citep[\pyTTV, ][]{Korth2020}. In brief, the transits from all the planets in a system are fitted together simultaneously by modeling them with the quadratic \citet{2002MandelAgol} transit model implemented in \pytransit \citep{Parviainen2015}, and fitted for all the transit centers $t_{c}$ for all planets, impact parameter $b$ for all planets, planet-to-star radius ratio \RpRs for all planets, quadratic limb darkening coefficients $(u,v)$, and mean stellar density \rhostar. The stellar variability is modeled as a GP with a matern 3/2 kernel using \celerite \citep{Foreman-Mackey2017}. For the search for variations and periodicities in TTVs, a model--linear, quadratic or sinusoidal--is fitted and subtracted from the transit centers to obtain the TTVs. The generalized Lomb-Scargle periodogram (GLS) from \citet{2009ZechmeisterGLS} is used to search for periodicities in the TTVs, to calculate the best-fitting parameters and their uncertainties, and to test the significance of the signal. To find out which of the aforementioned models is best-suited, the Bayesian Information Criterion (BIC) is calculated. The model with the lowest BIC is chosen as the best model and the significance of the other models with respect to the best model is calculated via the $\Delta \mathrm{BIC}$. 

\section{Results and discussion}
    \label{sec:results}
    
    \subsection{Overview} 
    
\begin{figure}
        \centering
            \includegraphics[clip,trim={120 20 120 20},width=\columnwidth]{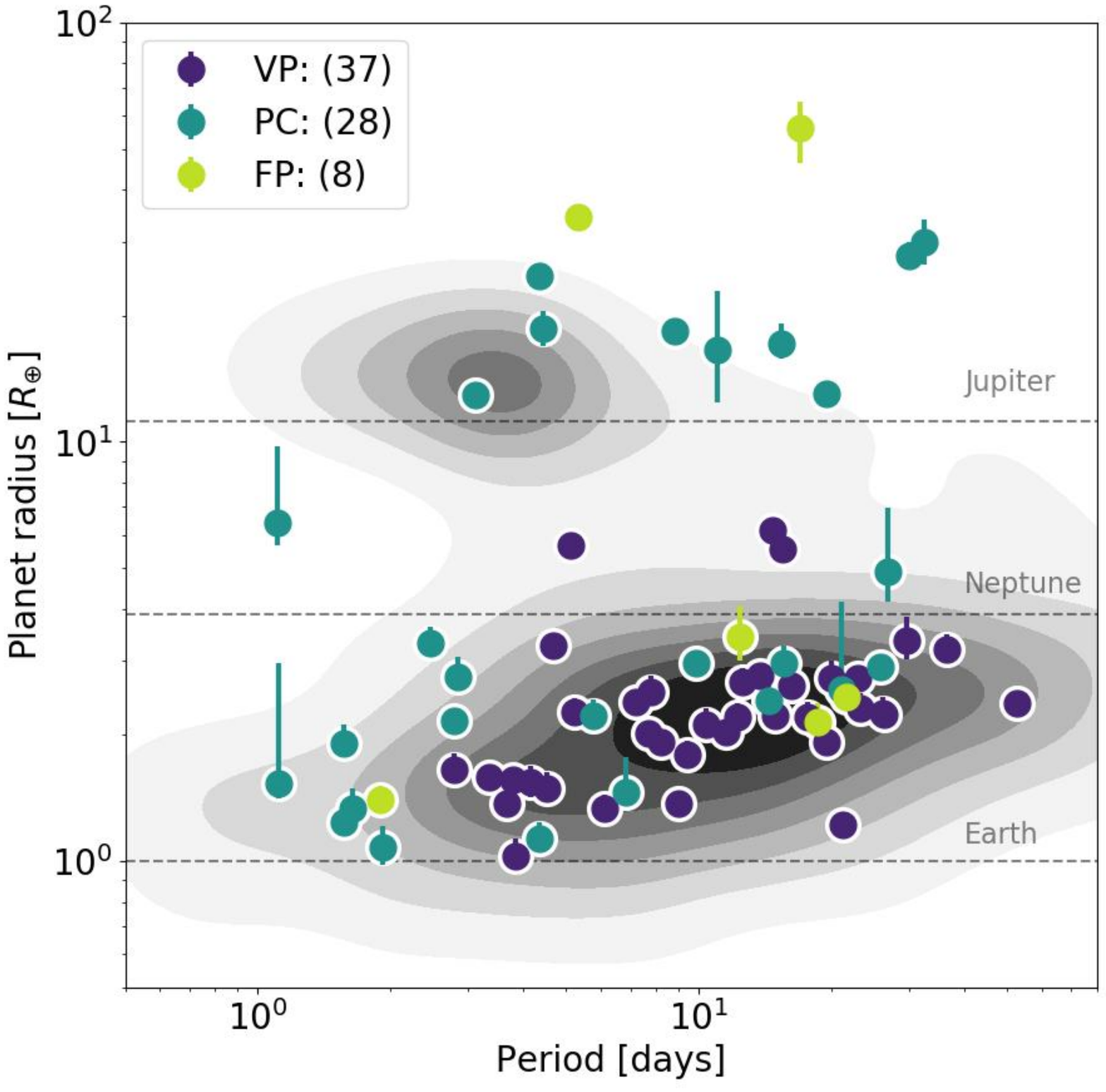}
            \caption{Distribution of validated planets (VP), planet candidates (PC), and false positives (FP) in this work, in the context of known planets (black contour lines). The VPs have a typical size of \medianRp and orbital periods between \minP and \maxP days. FPs with large radii are the result of eclipsing binary scenarios with little to no dilution from blended stars whereas FPs with small radii are plausible binaries with diluted eclipses hinted by \gaia.}
            \label{fig:PR}
    \end{figure}

    We now compare the validated planets, planet candidates, and false positives analyzed in this work to the population of known exoplanets. As shown in Figure~\ref{fig:PR}, the majority of validated planets in this work have small radii (median of \medianRp) with periods between \minP and \maxP d. 
    From Figure~\ref{fig:aper_grid}, it is clear that many of the statistically favored interpretations as planet candidates are consistent with the paucity or lack of nearby bright sources to the targets. 
    Meanwhile, the majority of the \numfp stars that are FPs have large radii and V-shaped transits (see Figure~\ref{fig:lc-grid}). Whereas those with small radii have hosts that are plausible binaries with diluted eclipses hinted at by \gaia. We also cross-matched our sample using \gaia DR2 source identifier with the \tess-\gaia v8 (TGv8) catalog \citep{2020CarilloTGv8} to determine their thin/thick disk membership probabilities. We found 10 matches, which all have >50\% membership probability in the thin disk population similar to the majority of known planetary systems and none in the thick disk.
    In the following, we discuss the unique and interesting systems in detail.
    
    \subsection{Long-period planets}
    
        The majority of the long period (\Porb>30 d) transiting planet population were discovered during the \kepler prime mission. 
        Here we report K2-185 (EPIC~211611158), a K-type star with 2 planets: a sub-Neptune with \Rp=2.4~\rearth, \Porb=52.7 d, and also a super-Earth with \Rp=1.2~\rearth and \Porb=10.6 d, already validated as \ktwo-185\,b by \citet{2018MayoC0to10}. The outer planet candidate was also detected by \citet{2019KruseC0to8} but left it as a candidate since only 2 transits were detected in C5. Here we clearly detected 3 additional transits in C16 \& C18 which finally allowed us to validate the signal to be of planetary origin. 
        Hence we found the second longest orbit with precisely measured period found by \ktwo only after EPIC~212737443\,c with \Porb=65.5 d based on 2 transits observed in C6 \citep{2019Herath}. 
        Despite its relatively long period, its equilibrium temperature, \Teq, of 477 K is still slightly higher than that of Mercury.\footnote{\url{https://ssd.jpl.nasa.gov/?planets}}. 
        We also detected a third candidate with \Porb=14.77 d present in all campaigns but we did not validate it due to its SNR=7, which is lower than our cutoff at SNR=10.
        Other known planets found by \ktwo with precisely measured periods greater than 50 d are \ktwo-118\,b \citep{2017DressingC1to7planets}, \ktwo-93\,c \citep[or HIP 41378 c, ][]{2016VanderburgK293, 2019Berardo2019}, and \ktwo-263\,b \citep{2018MortierK2263} with \Porb=50.9, 50.8, 50.8 d, respectively. K2-185 c is most similar to K2-263\,b based on its size and period.
        Similarly, K2-341 (EPIC~21197898) is a solar type star hosting another long period sub-Neptune with \Porb=36.6 d and \Teq=598 K. 
    
    \begin{figure*}
        \centering
        \includegraphics[trim={50 100 80 110},width=\textwidth]{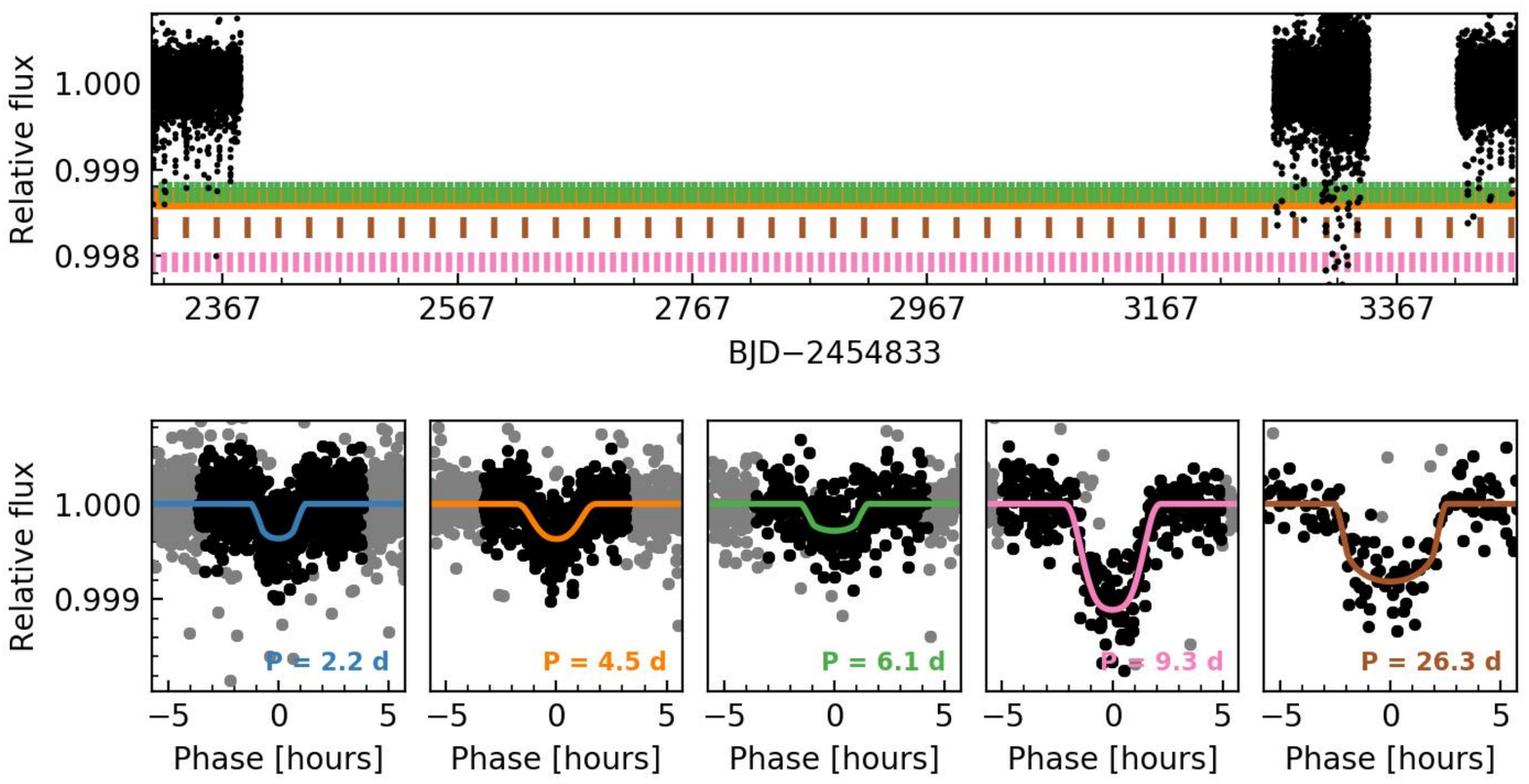}
        \caption{The \ktwo/C5, C16, and C18 lightcurves of the five planet system K2-268 (EPIC~211413752). The top panel shows the flattened lightcurves marking the locations of the individual transits, and the bottom panel shows the phase-folded lightcurves. The planets with shortest period (\Porb=2.2 d) and deepest deepest transit (depth=13 ppt, \Porb=9.3 d) have been previously validated. We detected three additional planet candidates with \Porb=4.5 d, 6.1 d, and 26.3 d which we validated after clearly detecting them in \ktwo/C5, C16 \& C18 data (see \S\ref{sec:multi} for details).}
        \label{fig:epic3752}
    \end{figure*}
    
    \subsection{Multi-planet systems} \label{sec:multi}
    
        The following briefly describes the architecture of the multi-planetary systems we validate in this work. All such systems are dynamically stable based on the criteria described in \S\ref{sec:stability}. We also did not measure rotation periods that coincide with the periods of the planets in these systems, further adding evidence to the legitimacy of the signals.
        
        \begin{itemize}
            \item K2-268 (EPIC~211413752) is a K dwarf hosting 5 detections, of which the shortest period (\Porb=2.15 d) and the deepest transit (depth=13 ppt, \Porb=9.33 d) had been validated as K2-268 b \& c by \citet{2018Livingston60planets}. 
            As previously reported by \citet{2018Livingston60planets}, there is a nearby AO companion ($r=4.7\arcsec$, $\Delta K_p$=5.9) detected with Gemini AO imaging and also with our WIYN speckle imaging and \gaia DR2. We confirm that we can indeed rule out the faint nearby star as the source of the signal following the analysis described in \S\ref{sec:dilution}. 
            Moreover, there are 3 additional candidates reported by \citet{2019KruseC0to8} which we also detected using \ktwo/C5 lightcurves (see Figure~\ref{fig:epic3752}) which we validate here after detecting each candidate in all campaigns i.e. \ktwo/C5, C16, \& C18.
            even though the combined differential photometric precision \citep[CDPP, ][]{2012ChristiansenCDPP} in C16 \& C18 (CDPP$\approx$120 ppm) are larger than in C5 (CDPP$\approx$90 ppm) for this target, which is comparable to the transit depths (0.2 ppt) of the undetected candidates. 
            \item K2-2331 (EPIC~211502222) is a solar-type star with a sub-Neptune and a super-Earth with \Rp=2.7 and 1.8~\rearth, and \Porb=23.0 and 9.4 d, respectively. The outer planet was detected by \citet{2018YuC16} while the inner planet is a new detection in this work. The planets reside on the opposite sides of the radius gap 
            \citep[1.7--2.0~\rearth; ][]{2017FultonGap, 2018VanEylenRadiusValley, 2020HardegreeUllmanK2host},
            a configuration favorable for testing the photoevaporation theory \citep{2020Owen}. 
            K2-331\,b is therefore a likely remnant core that lost its envelope either due to star-powered or core-powered mass-loss mechanisms \citep[e.g., ][]{2017OwenWu, 2019GuptaSchlichting}. 
            \item K2-352 (EPIC~251319382) is a solar-type star with 3 planet candidates with \Rp=2.2, 1.9, 1.4~\rearth and \Porb=14.87, 8.23, 3.67 d, respectively. The 2 outer planet candidates were detected by \citet{2018YuC16} in \ktwo/C16 and the innermost one is a new detection in this work. 
            \item K2-343 (EPIC~212072539) is an M dwarf that hosts a super-Earth and a sub-Neptune with \Rp=1.7~\rearth and 2.0~\rearth, and \Porb=2.8 d and 7.7 d, respectively. These candidates were initially reported by \citet{2019KruseC0to8}. Both planets have more than 1 ppt transit depths but their host star is relatively faint ($J$=12).
            \item K2-304 (EPIC~212297394) is a K dwarf with 2 planets with \Rp=2.2 and 1.5~\rearth and \Porb=5.2 and 2.3 d. Both candidates were also detected by \citet{2019KruseC0to8} and the inner planet was validated as K2-304\,b by \citet{2019HellerTLS2}.
            \item K2-348 (EPIC~212204403) is a K dwarf with 2 planets with \Rp=3.3 and 2.7~\rearth and \Porb=4.7 and 12.6 d, respectively. Both candidates were originally detected by \citet{2018YuC16} in \ktwo/C16. Both planets have more than 1 ppt transit depths and their host star is moderately bright at $J$=11 and $V$=12.5. 
        \end{itemize}
        
    \subsection{Sub-saturns around F stars}
    
        K2-333 (EPIC~211647930) and K2-334 (EPIC~211730024) are F-stars each hosting a warm sub-Saturn with radii of 6.2 and 5.7~\rearth and periods of 14.8 and 5.1 d, respectively. Apart from their rarity relative to the class of planets discussed in previous subsections, sub-Saturns are interesting because of the diversity in their core and envelope masses \citep{2017Petigura4SubSaturns}. Hence despite their similar radii, their expected masses can take a wide range of values from $\sim$6-60 \mearth. Both stars have moderate brightness (Vmag=11.5) which makes them amenable for RV follow-up, as long as the planets have massive cores that would induce detectable RV semi-amplitudes. These systems add to the small but growing number of sub-Saturns orbiting giant stars that will help to elucidate our understanding of this rare type of planetary system.
    
    \subsection{Planet candidates}
    
        We found \numpc PCs in our sample that did not meet all the criteria set in \S\ref{sec:disposition} for planet validation. 
        The majority of the PCs did not pass due to their FPP>1\%. A number of PCs have large radii above our \Rp=8\rearth cutoff which does not rule out the possibility of low-mass eclipsing binaries. Still, some remain as PCs due to the existence of nearby companions detected using \gaia or AO/speckle observations. Follow-up observations such as multi-color photometry can help to validate these as planets \citep[e.g., ][]{2020Parviainentoi519}. We also highlight below some interesting PCs due to their potential scientific impact once proven that they are indeed planets. 
        
        \subsubsection{Candidates with large radii}
        EPIC~211399359 is a K dwarf hosting a 14.7~\rearth companion on a 3.1 d orbit. Although \vespa computed an FPP$\ll$1\%, we do not validate it due to its size similar to eclipsing companions found to be false positives by \citet{2017Shporer3FP}. Traditional means to determine if the companion is indeed in the sub-stellar regime is to obtain RV measurements to constrain the companion mass. Due to its faintness (V=14.6) however, an alternative method to constrain the mass is to model the phase curve modulations \citep[e.g.,][]{2020Parviainentoi519} and potentially determine the nature of the companion. One possible complication however is that the star exhibits strong variability with \Prot=17.23 d. 
        
        All candidates with large radius (\Rp>8~\rearth) in our sample are indicated with "LR" in the notes column in Table~\ref{tab:planet}. Among the host stars with LR candidates, four stars also have a nearby companion detected in their AO/speckle images (indicated with AO in Table~\ref{tab:planet}). For example, we derived \Rp$\approx30$~\rearth for EPIC~211995398.01 after correcting for dilution due to a nearby star ($r=0.4\arcsec$, $\Delta K_p$=0.6)--unbeknown to \citet{2019KruseC0to8} who reported a radius 3 times smaller for the same candidate with \Porb=32.5 d.
        
        \subsubsection{Special case}
        EPIC~212178066 is a bright ($K_p$=6.8) F star with a sub-Neptune candidate detected and described in detail in \citet{2018YuC16}. Given its brightness, all the \gaia sources within the field of view shown in Figure~\ref{fig:aper_grid} are ruled out as potential NEBs. Moreover, DSS archival images taken in the 1950s are helpful in ruling out potential BEB scenarios given the star's large proper motion. However, since the star is saturated in \ktwo data, the transit depth derived from either \everest or \ktwosff lightcurves may not be very reliable. Despite this we report our derived companion radius of \Rp=2.9$\pm$0.7~\rearth consistent with the value reported in \citet{2018YuC16}. Assuming our derived values are correct, we attempted to run \vespa and report the values for EPIC~212178066 in Table~\ref{tab:planet} which should be taken with caution.
        
        \begin{figure*}
            \centering
            \includegraphics[clip,trim={20 200 200 50},width=0.9\textwidth]{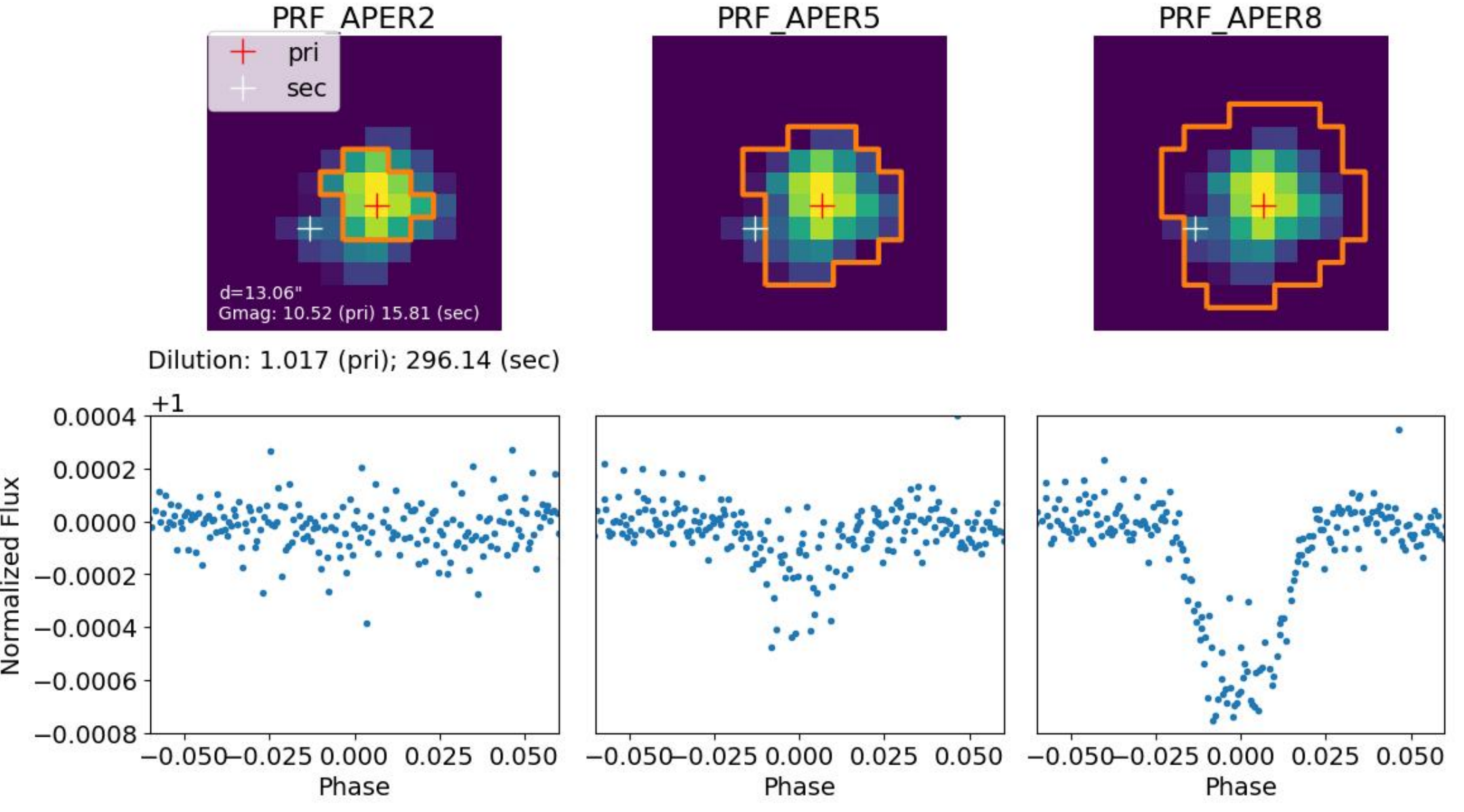}
            \caption{Pixel level multi-aperture analysis of EPIC~212099230 showing the transit depth is dependent on the \ktwosff aperture size (orange polygon) and the \ktwo signal actually originates from the fainter star (white cross) separated $10\arcsec$ southeast of the target (red cross).}
        \label{fig:PLA_9230}
        \end{figure*}
        
    \subsection{False positives}
    
        From vetting, we found \numstarsdave targets that exhibit secondary eclipses. We also found 
        EPIC~212099230 to be a false positive due to the apparent difference in transit depth as a function of aperture size (increasing from left to right) as shown in Figure~\ref{fig:PLA_9230}. The \Porb=7.1-day signal reported by \citet{2018PetiguraC5to8} and \citet{2018YuC16} using the \ktwophot pipeline lightcurves is detected only when the aperture (orange polygon) centered on the target (red cross) is large enough to include the nearby faint star (white cross). This indicates that the nearby faint star with $\Delta K_p$=5.25 separated by $10\arcsec$~is the actual source of the signal. We do not validate this signal due to the missing $JHKs$ photometry hindering us to derive the host star parameters using \isochrones. We note however that we derived a companion radius \Rp=8.3~\rearth assuming \rstar=0.99~\rsun (for \gaia DR2 source ID 665640392382991232) after taking into account the dilution caused by the brighter star.
        
        We also found 4 targets (EPIC~211335816, EPIC~211336288, EPIC~211541590, \& EPIC~212639319) that were previously reported planet candidates based on the analysis of a single \ktwo campaign but did not detect corresponding signal succeeding campaigns using both \everest and \ktwosff lightcurves. 
        We did not categorically identify these as false positives since it is possible to recover the signal using more advanced techniques which are outside the scope of this work. 
    
        \begin{figure}
            \centering
            \includegraphics[clip,trim={40 20 60 40},width=\columnwidth]{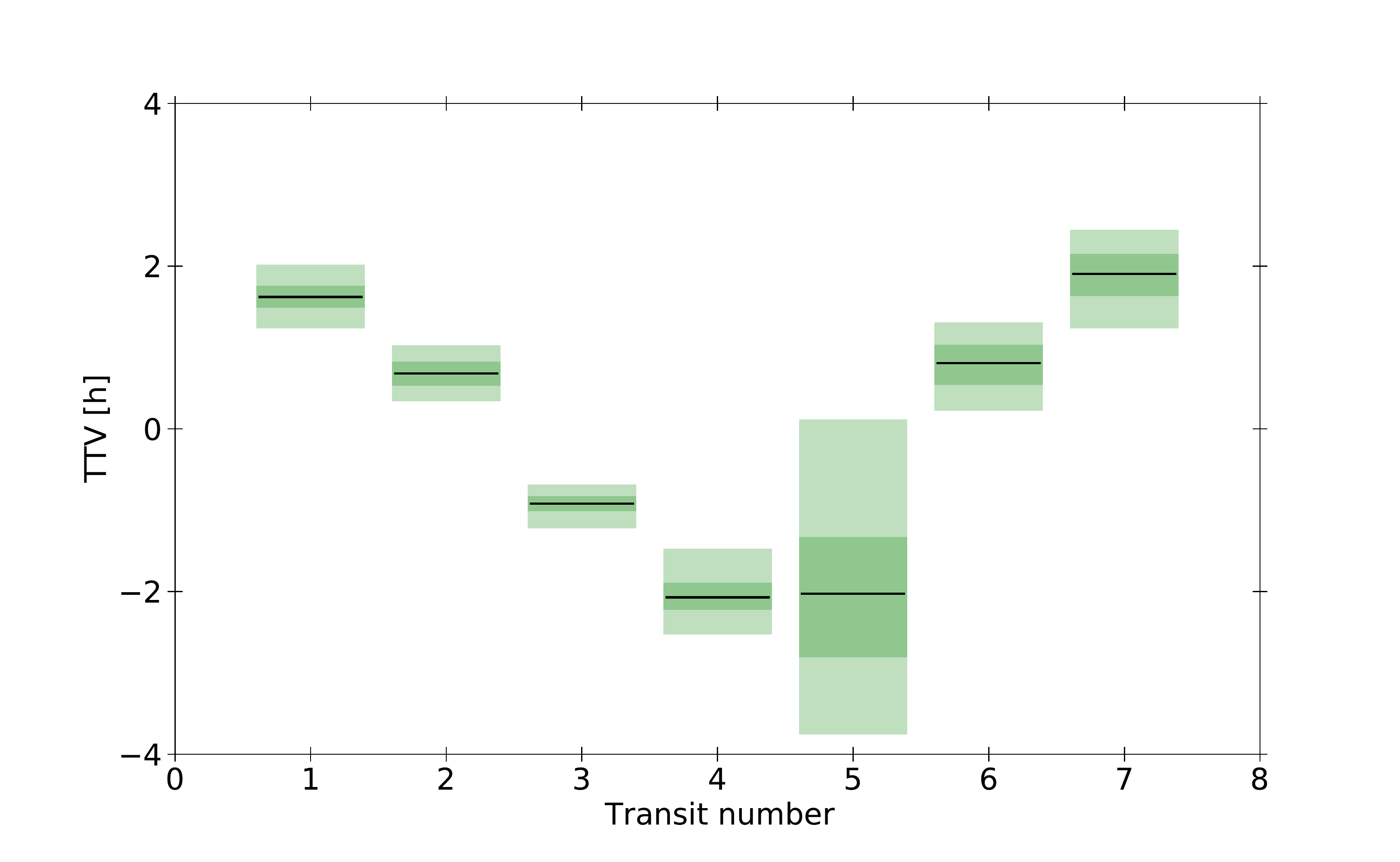}
            \caption{Transit timing variations in EPIC~212058012 where only one transiting planet is hitherto known. The black lines mark the median values and the 68\% and 99\% central posterior percentiles are indicated by the dark and light shaded area, respectively.}
            \label{fig:ttv}
        \end{figure}
        
    \subsection{Transit timing variation}
        Significant TTVs were detected for K2-342 (EPIC~212058012) where only one transiting planet EPIC~212058012.01 is currently known. As shown in Figure~\ref{fig:ttv}, we measured a peak-to-peak TTV amplitude of less than 120 minutes over the course of 8 orbits, hinting the existence of additional non-transiting planet(s) in the system. 
           
    \subsection{Stellar rotation periods}
        Reliable rotation periods are obtained for \numprot stars and tentative rotation periods for additional three stars. Their values are listed in Table~\ref{tab:prot}. Of these, nine stars host at least one VP. The rotation periods (and their harmonics) are not synchronized with the orbital periods of these planets except K2-331 (EPIC~211502222; \Prot$\sim$\Porb) and K2-350 (EPIC~212440430; \Prot$\sim$0.5\Porb).
        
\begin{figure}
            \includegraphics[clip,trim={30 30 60 10},width=\columnwidth]{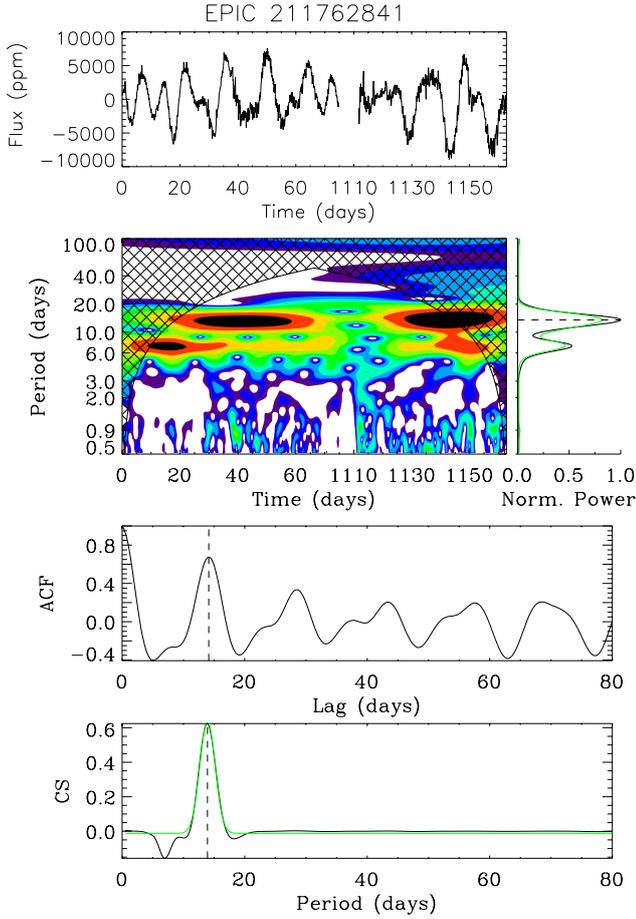}
            \caption{Stellar rotation analysis for EPIC~211762841 observed in C5 and C18 where the 3-year gap in between is removed for visualization purposes in the top panel (the lightcurve) and in the second panel (the time-period analysis). The ACF and the CS are shown in the two lower panels. The hashed region of the wavelet analysis corresponds to the region that cannot be studied with the current length of the time series. The projection of the wavelet power spectrum into the period axis is shown at the right hand side of the second panel. The vertical dashed lines correspond to the retrieved rotation period of $13.15 \pm 1.28$ d with a clear second harmonic at around 6.5 d. The green line in the CS represents the Gaussian fit to the main rotation peak. The ACF shows a ``double dip'' structure suggesting two active region are separated close to $180^{\circ}$ as explained in the text. }
            \label{fig:prot}
        \end{figure}

        Figure~\ref{fig:prot} summarizes the rotation analysis for EPIC~211762841. From various methods, we found a rotation period of $13.15 \pm 1.28 $ d with a clear second harmonic at around 6.5 d seen in the time-period plot but not in ACF. This star is classified as a ``double dip'' \citep{2014MacquillanACF} meaning that two peaks are visible in the ACF. This behaviour is typical of stars where two active regions are located around $180^{\circ}$ apart. Indeed, in the top panel two active regions are alternately seen in the first 30 days.
        
        The magnetic activity proxy, $S_{\rm ph}$ computed as the standard deviation on subseries of 5\,$\times P_{\rm rot}$ as described in \citet{2014MathurA, 2014MathurB} is also provided in the table. For the Sun, $S_{\rm ph}$ values typically range between 67.4 and 314.5\,ppm \citep{2019Mathur}, corresponding respectively to the minimum and maximum of the magnetic activity cycle. Note that among our sample of stars with measured rotation periods, all except three of them have magnetic activity levels above the Sun at maximum activity.
        
        \begin{table}
            \centering
            \caption{\Prot and $S_\mathrm{ph}$ for \numprot stars in our sample. Noted are those with tentative rotation periods (T) and those stars that host planets validated in this work (VP).}
\begin{tabular}{llll}
\hline
EPIC		&	\Prot & $S_{ph}$ & Note  \\	
\hline
211314705 &  $10.39\pm0.81$ &    $2135.89\pm122.63$ &        \\ 
211342524 &  $12.19\pm1.06$ &    $ 1681.88\pm74.84$ &        \\ 
211357309 &  $16.41\pm1.25$ &    $ 1067.07\pm47.57$ &      T \\ 
211399359 &  $17.23\pm1.67$ &    $3817.92\pm171.21$ &       \\ 
211413752 &  $24.87\pm2.64$ &    $  898.20\pm33.88$ &      VP \\ 
211502222 &   $9.98\pm0.71$ &    $  208.41\pm11.87$ &      VP \\ 
211578235 &  $14.69\pm1.13$ &    $4648.31\pm206.41$ &        \\ 
211579112 &  $31.71\pm3.60$ &    $4445.99\pm265.41$ &        \\ 
211611158 &  $17.35\pm2.87$ &    $  903.46\pm43.65$ &      VP \\ 
211645912 &  $10.83\pm0.86$ &    $6021.45\pm267.27$ &        \\ 
211731298 &  $11.85\pm0.99$ &    $4979.20\pm239.15$ &       VP \\ 
211741619 &   $9.39\pm0.89$ &    $ 1594.10\pm91.51$ &        \\
211762841 &  $13.15\pm1.28$ &    $2954.98\pm131.30$ &        \\
211796070 &  $21.66\pm2.59$ &    $  540.95\pm24.24$ &        \\
211799258 &  $26.48\pm3.62$ &    $7293.34\pm323.67$ &        \\
211800191 &   $2.19\pm0.13$ &    $  233.98\pm20.68$ &    \\
211843564 &  $18.47\pm1.49$ &    $3910.77\pm173.68$ &        \\
211886472 &  $19.79\pm1.44$ &    $  301.09\pm12.22$ &        \\
211897691 &  $11.13\pm0.83$ &    $5909.71\pm239.18$ &        \\
211965883 &  $10.61\pm0.82$ &    $3341.08\pm191.71$ &        \\
211987231 &   $7.30\pm0.54$ &    $1754.43\pm100.89$ &        \\
211988320 &  $24.53\pm1.64$ &    $  961.51\pm55.35$ &      VP  \\ 
211997641 &   $3.43\pm0.24$ &    $5469.70\pm387.92$ &    \\
212041476 &  $27.61\pm3.25$ &    $  992.11\pm55.44$ &      VP \\ 
212066407 &   $1.61\pm0.12$ &    $  129.69\pm13.58$ &    \\
212088059 &  $18.47\pm1.60$ &    $5285.10\pm188.854$ &      VP \\ 
212096658 &  $30.24\pm3.26$ &    $  625.50\pm22.11$ &        \\ 
212138198 &  $16.99\pm2.04$ &    $ 1146.94\pm51.11$ & T \\
212315941 &  $12.70\pm0.97$ &    $ 1733.76\pm96.92$ &        \\
212330265 &  $16.76\pm1.75$ &    $5356.76\pm656.14$ &        \\
212428509 &   $5.34\pm0.36$ &    $  335.27\pm18.84$ &        \\ 
212440430 &   $8.22\pm0.47$ &    $  554.50\pm25.42$ &      \\ 
212543933 &  $25.23\pm1.82$ &    $  351.01\pm19.85$ &    VP \\ 
212570977 &  $12.70\pm1.16$ &    $6097.36\pm250.89$ &       \\ 
212586030 &  $16.61\pm1.46$ &    $  365.39\pm15.24$ &        \\
212628098 &   $4.02\pm0.28$ &  $20639.54\pm1369.17$ &        \\
212703473 &  $20.49\pm1.78$ &      $958.19\pm53.76$ &        \\
212773272 &   $5.27\pm0.41$ &   $12122.11\pm675.84$ &        \\
212797028 &  $25.05\pm2.77$ &    $ 2010.69\pm82.87$ &    T   \\
251288417 &  $30.84\pm3.52$ &   $14425.97\pm802.14$ &        \\
251319382 &  $17.59\pm1.14$ &    $  485.29\pm27.24$ &   VP \\ 
\hline
\end{tabular}
            \label{tab:prot}
        \end{table}
        
    \subsection{Ephemeris improvement}
    
        The long baselines of the photometric data from \ktwo enable us to measure the orbital period very precisely. For the \nummulticamp stars observed in multiple campaigns, we measured a factor of 21$\pm$19 improvement in the precision of the period as a result of analysing targets in multiple campaigns, as compared to a single campaign (i.e C5 or C6 data set only). This orbital precision improvement is comparable to the values reported by \citet{2018Livingston60planets} which is about 10-40$\times$ for a subset of their targets observed in C5 \& C16. With the addition of C18 data, the highest orbital precision improvement we achieved is about 80$\times$ for K2-339 (EPIC~211897691), which was observed in C5, C16 and C18. 
        This factor depends on the baseline of the observation such that even a single transit observed in the more recent campaign would improve the precision. The precise ephemerides we report therefore significantly reduce the uncertainty in prediction of future times of transit, which is valuable for planning ground-based follow-up observations.

\section{Summary} \label{sec:summary}

    We analysed \numstars stars in Cancer and Virgo constellations observed by \ktwo during campaigns 5, 16, \& 18, and campaigns 6 \& 17, respectively, together with a suite of follow up observations including AO/speckle imaging, and reconnaissance spectroscopy. 
    The long baselines of the photometric data from \ktwo enabled us to measure the transit ephemeris very precisely, revisit single transit candidates identified in earlier campaigns, and search for additional transiting planets not detectable in previous works. 
    The validated planets have a median radius of \medianRp and \Porb between \minP and \maxP d, and enhance the currently known population of long period (\Porb>20 d) planets from \ktwo. 
    
    Interesting systems include (a) K2-185~c: a sub-Neptune with the second longest orbit with precisely measured period observed by \ktwo; (b) K2-333~b and K2-334~b: both sub-Saturns orbiting an F star which are interesting due to their rarity and diversity of bulk densities; (c) and several multi-planet systems in a variety of architectures, including K2-268 with 5 planets.
    We also report rotation periods between \minProt and \maxProt d in \numprot stars in our sample--\numprotwithplanets of which host planets. We also searched for TTVs and detected evidence for additional planet(s) in K2-342 where only one transiting planet is hitherto detected.
    These results show that there is still a wealth of interesting planets in \ktwo data that can be validated using minimal follow-up data taking advantage of extensive analyses presented in previous catalogs.

\section*{Acknowledgements}

    This work was carried out as part of the KESPRINT consortium. The WIYN/NESSI observations were conducted as part of an approved NOAO observing program (P.I. Livingston, proposal ID 2017A-0377). Data presented herein were obtained at the WIYN Observatory from telescope time allocated to NN-EXPLORE through the scientific partnership of the National Aeronautics and Space Administration, the National Science Foundation, and the National Optical Astronomy Observatory. 
    NESSI was funded by the NASA Exoplanet Exploration Program and the NASA Ames Research Center. NESSI was built at the Ames Research Center by Steve B. Howell, Nic Scott, Elliott P. Horch, and Emmett Quigley. The authors are honored to be permitted to conduct observations on Iolkam Du'ag (Kitt Peak), a mountain within the Tohono O'odham Nation with particular significance to the Tohono O'odham people. 
    This work is supported by JSPS KAKENHI grant numbers 18H05442, 15H02063, 17H04574, 18H01265, 18H05439, and 20K14518, and JST PRESTO grant number JPMJPR1775. 
    This work is also supported by a NASA WIYN PI Data Award, administered by the NASA Exoplanet Science Institute. MF gratefully acknowledges the support of the Swedish National Space Agency (DNR 65/19, 174/18). 
    S.M. acknowledges 674 support by the Spanish Ministry with the Ramon y Cajal fellowship number RYC-2015-17697. R.A.G. acknowledges the support of the CNES PLATO grant. JK gratefully acknowledge the support of the Swedish National Space Agency (SNSA; DNR 2020-00104)
    This research has made use of the Exoplanet Followup Observation Program website, which is operated by the California Institute of Technology, under contract with the National Aeronautics and Space Administration under the
    Exoplanet Exploration Program.
    This research has made use of the NASA Exoplanet Archive, which is operated by the California Institute of Technology, under contract with the National Aeronautics and Space Administration under the Exoplanet Exploration
    Program.
    This work has made use of data from the European Space Agency (ESA) mission Gaia (https://www.cosmos. esa.int/gaia), processed by the Gaia Data Processing and Analysis Consortium (DPAC, https://www.cosmos.esa.int/web/gaia/dpac/consortium). Funding for the DPAC has been provided by national institutions, in particular the institutions participating in the Gaia Multilateral Agreement. 
    The Digitized Sky Surveys were produced at the Space Telescope Science Institute under U.S. Government grant NAG W-2166. The images of these surveys are based on photographic data obtained using the Oschin Schmidt Telescope on Palomar Mountain and the UK Schmidt Telescope. The plates were processed into the present compressed digital form with the permission of these institutions.
    KWFL acknowledges the support by DFG grants RA714/14-1 within the DFG Schwerpunkt SPP 1992, Exploring the Diversity of Extrasolar Planets.
    The simulations were run on the CfCA Calculation Server at NAOJ. A.A.T. acknowledges support from JSPS KAKENHI Grant Numbers 17F17764 and 17H06360.

\section*{Data availability}
    The data underlying this article were accessed from MAST (\url{https://archive.stsci.edu/hlsp/}) with specific links mentioned in the article. 
    The tables presented in this work will also be made available at the CDS (\url{http://cdsarc.u-strasbg.fr/}).



\bibliographystyle{mnras}
\bibliography{ref.bib} 




\appendix
Affiliations \newline
$^{1}$Department of Astronomy, University of Tokyo, 7-3-1 Hongo, Bunkyo-ku, Tokyo 113-0033, Japan\\
$^{2}$Department of Astronomy and McDonald Observatory, University of Texas at Austin, 2515 Speedway, Stop C1400, Austin TX 78712, USA\\
$^{3}$Center for Planetary Systems Habitability, The University of Texas at Austin, 2305 Speedway Stop C1160, Austin, TX 78712-1692, USA\\
$^{4}$Department of Earth and Planetary Sciences, Tokyo Institute of Technology, 2-12-1 Ookayama, Meguro-ku, Tokyo 152-8551, Japan\\
$^{5}$AIM, CEA, CNRS, Universite Paris-Saclay, Universite Paris Diderot,  Sorbonne Paris Cite, F-91191 Gif-sur-Yvette, France\\
$^{6}$Instituto de Astrofisica de Canarias (IAC), Tenerife, La Laguna E-38205, Spain\\
$^{7}$Departamento de Astrofisica, Universidad de La Laguna, Tenerife, La Laguna E-38206, Spain\\
$^{8}$Center for Astronomy and Astrophysics, Technical University Berlin, Hardenbergstr. 36, Berlin D-10623, Germany\\
$^{9}$Department of Space, Earth and Environment, Astronomy and Plasma Physics, Chalmers University of Technology, Gothenburg SE-412 96, Sweden\\
$^{10}$Department of Earth Science and Astronomy, College of Arts and Sciences, The University of Tokyo, 3-8-1 Komaba, Meguro-ku, Tokyo 153-8902, Japan\\
$^{11}$Division of Geological and Planetary Sciences, California Institute of Technology, 1200 East California Blvd, Pasadena CA 91125, USA\\
$^{12}$Instituto Universitario de Ciencias y Tecnologias Espaciales de Asturias (ICTEA), C.Independencia 13, E-33004, Spain\\
$^{13}$Leiden Observatory, Leiden University, Leiden NL-2333CA, the Netherlands\\
$^{14}$Onsala Space Observatory, Department of Space, Earth and Environment, Chalmers University of Technology, Onsala SE-439 92, Sweden\\
$^{15}$Komaba Institute for Science, The University of Tokyo, 3-8-1 Komaba, Meguro, Tokyo 153-8902, Japan\\
$^{16}$Dipartimento di Fisica, Universita di Torino, Via P. Giuria 1, Torino I-10125, Italy\\
$^{17}$Astronomical Institute, Czech Academy of Sciences, Fricova 298, Ondrejov CZ-25165, Czech Republic\\
$^{18}$Division of Geological and Planetary Sciences, California Institute of Technology, 1200 East California Blvd, Pasadena CA 91125, USA\\
$^{19}$Department of Astronomy, University of Maryland, College Park, College Park MD 20742, USA\\
$^{20}$Department of Astronomy, University of California Berkeley, Berkeley CA 94720, USA\\
$^{21}$Astrobiology Center, 2-21-1 Osawa, Mitaka, Tokyo 181-8588, Japan\\
$^{22}$Japan Science and Technology Agency, PRESTO, 3-8-1 Komaba, Meguro, Tokyo 153-8902, Japan\\
$^{23}$Department of Space and Climate Physics, Mullard Space Science Laboratory, Holmbury St. Mary, Dorking, Surrey RH5 6NT, UK\\
$^{24}$National Astronomical Observatory of Japan, 2-21-1 Osawa, Mitaka, Tokyo 181-8588, Japan\\


\section{\vespa likelihoods}
    \begin{table*}
    \scriptsize
    \caption{\vespa likelihoods. \label{tab:fpp}} 
    \begin{tabular}{lrrrrrrrr}
\hline
ID &  $\mathrm{L_{beb}}^{a}$ &  $\mathrm{L_{beb}Px2}^{a}$ &  $\mathrm{L_{eb}}^{b}$ &  $\mathrm{L_{eb}Px2}^{b}$ &  $\mathrm{L_{heb}}^{c}$ &  $\mathrm{L_{heb}Px2}^{c}$ &  $\mathrm{L_{pl}}^{d}$ &      FPP \\
\hline
211314705.01 &     0.00e+00 &         0.00e+00 &    2.80e-05 &        1.31e-06 &     2.45e-06 &         3.75e-07 &    7.26e-03 & 4.41e-03 \\
211357309.01 &     6.77e-04 &         3.47e-04 &    0.00e+00 &        0.00e+00 &     0.00e+00 &         0.00e+00 &    2.10e-02 & 4.65e-02 \\
211383821.01 &     3.74e-04 &         1.13e-03 &    4.41e-04 &        1.89e-04 &     7.57e-18 &         3.85e-05 &    6.89e-02 & 3.06e-02 \\
211399359.01 &     0.00e+00 &         0.00e+00 &    2.58e-17 &        8.80e-15 &     1.05e-96 &         6.95e-47 &    4.45e-02 & 1.99e-13 \\
211401787.01 &     4.08e-07 &        2.78e-131 &    1.40e-06 &        2.31e-17 &     2.15e-07 &         9.35e-11 &    9.88e-03 & 2.04e-04 \\
211413752.03 &     7.72e-08 &         4.13e-15 &    1.51e-05 &        2.92e-07 &     2.26e-13 &         5.41e-19 &    5.12e-03 & 3.01e-03 \\
211413752.04 &     3.60e-05 &         2.84e-06 &    8.14e-04 &        5.88e-05 &     7.47e-06 &         3.39e-06 &    1.87e-02 & 4.69e-02 \\
211413752.05 &     9.13e-07 &         6.85e-14 &    8.74e-08 &        1.46e-08 &     4.86e-07 &         1.43e-09 &    3.44e-03 & 4.37e-04 \\
211439059.01 &     1.55e-06 &         1.74e-19 &    3.50e-06 &        1.67e-07 &     2.87e-07 &         1.82e-08 &    1.49e-03 & 3.69e-03 \\
211490999.01 &     2.10e-05 &         4.38e-13 &    3.53e-04 &        1.74e-08 &     3.10e-06 &         1.30e-21 &    2.91e-02 & 1.28e-02 \\
211502222.01 &    3.04e-136 &         0.00e+00 &    3.43e-17 &        1.92e-18 &     0.00e+00 &        1.30e-146 &    1.38e-06 & 2.62e-11 \\
211502222.02 &     2.57e-06 &         6.71e-09 &    1.46e-05 &        1.66e-07 &     6.17e-07 &         2.63e-09 &    5.83e-03 & 3.07e-03 \\
211578235.01 &     2.46e-03 &         2.38e-04 &    7.38e-03 &        2.44e-04 &     7.65e-04 &         1.62e-05 &    4.45e-03 & 7.14e-01 \\
211579112.01 &     3.54e-11 &         6.45e-12 &    5.52e-06 &        1.58e-07 &     2.97e-07 &         7.04e-08 &    2.23e-03 & 2.70e-03 \\
211611158.02 &     1.15e-05 &         1.11e-29 &    6.40e-05 &        1.11e-09 &     5.99e-20 &         8.72e-17 &    4.43e-03 & 1.68e-02 \\
211647930.01 &     1.01e-09 &         2.28e-88 &    5.96e-24 &        8.14e-48 &     3.08e-08 &         4.80e-11 &    4.27e-03 & 7.46e-06 \\
211694226.01 &     6.88e-04 &         2.08e-03 &    2.16e-04 &        7.91e-05 &     1.62e-05 &         2.09e-05 &    3.18e-02 & 8.88e-02 \\
211730024.01 &     3.12e-22 &        1.11e-100 &    8.89e-08 &        9.20e-10 &     3.10e-15 &         2.82e-11 &    1.15e-02 & 7.83e-06 \\
211743874.01 &     5.09e-07 &         1.15e-30 &    5.97e-06 &        8.43e-08 &     7.55e-08 &         7.38e-10 &    4.01e-03 & 1.65e-03 \\
211762841.01 &     1.63e-03 &         6.82e-03 &    1.69e-03 &        3.98e-04 &     1.39e-05 &         6.48e-05 &    7.18e-02 & 1.29e-01 \\
211763214.01 &     0.00e+00 &         0.00e+00 &    1.90e-21 &        1.40e-15 &     1.41e-17 &         1.79e-36 &    9.87e-04 & 1.43e-12 \\
211770696.01 &     8.75e-08 &         5.38e-80 &   1.27e-131 &        1.29e-53 &     8.17e-12 &         8.61e-25 &    1.14e-03 & 7.65e-05 \\
211779390.01 &     2.91e-05 &         4.57e-06 &    2.54e-05 &        3.43e-07 &     5.66e-07 &         1.72e-06 &    1.63e-02 & 3.77e-03 \\
211796070.01 &     4.00e-12 &         1.16e-12 &    1.90e-12 &        2.30e-12 &     6.88e-13 &         2.19e-12 &    8.81e-49 & 1.00e+00 \\
211797637.01 &     7.91e-04 &         5.53e-03 &    8.93e-04 &        3.46e-04 &     4.11e-08 &         1.09e-06 &    1.45e-02 & 3.43e-01 \\
211799258.01 &     0.00e+00 &         0.00e+00 &    5.25e-03 &        3.84e-03 &     9.89e-04 &         3.18e-04 &    4.17e-03 & 7.14e-01 \\
211800191.01 &     1.44e-05 &         0.00e+00 &    1.21e-03 &        3.33e-04 &     1.92e-05 &         2.77e-06 &    1.96e-02 & 7.47e-02 \\
211817229.01 &     1.19e-03 &         3.81e-03 &    2.41e-06 &        7.41e-06 &     3.73e-06 &         2.85e-05 &    4.89e-02 & 9.36e-02 \\
211897691.01 &     1.14e-05 &         4.12e-05 &    1.08e-05 &        2.96e-05 &     7.92e-07 &         3.94e-07 &    1.18e-03 & 7.41e-02 \\
211897691.02 &     3.33e-05 &         6.57e-05 &    3.36e-05 &        2.51e-05 &     1.77e-06 &         1.33e-06 &    1.42e-03 & 1.02e-01 \\
211923431.01 &     6.27e-09 &         7.11e-10 &    3.97e-07 &        2.96e-15 &     2.31e-07 &         1.69e-09 &    2.17e-03 & 2.94e-04 \\
211939692.04 &     3.42e-08 &         4.74e-20 &    1.28e-04 &        2.17e-06 &     4.52e-07 &         5.82e-08 &    2.12e-05 & 8.60e-01 \\
211965883.01 &     4.69e-05 &         8.29e-13 &    4.74e-04 &        1.19e-04 &     4.31e-05 &         4.71e-06 &    2.75e-03 & 2.00e-01 \\
211978988.01 &     2.18e-06 &        3.24e-123 &    1.79e-05 &        5.19e-07 &     1.34e-12 &         3.46e-27 &    1.74e-02 & 1.18e-03 \\
211987231.01 &     8.84e-86 &        3.45e-303 &    9.06e-03 &        2.37e-04 &     2.93e-04 &         2.53e-06 &    1.01e-04 & 9.90e-01 \\
211995398.01 &     1.42e-07 &         3.03e-09 &    3.20e-09 &        5.94e-13 &     6.17e-59 &         3.14e-42 &    2.75e-03 & 5.40e-05 \\
211997641.01 &     0.00e+00 &         0.00e+00 &    0.00e+00 &        2.29e-02 &     3.61e-04 &         5.31e-04 &    1.14e-04 & 9.95e-01 \\
212006318.01 &     8.30e-06 &         6.26e-13 &    8.15e-05 &        1.11e-06 &     4.57e-06 &         4.11e-09 &    4.56e-03 & 2.05e-02 \\
212009150.01 &     1.14e-05 &         4.58e-06 &    2.68e-06 &        4.60e-14 &     6.37e-08 &         5.92e-08 &    1.72e-04 & 9.81e-02 \\
212040382.01 &     3.30e-99 &         0.00e+00 &    3.15e-04 &        2.79e-22 &     3.86e-18 &         1.92e-13 &    1.82e-02 & 1.70e-02 \\
212041476.01 &     2.07e-05 &         1.45e-06 &    7.51e-06 &        1.49e-07 &     1.17e-05 &         4.09e-07 &    7.89e-02 & 5.31e-04 \\
212058012.01 &     2.24e-06 &         5.49e-32 &    2.09e-04 &        1.84e-09 &     9.24e-09 &         8.13e-13 &    3.86e-02 & 5.43e-03 \\
212072539.01 &     1.84e-07 &         2.30e-43 &   1.05e-195 &       7.94e-112 &     1.18e-55 &         6.15e-41 &    2.07e-03 & 8.88e-05 \\
212072539.02 &     2.58e-04 &         1.01e-03 &    1.91e-04 &        1.08e-04 &     4.41e-05 &         2.16e-05 &    3.67e-02 & 4.27e-02 \\
212081533.01 &     0.00e+00 &         1.62e-54 &    1.63e-17 &        6.43e-21 &     5.45e-06 &         5.58e-14 &    3.44e-02 & 1.59e-04 \\
212088059.01 &     2.21e-05 &         1.36e-34 &    7.45e-08 &        5.42e-08 &     4.04e-08 &         2.35e-08 &    2.47e-02 & 9.00e-04 \\
212132195.01 &     0.00e+00 &         0.00e+00 &    1.13e-06 &        1.88e-07 &     2.50e-18 &         3.77e-15 &    1.38e-02 & 9.57e-05 \\
212161956.01 &     8.03e-06 &         4.74e-08 &    1.97e-05 &        1.08e-06 &     5.50e-06 &         7.07e-08 &    1.55e-02 & 2.22e-03 \\
212178066.01 &     1.00e-07 &         1.32e-23 &    5.81e-06 &        7.82e-08 &     4.83e-08 &         2.22e-12 &    1.55e-03 & 3.88e-03 \\
212204403.01 &     1.76e-04 &         1.35e-36 &    1.25e-06 &        1.74e-11 &     1.16e-27 &         2.24e-27 &    7.94e-02 & 2.23e-03 \\
212204403.02 &     3.72e-06 &         4.18e-23 &    2.90e-06 &        6.37e-08 &     1.64e-14 &         1.01e-16 &    6.10e-03 & 1.09e-03 \\
212278644.01 &     1.43e-03 &         6.70e-04 &    2.79e-04 &        6.85e-06 &     5.12e-05 &         2.12e-07 &    0.00e+00 & 1.00e+00 \\
212297394.01 &     3.85e-05 &         2.54e-39 &    4.22e-06 &        1.50e-06 &     8.13e-07 &         1.23e-09 &    1.54e-02 & 2.92e-03 \\
212420823.01 &     1.75e-06 &         3.52e-37 &    3.18e-06 &        7.20e-10 &     8.77e-08 &         4.70e-09 &    3.19e-03 & 1.57e-03 \\
212428509.01 &     5.81e-05 &         0.00e+00 &    1.71e-02 &        1.19e-03 &     9.56e-04 &         2.22e-06 &    1.27e-04 & 9.93e-01 \\
212435047.01 &     9.99e-04 &        8.63e-218 &    1.40e-03 &        2.12e-04 &     0.00e+00 &         3.40e-23 &    1.70e-02 & 1.33e-01 \\
212440430.01 &     2.02e-05 &         4.82e-08 &    2.45e-05 &        6.33e-08 &     1.35e-08 &         1.38e-12 &    1.42e-02 & 3.14e-03 \\
212440430.02 &     2.90e-05 &         1.30e-15 &    4.73e-06 &        3.79e-08 &     1.36e-06 &         1.02e-08 &    7.97e-03 & 4.39e-03 \\
212495601.01 &     1.37e-06 &         5.23e-42 &    3.01e-06 &        8.27e-09 &     2.94e-15 &         6.04e-13 &    0.00e+00 & 1.00e+00 \\
212543933.01 &     1.29e-05 &         3.37e-10 &    1.12e-05 &        7.01e-07 &     1.11e-07 &         1.19e-09 &    1.73e-02 & 1.43e-03 \\
212570977.01 &     2.74e-56 &         0.00e+00 &    1.98e-03 &        7.06e-12 &     5.32e-05 &         1.08e-11 &    8.40e-03 & 1.95e-01 \\
212587672.01 &     5.03e-06 &         3.45e-46 &    6.50e-05 &        9.03e-07 &     4.69e-08 &         3.14e-11 &    9.20e-03 & 7.66e-03 \\
212628098.01 &     3.11e-52 &         0.00e+00 &    1.21e-03 &        3.03e-03 &     3.26e-04 &         2.90e-04 &    3.08e-03 & 6.12e-01 \\
212628477.01 &     9.23e-06 &         3.94e-85 &    5.83e-04 &        7.63e-04 &     1.17e-04 &         1.68e-05 &    1.32e-03 & 5.31e-01 \\
212634172.01 &     6.54e-05 &         3.36e-03 &    2.67e-03 &        9.14e-04 &     3.52e-04 &         3.04e-04 &    3.25e-02 & 1.91e-01 \\
212661144.01 &     2.30e-04 &        2.03e-127 &    3.32e-04 &        1.35e-05 &     7.72e-05 &         8.01e-06 &    6.53e-02 & 1.00e-02 \\
212690867.01 &     1.39e-03 &         7.96e-04 &    3.02e-04 &        4.80e-05 &     4.50e-05 &         6.43e-06 &    1.79e-03 & 5.92e-01 \\
212797028.01 &     2.97e-05 &        5.45e-283 &    5.44e-05 &        1.97e-05 &     1.19e-04 &         1.83e-06 &    1.81e-04 & 5.53e-01 \\
251319382.01 &     6.66e-07 &         1.22e-11 &    4.27e-06 &        7.25e-12 &     3.74e-08 &         3.30e-14 &    2.47e-02 & 2.01e-04 \\
251319382.02 &     7.62e-07 &        5.55e-157 &    3.53e-05 &        4.27e-10 &     1.30e-08 &         3.48e-30 &    2.72e-02 & 1.32e-03 \\
251319382.03 &     1.69e-05 &         9.16e-06 &    2.07e-04 &        4.27e-05 &     7.05e-07 &         1.18e-06 &    6.40e-03 & 4.15e-02 \\
251554286.01 &     2.94e-06 &        2.03e-276 &    3.74e-05 &        4.43e-08 &     3.86e-11 &         5.34e-21 &    1.27e-02 & 3.16e-03 \\
\hline
\end{tabular}
        \vspace{1ex}
        {\raggedright 
        \newline
        (a) Likelihood that the signal is due to a BEB at the measured period or twice that. \newline
        (b) Likelihood that the signal is due to an eclipsing binary at the measured period or twice that. \newline
        (c) Likelihood that the signal is due to a hierarchical star system with an eclipsing component at the measured period or twice that. \newline
        (d) Likelihood that the signal is due to a planet.
         \par}
    \end{table*}

\bsp	
\label{lastpage}
\end{document}